\documentclass{egpubl} 
\usepackage{eg2025}
 
\STAR                   
\usepackage[T1]{fontenc}
\usepackage{dfadobe}  

\usepackage{cite}  

\usepackage{tikz}
\usetikzlibrary{calc}

\usetikzlibrary{patterns}
\usepackage{amsmath}
\usepackage{amssymb}
\usepackage{booktabs}
\usepackage{float}

\usepackage{lscape} 
\usepackage{rotating}
\usepackage{tabularx}
\usepackage{xcolor}
\usepackage{amssymb}
\usepackage{scalerel}
\usepackage{stackengine}

\usepackage{pifont}

\usepackage[table]{xcolor}
\definecolor{lightred}{rgb}{1, 0.92, 0.92}
\definecolor{lightblue}{rgb}{0.88, 0.95, 1}
\definecolor{lightyellow}{rgb}{1, 1, 0.9}
\definecolor{lightgreen2}{rgb}{0.9, 1, 0.9}
\definecolor{lightpink}{rgb}{1, 0.94, 0.95}
\definecolor{lightpurple}{rgb}{0.9, 0.8, 0.95}
\definecolor{lightgray}{rgb}{0.96, 0.96, 0.96}
\usepackage{booktabs} 
\usepackage{multirow} 
\usepackage{array}    
\usepackage{ulem}

\usepackage{placeins} 

\usepackage[table,xcdraw]{xcolor} 

\usepackage{tikz}
\usetikzlibrary{shapes.geometric}

\usepackage{tikz}
\usetikzlibrary{shapes.geometric}

\usepackage[most]{tcolorbox}
\newtcbox{\mymath}[1][]{%
    nobeforeafter, math upper, tcbox raise base,
    enhanced, colframe=blue!30!black,
    colback=blue!30, boxrule=1pt,
    #1}
\usepackage{empheq} 
\definecolor{lightgreen}{HTML}{90EE90}
\usepackage{adjustbox}

\usepackage{bm}

\definecolor{pinkish}{HTML}{F5B0B0}
\definecolor{blueish}{HTML}{DAE3F3}
\definecolor{lightpurple}{HTML}{F3DFF5}
\definecolor{lightgreen}{HTML}{E2F0D9}
\definecolor{lightgray}{HTML}{E7E6E6}
\definecolor{lightyellow}{HTML}{FFF2CC}
\BibtexOrBiblatex
\electronicVersion
\PrintedOrElectronic

\newcommand{\neuralsketchmethods}{\textsc{DS-3DM}}
\newcommand{\designspace}{\textsc{MORPHEUS}}

\title[Deep Sketch-Based 3D Modeling: A Survey\\]%
      {Deep Sketch-Based 3D Modeling: A Survey\\}

\author[A. Tono et al.]
{\parbox{\textwidth}{\centering Alberto Tono$^{1,2,*}$\orcid{0000-0003-0066-7406}
        $\quad$ Jiajun Wu$^{1}$\orcid{0000-0002-4176-343X}
        $\quad$ Gordon Wetzstein$^{1}$\orcid{0000-0002-3450-0447}
        $\quad$ Iro Armeni$^{1}$\orcid{0000-0002-4230-5916}
        $\quad$ Hariharan Subramonyam$^{1}$\orcid{0000-0002-3450-0447}
        $\quad$ James Landay$^{1}$\orcid{0000-0003-1520-8894}
        $\quad$ Martin Fischer$^{1}$\orcid{0000-0002-5071-017X} 
        }
        \\
{\parbox{\textwidth}{\centering $^1$ Stanford University, USA, atono, jiajunw, gordonwz, iarmeni, harihars, landay, fischer@stanford.edu\\ 
$^2$ Computational Design Institute, USA, alberto.tono@cd.institute\\
$^*$ Corresponding author: atono@stanford.edu}}
\vspace{-30pt}
}

\begin{document}

\teaser{
 \includegraphics[width=0.9\linewidth]{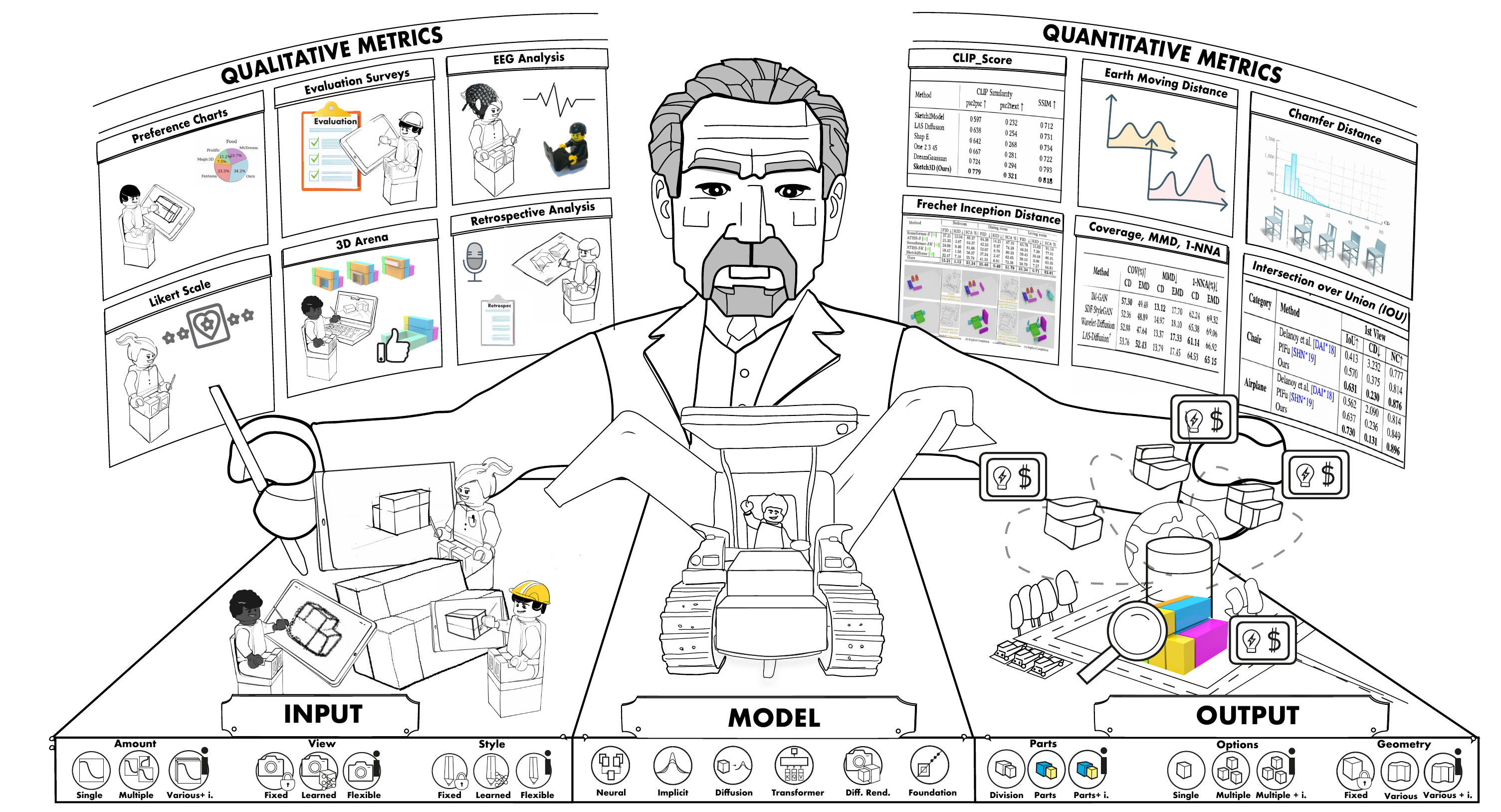}
 \centering
  \caption{Illustrates our design space $\designspace$ organized around the Input-Model-Output (IMO) framework used to survey deep sketch-based 3D modeling ($\neuralsketchmethods$) methods. The Input is categorized into the amount of sketches, their type of viewpoint, and sketch styles with their respective subcategories. The Model is categorized based on the architecture type: neural networks, implicit functions, diffusion models, transformers, differentiable renderers, and foundation models. The Output is categorized by the 3D output compositions, such as part-based semantic, number of options, and geometric topology as well as their corresponding subcategories. The background displays the qualitative and quantitative metrics used to evaluate $\neuralsketchmethods$. The figure encapsulates the paper's goal: enabling a user-controlled process driven by human-centric metrics to support informed design.}
\label{fig:teaser}
}
\newcommand{\undefinedpagestyle}{}
\maketitle
\begin{abstract}
   In the past decade, advances in artificial intelligence have revolutionized sketch-based 3D modeling, leading to a new paradigm known as Deep Sketch-Based 3D Modeling ($\neuralsketchmethods$). $\neuralsketchmethods$ offers data-driven methods that address the long-standing challenges of sketch abstraction and ambiguity.
 $\neuralsketchmethods$ keeps humans at the center of the creative process by enhancing the flexibility, usability, faithfulness, and adaptability of sketch-based 3D modeling interfaces. This paper contributes a comprehensive survey of the latest \neuralsketchmethods\ within a novel design space: \textbf{MORPHEUS}. Built upon the Input-Model-Output (IMO) framework, \textbf{MORPHEUS} categorizes \textbf{M}odels outputting \textbf{O}ptions of 3D \textbf{R}epresentations and \textbf{P}arts, derived from \textbf{H}uman-inputs (varying in quantity and modality), and \textbf{E}valuated across diverse \textbf{U}ser-views and \textbf{S}tyles. Throughout $\designspace$ we highlight limitations and identify opportunities for interdisciplinary research in Computer Vision, Computer Graphics, and Human-Computer Interaction, revealing a need for controllability and information-rich outputs. These opportunities align design processes more closely with user' intent, responding to the growing importance of user-centered approaches.

\ccsdesc[500]{Computing methodologies~Computer vision representations}
\ccsdesc[500]{Human-centered computing~User interface toolkits}
\ccsdesc[500]{Computing methodologies~Graphics input devices}

\printccsdesc   
\end{abstract}  
\section{Introduction}
\label{introduction}
In the \href{https://youtu.be/tHrtb4y4SF8}{iconic scene} from \textit{The Matrix Reloaded} movie, when Neo (Thomas A. Anderson) meets `The Architect,' the camera transitions from a panoramic view of the galaxy to a pencil. This cinematic moment encapsulates the essence of sketch modeling: a universal method capable of translating the Architect's intent from the physical world to the digital one. Indeed, 2D sketches are a simple yet effective tool for rapidly communicating complex and abstract concepts \cite{sketchrnn}. Today, the challenge lies in bridging the gap between the physical and digital medium, especially for 3D modeling, by removing intrinsic sketch ambiguities and ensuring that the user's intent is accurately captured and conveyed to the final 3D digital representation. Addressing this gap will be a significant step toward democratizing the design process, thereby empowering individuals to articulate their concepts through sketches. A critical component of this effort involves evaluating the alignment between the user's intent and the output. This alignment requires the development and iteration of human-centric metrics that assess how well the generated 3D models reflect the user's original vision. Key questions arise: How can we quantify the fidelity of this alignment? What specific metrics should be used to evaluate both the semantic richness and geometric accuracy of the output? How do we ensure that the information embedded in the 3D model, such as annotations, context, or material properties, enhances usability and supports decision-making? Addressing these questions is essential for ensuring an intuitive and effective transition from physical sketches to digital 3D representations while fostering a more inclusive and user-centered approach to 3D modeling. To support this endeavor, we provide insights into these metrics and potential research directions in Section~\ref{geometry}, and in Table~\ref{tab:evaluation_metrics}.

\begin{figure}
    \centering
    \includegraphics[width=\linewidth]{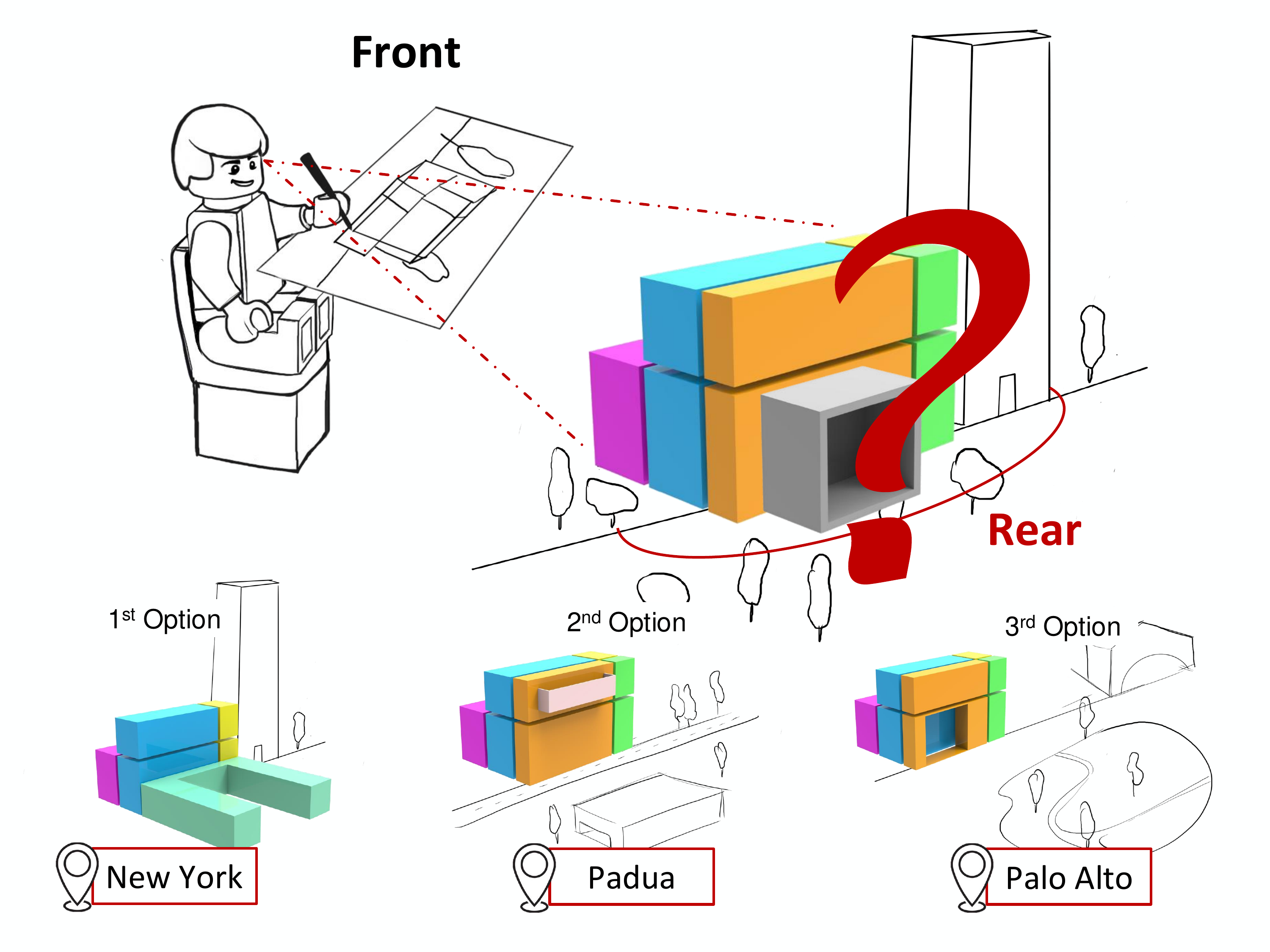}
    \caption{The initial sketch in the architectural design process contains partial massing, volumetric, and geometric information \cite{TONOVitruvio22}. This early representation is incomplete because it represents a building only from its front view, conveying only partial information lacking a comprehensive 3D understanding. This single perspective leaves out other buildings' details, particularly the rear \cite{Front2Back2020}. Moreover, this missing information is influenced by \textit{extrinsic} and \textit{intrinsic} factors. Extrinsic factors rely on contextual information such as location, surrounding neighborhood, and local climate. For example, the building will have different shapes if it is designed in New York, Padua, or Palo Alto. Intrinsic factors, such as building typology (office, school, house), design constraints, materials, textures, and appearance reside in the architect's vision of how to translate client requirements into physical spaces.}
    \label{fig:PartialInformation}
\end{figure}

Sketch-Based Interfaces for Modeling (SBIM) \cite{landay2001sketchinterface, landay1995interactive, landay1995just, DigitalClay2000, sutherland1963sketchpad, teddy1999, SKETCH1996Zeleznik,directmodeling2002,sketchmodeling2009, dingsketchmodelingsurvey2016, bonnici2019, SVRsurvey, li2025meshpad} predominantly utilize 2D hand-drawn sketches as input, with limited exploration of 3D sketching interfaces \cite{3D_VR_luo20233d, gu2025vrsketch2gaussian3dvrsketch, XIAO2024VResinspatialmemory}. The input modalities include hand-drawn doodles (abstract drawings made by amateur designers in a few seconds), scaffolding-based sketches \cite{How2SketchHennessey16}, one-, two-, and three-point perspectives \cite{TONOVitruvio22}, axonometric sketches \cite{Free2CAD}, freehand sketches \cite{freehandreconstruction}, scribbles \cite{sanghi2023sketchashapezeroshotsketchto3dshape}, contour-based sketches \cite{Contourbased3DModelingAobo20}, and design sketches \cite{zhong2020towards}. These modalities leverage the human's innate ability to communicate design ideas \cite{deng2009imagenet} visually, which is deeply rooted in human cognition and culture \cite{sketchmodeling2009}. However, while sketches are a powerful tool for expression, they capture only partial representations of the users' comprehensive design intent, as illustrated in Figure~\ref{fig:PartialInformation}. These inherently ambiguous and incomplete representations impede current SBIM methods from accurately translating users' intent into precise 3D geometric outcomes. As researchers, we might wonder how SBIM can solve this problem, capturing the unsketched knowledge and adequately communicating the user’s intent to an information-rich 3D representation \cite{Front2Back2020}. To address this, modern approaches are increasingly relying on data-driven methods \cite{surveyondatadrivedictionarybasedmodel2018} to infer missing information from datasets of paired shapes and sketches. To ensure that designers' intentions are effectively translated and understood by all, this paper explores how SBIM can capture unsketched knowledge and predict the user’s intent. 

Some SBIMs ensure that the designers' intentions are translated in the 3D space by learning the relationship between sketches and 3D shapes, modeled as a conditional distribution between these two modalities in the training dataset. This learned mapping enables the model to infer, predict, and complete the overall form of an object from partial sketch inputs \cite{Occnet2019, TONOVitruvio22}. For example, given a frontal sketch of a couch, the model can use prior knowledge to generate the unseen sides of the couch—similar to how humans make accurate and informed predictions about unseen parts \cite{howhumanssketchobjects2012eitz, How2SketchHennessey16}. While these methods excel at capturing the overall semantics of a simple black-and-white sketch, they struggle to infer additional details—such as colors, materials, or textures—that are absent from the drawing. Critical to the final design, these visual elements remain beyond the reach of such models when not explicitly provided. 

Other SBIMs tackle these representational limitations through sketch-guided text-to-3D generative models \cite{lee2024textto3d, li2024generativeaimeets3d}, enabling users to provide additional details through text or speech, such as colors and materials. However, while these models often produce precise and geometrically accurate representations, they still struggle with language-driven ambiguities and need improvement in effectively conveying intricate design ideas \cite{liu2024sketchdreamsketchbasedtextto3dgeneration, chen2024sketch2nerfmultiviewsketchguidedtextto3d, zheng2024sketch3d}. Accurately capturing intent and resolving ambiguities relies on a harmonious interaction between humans and machines. Even a perfectly calibrated sketch can still harbor multiple ambiguities, making a seamless interaction crucial for bridging the gap between creative intent and precise execution. Since not everyone possesses advanced sketching skills, there is a need for methods that augment the human ability to communicate 3D representations through a level of abstraction \cite{koley2023picturesketchphotorealisticimage} usually captured by simple and quick doodles \cite{bandyopadhyay2024doodle, sketchrnn}. 

The concept of augmenting human capabilities through interfaces predates SketchPad \cite{sutherland1963sketchpad} and can be traced back to 1962 with Douglas Engelbart’s seminal work, where he articulated the need for a \textit{"clerk"} that enhances human intelligence by supporting informed decision-making and aligning with the user's intent \cite{engelbart1962augmenting}. In his example, the \textit{"clerk"} needed to support an architect or builder during their design process, hence our analogy and focus throughout the paper on the building environment.

Even if Engelbart perfectly anticipated the advent of the personal computer, he overlooked the potential of novel SBIM for 3D content generation. In fact, the growing demand for 3D user-generated content for gaming and design platforms has underscored the urgent need for advancements in real-time multimodal generation \cite{cheng2023sdfusionmultimodal3dshape, dreamfields_zeroshot}. To answer this need for novel multimodal generative models, this report explores novel SBIM technologies, providing researchers with future research directions and equipping the industry to commercialize these advancements effectively by aligning design processes more closely with user intent. The information needed for this alignment varies across sectors and design phases—for instance, precise spatial geometry in architecture, dynamic character models in gaming, or ergonomic and functional details in industrial design—underscoring the need for flexible, user-centered approaches tailored to diverse applications. To achieve this flexibility, we examine methods that combine sketch-based and learning-based approaches with a particular focus on sketch-to-3D object generation, called throughout the literature \textbf{Deep Sketch-Based 3D Modeling ($\neuralsketchmethods$)}\cite{deepsketchmodeling, zhong2020towards}. To facilitate a conversation around this topic, we introduce $\designspace$, a design space \cite{designspace90} for $\neuralsketchmethods$ methods \cite{reviewpracticeevaluationvisualization2013}. Built upon the Input-Model-Output (IMO) framework, \textbf{MORPHEUS} categorizes \textbf{M}odels outputting \textbf{O}ptions of 3D \textbf{R}epresentations and \textbf{P}arts, derived from \textbf{H}uman-inputs (varying in quantity and modality), and \textbf{E}valuated across diverse \textbf{U}ser-views and \textbf{S}tyles. 
$\designspace$ is a simple and structured design space that identifies the key components of these methods. While previous sketch-based surveys \cite{sketchmodeling2009, sketch3dmodeling2009} provided early insights into 3D sketch interfaces, they lacked sections on single-image 3D reconstruction and generation models, since large sketch and 3D datasets were not yet available at that time. This resulted in coverage limited to computationally intensive geometric approaches integrated within traditional WIMP (Window, Icon, Menu, Pointer)-based interfaces. More recent surveys \cite{SOTASketchProcessing25Liu} have focused primarily on 2D sketch processing, briefly covering mid-air 3D strokes for AR/VR applications. In contrast, we introduce these components and group these methods into categories and related subcategories, as detailed in Section~\ref{desiderata}. Overall, Section~\ref{input} explores the diversity of sketch input modalities, such as doodles, freehand drawings, and scribbles, showing the ongoing efforts within the community to expand the flexibility of the input sketches that can be drawn by different users with different styles and from different viewpoints. Section~\ref{aimodel} categorizes the $\neuralsketchmethods$ methods following a simple division and analysis based on the type of AI model architecture. Section~\ref{output} presents a set of qualitative and quantitative metrics, categorizing outputs based on their adaptability and alignment with user intent. Specifically, these metrics evaluate the method's capability to produce geometrically accurate and topologically appropriate shapes that meet user requirements, as well as its ability to generate multiple output options accompanied by relevant information to facilitate user selection. Finally, Section~\ref{discussion} summarizes unresolved challenges presented throughout the report and highlights future research directions. Our design space and related insights, highlighting the intersection of human-computer interaction (HCI) \cite{sketchguided3dgenAI24} and computer graphics, play a critical role in developing novel $\neuralsketchmethods$ to allow users to fully convey their design intents.

\section{Scope}
Since research on $\neuralsketchmethods$ is interdisciplinary, this report serves a diverse audience with expertise in 3D content creation. Sketch-modeling touches many industries such as automotive \cite{sketch2mesh, Piecewisesmoothsketch2022unstructured,3dGANandlazylearning2022sketch}, architecture \cite{TONOVitruvio22,componet2021,Encodedmemory,federova, nam20223dldm,YangSketchto3DHouses23, deng2022sketch2pqfreeformplanarquadrilateral,XIAO2024VResinspatialmemory}, industrial design \cite{dreamsketchgenerativedesign2017, Bitmap2MachineMadeShapes,productdesignsketch2024, sanghi2024waveletlatentdiffusionwala,Img2CAD2025Chen,Deep3DSketchImChen2024}, interior design \cite{chen2023reality3dsketchrapid3dmodeling}, fashion \cite{zhang2024texcontrolsketchbasedtwostagefashion,FashionDiffHan2023,interactivedesigncrossdomain2023, fashiondesigntransfer2021}, comics \cite{DemoCaricature2024Sketch,gao2023controllablevisualtactilesynthesis}, media and entertainment \cite{xu2024sketch2sceneautomaticgenerationinteractive, Shen2024neuralcanvas, doodleyourmotion24}, biology \cite{olivierbiosketch2023}, and many others \cite{MUKTI2024vesselsoceanboat, WANG2024102327VQCAD, Sketch2Seq2025Sun}. Specifically, designers use sketches to design 3D representations such as garments \cite{GarmentIdeation2022, garmentcontrollabledomainfeatures2023, guo2025craftdesigningcreativefunctional, guo2025higarmentcrossmodalharmonybased}, digital avatars \cite{DeepSketch2FaceHan2017, wang2024survey3dhumanavatar, SketchMannequin2022, luo2023rabitparametricmodeling3d, wang2025stickmotiongenerating3dhuman}, animals \cite{robustflow, SimpModeling2021Luo3Danimal, luo2023rabitparametricmodeling3d}, heads \cite{SAniHeadanimalsketching2020collaborative}, hair \cite{DeepSketchHairShen_2021, zhang2025stranddesignerpracticalstrandgeneration}, smoke \cite{deepreconstruction3dsmoke2022}, faces \cite{DeepSketch2FaceHan2017, ling2022structureawareeditablemorphablemodel, luo2023sketchmetafacelearningbasedsketchinginterface, sketchfacenerf2023, wang2024s2tdfacereconstructdetailed3d}, trees \cite{Manfredi2023TreeSketchNet}, and many other shapes. Although our readers may come from diverse backgrounds, we anticipate that they possess a foundational understanding of deep learning, computer vision, computer graphics, and human-computer interaction. We invite the readers to review previous 3D content generation surveys \cite{li2024advances3dgenerationsurvey, liu2024comprehensivesurvey3dcontent, Xia_2023surveydeepgen3dawareimagesynthesis, 3d-arena}, for image \cite{Xia_2023surveydeepgen3dawareimagesynthesis, li2024generativeaimeets3d, shi2023deepgenerativemodels3d} or text to 3D generation \cite{learninggenerativemodels3Dstructures20, chao2023textguidedimageandshapeeditinggeneration, cheng2023sdfusionmultimodal3dshape, hui2022neuralwaveletdomaindiffusion3d, lee2024textto3d}.

To keep this report clear and concise, we have intentionally limited its scope. This report focuses exclusively on $\neuralsketchmethods$ for 3D static objects, as presented in Table \ref{datasets_table}, which shows datasets used in this line of research \cite{chang2015shapenet, deitke2022objaverseuniverseannotated3d}. We deliberately exclude animated shapes (animals, humans, faces, smoke, hair, clothing) and 3D scene generation, as these domains introduce additional complexity requiring physics simulations, temporal dynamics, spatial relationships, and specialized frameworks that warrant separate analysis. These 3D static objects are paired with user sketches drawn on a flat surface: paper or tablet. We do not consider non-learning based methods \cite{Structuralsketcher04, DigitalClay2000, teddy1999,SmoothSketch2006, FiberMesh2007NealenIgarashi, SketchedConv2004, Interactiveimplicitspherical2004, ShapeShop2006, freeformimplicit2002, True2Form2014, interactivesketchsiggraph07, BendSketch17, Geosemanticsnappingsketchbasedmodeling2013, symmetrydriven3Dsketch22, Piecewisesmoothsketch2022unstructured} such as inflatable techniques \cite{Dvoroznak20MonsterMash} or skeleton-based \cite{ma2021realtimeskeletonizationsketchbasedmodeling, interactiveliquidsplashmodeling2020} approaches. We also do not consider 3D immersive reality sketch-modeling \cite{chatterjee2024freeformshapemodelingxr, 3D_VR_luo20233d, 3Doodle2024} or purely retrieval-based methods \cite{finegrainedsketchretrieval, noisetolerantsketchretrieval, sketchbasedshaperetrieval2012, Berardi_2023_ICCVZeroShotSketchretrieval, sketch3dretrievalcnn, DemSketchto3DShapeRetrievalPivoting23, unsupervisedsketchto3d2018, xu2022domaindisentangledgenerativeadversarial}. 

 $\designspace$ offers a simple and unified design space \cite{designspace90} to survey $\neuralsketchmethods$ methods. We grouped each method into one of the following: reconstruction, generation, and editing \cite{hertz2022spaghettieditingimplicitshapes}, as outlined in the Supplementary material in Table \ref{tab:model table_supplementary}; however, discussions and considerations specific to editing are beyond the scope of this report. The overarching goal of $\neuralsketchmethods$ is to democratize 3D content creation through controlled and informed design processes.  $\designspace$ achieves this by: 1) identifying $\neuralsketchmethods$' limitations and 2) highlighting future research areas. Our proposed design space provides a simple yet unified framework to survey $\neuralsketchmethods$, facilitating both industry and academic use cases. For industry practitioners, this design space can guide the selection of methods most suited to their applications. Meanwhile, academic researchers can use it to identify limitations in existing approaches and explore potential directions for future research.

A central focus of this design space is the role of information in enriching 3D content creation. Information extends beyond visual attributes like color, texture, or geometry to encompass complex processes such as cost estimation, meshing, or aerodynamic performance \cite{engelbart1962}. Given that $\neuralsketchmethods$ methods are often highly application-specific, our analysis reveals a growing trend toward incorporating contextual factors, external influences that shape a design’s functionality and environment. Examples include urban surroundings for buildings, road conditions for vehicles, or placement settings for furniture.

Embedding such insights into $\neuralsketchmethods$ ensures the creation of functional and contextually relevant designs. These external considerations are critical for tailoring the final design to its intended use case, carefully considering the surrounding environment \cite{shuai2024surveymultimodalguidedimageediting, zhan2024conditionalimagesynthesisdiffusion, cao2024controllablegenerationtexttoimagediffusion, chowdhury2022fscocounderstandingfreehandsketches}.

\section{A Design Space for Deep Sketch-Based 3D Modeling}
\label{desiderata}

\begin{figure}[H]
    \centering
    \includegraphics[width=\linewidth]{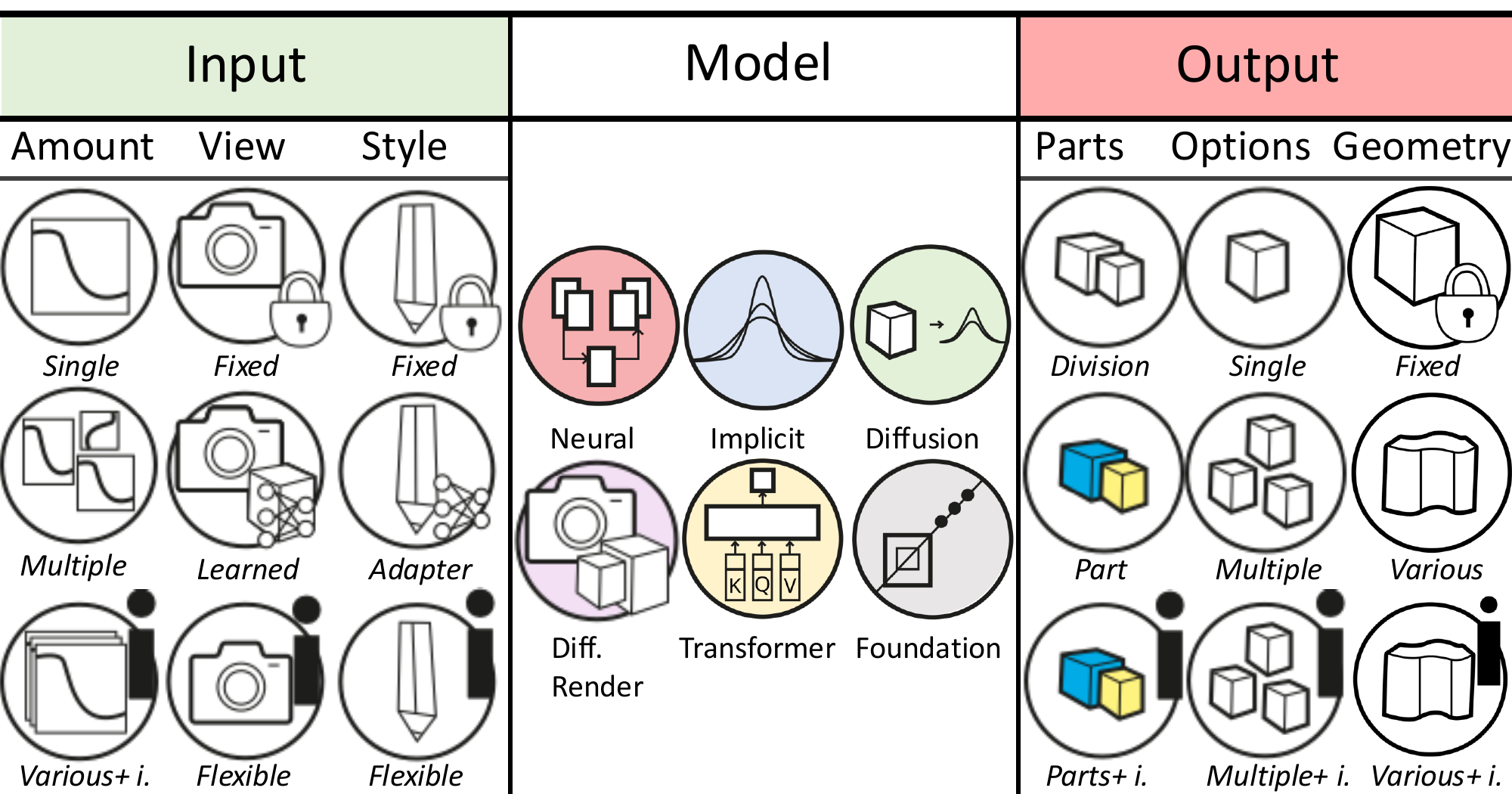}
    \caption{Illustrates $\designspace$ and its overall structure. The input is divided into three main aspects: (1) the quantity of sketches, including multiple sketches, single sketches, and single or multiple sketches with additional information (e.g., text); (2) the viewpoint, specifying whether a fixed viewpoint, learned camera parameters, or a view-independent approach is used; and (3) the style, categorized as fixed sketching styles, style adapters, or flexible styles. The model section highlights six key techniques—neural models, implicit representations, diffusion models, differentiable renderers, transformers, and foundation models—with methods often combining multiple techniques. The output is divided into: (1) parts and semantics, encompassing individual element divisions, part-based segmentation, and parts with related semantics (e.g., material properties); (2) geometric genus, which ranges from fixed representations to flexible genus and enriched geometry; and (3) options, indicating whether the method produces a single output, multiple outputs, or multiple outputs with additional information.}
    \label{fig:overalldivisionframework}
\end{figure}

 $\designspace$ assists readers in navigating the rapidly evolving field of $\neuralsketchmethods$. $\designspace$ is designed to serve multiple audiences: researchers seeking to understand current limitations and identify future opportunities, and industry leaders aiming to align their solutions with state-of-the-art methodologies. With $\designspace$, we present a comprehensive analysis of $\neuralsketchmethods$, structured within the input-model-output framework. $\designspace$ introduces a simple categorization detailed in Sections~\ref{input}, \ref{aimodel}, and \ref{output}, and visually summarized in Figures~\ref{fig:teaser} and \ref{fig:overalldivisionframework}. This categorization emphasizes both human-centric metrics and informed design, providing a structured foundation to guide future developments in sketch-based methods.

\begin{itemize}
    \item \textbf{Input}: The input section (Section~\ref{input}) emphasizes the diverse modalities and categorizes approaches based on their flexibility to handle various user-provided data. This reflects the research community's growing interest in aligning input processing capabilities with diverse user intents, minimizing constraints on the amount of sketches, their type of viewpoints, and their sketching style. 
    \item \textbf{Model}: The model section (Section~\ref{aimodel}) investigates the trade-offs between different architectures, providing a chronological visualization and a coherent explanation of such evaluation over time. This section maps the transformation from input to output, forming the paper's conceptual core. The architectural design—specifically the encoder, decoder, and loss functions—is critical in translating user intent into precise computational representations, bridging the gap between conceptual design and generated outcomes.
    \item \textbf{Output}: The output section (Section~\ref{output}) highlights the growing emphasis on diversity, adaptability, and user-centric evaluation of $\neuralsketchmethods$ methods. This section explores the methods' ability to provide users with informative 3D part-based representations, with various numbers of options to generate diverse and complex geometries. These outputs are evaluated using both qualitative and quantitative metrics, which are currently evolving to better reflect user needs. A unifying theme is the drive to integrate perceptual and task-based metrics, bridging the gap between subjective user satisfaction and objective performance measurements.
\end{itemize}

Each section has a table that facilitates a structured discussion and streamlined comparisons. These tables (Tables \ref{table:input}, \ref{tab:model table} and \ref{awareevaluation}) highlight trends, gaps, and future opportunities by organizing the design space with detailed input and output subcategories. This structured framework facilitates novel direction identification and encourages methods that better support informed, user-centered design processes.




\section{Input: $\boldsymbol{I}_{sketch}$} \label{input}

\begin{table}[htbp]
\centering
\setlength{\tabcolsep}{1pt} 
\renewcommand{\arraystretch}{1} 
\small
\begin{tikzpicture}
\node (table) {
\begin{tabular}{p{3cm}|p{0.5cm}p{0.5cm}p{0.5cm}|p{0.5cm}p{0.5cm}p{0.5cm}|p{0.5cm}p{0.5cm}p{0.5cm}}
    \rowcolor{green!5}
    \toprule
    \textbf{Paper} & \multicolumn{3}{c|}{\textbf{Amount}} & \multicolumn{3}{c|}{\textbf{View}} & \multicolumn{3}{c}{\textbf{Style}} \\ 
    \rowcolor{green!5}
    & \raisebox{-0.3\baselineskip}{\includegraphics[width=0.45cm]{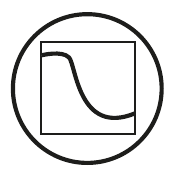}} & \raisebox{-0.3\baselineskip}{\includegraphics[width=0.45cm]{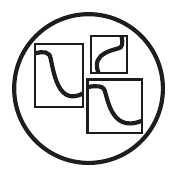}} & \raisebox{-0.3\baselineskip}{\includegraphics[width=0.45cm]{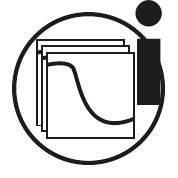}} & \raisebox{-0.3\baselineskip}{\includegraphics[width=0.45cm]{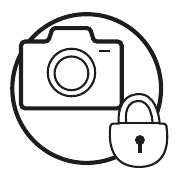}} & \raisebox{-0.3\baselineskip}{\includegraphics[width=0.45cm]{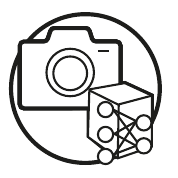}} & \raisebox{-0.3\baselineskip}{\includegraphics[width=0.45cm]{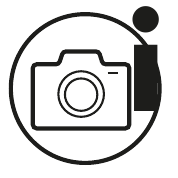}} & \raisebox{-0.3\baselineskip}{\includegraphics[width=0.45cm]{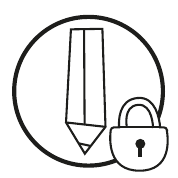}} & \raisebox{-0.3\baselineskip}{\includegraphics[width=0.45cm]{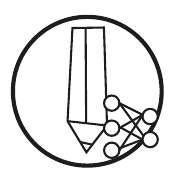}} & \raisebox{-0.3\baselineskip}{\includegraphics[width=0.45cm]{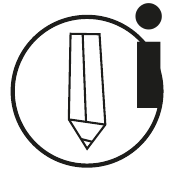}} \\ 
    \midrule
Nishida et al. \cite{Nishida2016} &  &  & \cellcolor{gray!25} &  &  &  & \cellcolor{gray!25} &  &  \\ 
Delanoy et al. \cite{Delanoy20203DJohanna} & \cellcolor{gray!25} &  &  & \cellcolor{gray!25} &  &  &  &  & \\ 
ShapeMVD \cite{3dshapereconstructionmvcnn} & \cellcolor{gray!25} & \cellcolor{gray!25} &  & \cellcolor{gray!25} &  &  & \cellcolor{gray!25} &  &  \\ 
Contour3D \cite{Contourbased3DModelingAobo20} & &  \cellcolor{gray!25} &  & \cellcolor{gray!25} &  &  & \cellcolor{gray!25} &  &  \\ 
DeepSketch \cite{zhong2020towards} &  \cellcolor{gray!25}  & &  &  & \cellcolor{gray!25} &  &  & \cellcolor{gray!25} &  \\ 
Sketch2CAD \cite{Li2020Sketch2CAD} & \cellcolor{gray!25} & \cellcolor{gray!25} &  & \cellcolor{gray!25} &  &  & \cellcolor{gray!25} &  &  \\ 
SketchDiff \cite{sketchmodelingdiffrenderer} &  &  & \cellcolor{gray!25} & \cellcolor{gray!25} &  &  & \cellcolor{gray!25} &  &  \\ 
FreeHandRec \cite{freehandreconstruction} &  &  & \cellcolor{gray!25} &  & \cellcolor{gray!25} &  &  &  & \cellcolor{gray!25} \\ 
Sketch2Model \cite{sketch2model} &  &  & \cellcolor{gray!25} &  & \cellcolor{gray!25} &  &  & \cellcolor{gray!25} &  \\ 
Sketch2Mesh \cite{sketch2mesh} &  &  & \cellcolor{gray!25} &  & \cellcolor{gray!25} &  & \cellcolor{gray!25} &  &  \\ 
Free2CAD \cite{Free2CAD} & \cellcolor{gray!25} & \cellcolor{gray!25} &  & \cellcolor{gray!25} &  &  & \cellcolor{gray!25} &  &  \\ 
SS2Mesh \cite{bhardwaj2022singlesketch2meshgenerating3d} &  &  & \cellcolor{gray!25} & \cellcolor{gray!25} &  &  &  & \cellcolor{gray!25} &  \\ 
GeoCode \cite{pearl2022geocodeinterpretableshapeprograms} & \cellcolor{gray!25} &  &  & \cellcolor{gray!25} &  &  & \cellcolor{gray!25} &  &  \\ 
SketchSampler \cite{sketchsampler2022eccv} & \cellcolor{gray!25} &  &  & \cellcolor{gray!25} &  &  &  &  & \cellcolor{gray!25} \\ 
LAS-Diffusion \cite{zheng2023lasdiffusion} &  &  & \cellcolor{gray!25} &  & \cellcolor{gray!25} &  &  & \cellcolor{gray!25} &  \\ 
Sketch-A-Shape \cite{sanghi2023sketchashapezeroshotsketchto3dshape} &  &  & \cellcolor{gray!25} &  &  & \cellcolor{gray!25} &  &  & \cellcolor{gray!25} \\ 
SKED \cite{Mikaeili_2023_sked} &  & \cellcolor{gray!25}  &  & \cellcolor{gray!25} &  &  &  &  & \cellcolor{gray!25} \\ 
CLIPXPlore \cite{CLIPXPlore2023sketchshape} &  &  & \cellcolor{gray!25} &  &  & \cellcolor{gray!25} &  & \cellcolor{gray!25} &  \\ 
D3DSketch+ \cite{chen2023deep3dsketchrapid3dmodeling} &  &  & \cellcolor{gray!25} & \cellcolor{gray!25} &  &  & \cellcolor{gray!25} &  &  \\ 
Control3D \cite{Control3D2023} &  &  & \cellcolor{gray!25} &  &  & \cellcolor{gray!25} &  &  & \cellcolor{gray!25} \\ 
Re3DSketch \cite{chen2023reality3dsketchrapid3dmodeling} &  &  & \cellcolor{gray!25} &  & \cellcolor{gray!25} &  & \cellcolor{gray!25} &  &  \\ 
Sketch2Point \cite{diffrefforsketchtopointmodeling2023Di} &  &  & \cellcolor{gray!25} &  &  & \cellcolor{gray!25} &  & \cellcolor{gray!25} &  \\ 
GA-Sketching \cite{zhou2023gasketchingshapemodelingmultiview} &  &  & \cellcolor{gray!25} &  &  & \cellcolor{gray!25} &  & \cellcolor{gray!25} &  \\ 
S2PointCol \cite{Wu2023sketch2pointcolored} &  &  & \cellcolor{gray!25} &  &  & \cellcolor{gray!25} &  &  & \cellcolor{gray!25} \\ 
Sketch2Vox \cite{sketch2vox} & \cellcolor{gray!25}  &  &  &  & \cellcolor{gray!25} & &  & \cellcolor{gray!25} &  \\ 
SketchDream \cite{liu2024sketchdreamsketchbasedtextto3dgeneration} &  &  & \cellcolor{gray!25} &  & \cellcolor{gray!25} &   & \cellcolor{gray!25}  &  &  \\ 
SENS \cite{sens2024binningerpartawaresketchimplicit} &  &  & \cellcolor{gray!25} &  &  & \cellcolor{gray!25} &  &  & \cellcolor{gray!25} \\ 
DY3D \cite{bandyopadhyay2024doodle} &  &  & \cellcolor{gray!25} &  &  & \cellcolor{gray!25} &  &  & \cellcolor{gray!25} \\ 
Vitruvio \cite{TONOVitruvio22} & \cellcolor{gray!25}  &  & &  & \cellcolor{gray!25} &  &  & \cellcolor{gray!25} &  \\ 
MVControl \cite{li2024controllabletextto3dgenerationsurfacealigned} &  &  & \cellcolor{gray!25} &  & \cellcolor{gray!25} &  &  & \cellcolor{gray!25} &  \\ 
SHLine \cite{sketchhiddenline2024} &  & \cellcolor{gray!25}  &  &  & \cellcolor{gray!25} &  &  & \cellcolor{gray!25} &  \\ 
M3DSketch \cite{zang2024magic3dsketchcreatecolorful3d} &  &  & \cellcolor{gray!25} &  & \cellcolor{gray!25} &  & \cellcolor{gray!25} &  &  \\ 
Sketch2NeRF \cite{chen2024sketch2nerfmultiviewsketchguidedtextto3d} &  &  & \cellcolor{gray!25} &  &  & \cellcolor{gray!25} &  &  & \cellcolor{gray!25} \\ 
DualShape \cite{dualshape2024} &  & \cellcolor{gray!25} &  & \cellcolor{gray!25} &  &  & \cellcolor{gray!25} &  &  \\ 
Sketch3D \cite{zheng2024sketch3d} &  &  & \cellcolor{gray!25} &  &  & \cellcolor{gray!25} &  &  & \cellcolor{gray!25} \\ 
\bottomrule
\end{tabular}%
};

\draw[red, line width=4pt, opacity=0.22, ->, >=stealth] 
  (table.north west) + (3.5cm,-1.2cm) -- ++(4.7cm,-13.4cm);

\draw[red, line width=4pt, opacity=0.22, ->, >=stealth] 
  (table.north west) + (5.1cm,-1.2cm) -- ++(6.3cm,-13.4cm);

\end{tikzpicture}

\caption{Categorization of $\neuralsketchmethods$ methods' Input to ensure its flexibility. As a disclaimer, we do recognize that this categorization could not properly capture the intent of the paper;\raisebox{-0.3\baselineskip}{\includegraphics[width=0.45cm]{images/flexiblesketch1.pdf}}: multiple sketches. \raisebox{-0.3\baselineskip}{\includegraphics[width=0.45cm]{images/flexiblesketch2.pdf}}: single sketches. \raisebox{-0.3\baselineskip}{\includegraphics[width=0.45cm]{images/flexiblesketch3.pdf}}: single or multiple sketch and information, \raisebox{-0.3\baselineskip}{\includegraphics[width=0.45cm]{images/flexibleview1.pdf}} : explicitly using camera parameters, \raisebox{-0.3\baselineskip}{\includegraphics[width=0.45cm]{images/flexibleview2.pdf}} : differentiable render (learnable camera parameters), \raisebox{-0.3\baselineskip}{\includegraphics[width=0.45cm]{images/flexibleview3.pdf}} : view-independent, and \raisebox{-0.3\baselineskip}{\includegraphics[width=0.45cm]{images/flexiblestyle1.pdf}}: the method has specific assumptions about the style (canny, suggestive contours, or others). \raisebox{-0.3\baselineskip}{\includegraphics[width=0.45cm]{images/flexiblestyle2.pdf}}: style specific, it has a specific method that accounts for style.\raisebox{-0.3\baselineskip}{\includegraphics[width=0.45cm]{images/flexiblestyle3.pdf}}: style independent. The "i"s on the icons signify the type of information described in Section~\ref{desiderata}. Specifically, when referring to information provided in the input sketch, it indicates that the method can accommodate additional prompt-based inputs, such as text or speech modalities, alongside the sketch itself. The red arrows show the trends that we discussed in our paper.}
\label{table:input}
\end{table}
In this section, we present the findings from a comprehensive analysis of $\neuralsketchmethods$ methods. $\neuralsketchmethods$ methods aim to improve the user experience in generating a desired 3D output ($\hat{X}_{\text{shape}}$) by minimizing the need for complex 3D spatial input. Instead, they tend to rely on just a single sketch ($I_{\text{sketch}}$) with additional text (see Section \ref{amount}) and more flexibility to both the sketch's viewpoint ($\Theta V$) and the user's style ($\Psi S$), as described in Figure~\ref{fig:flexibleimage} and respective Sections \ref{view} and \ref{style}. This trend became visible after analyzing Table~\ref{table:input}, which organizes these key aspects of input handling, facilitating discussions throughout the literature. Therefore, our discussion is structured to follow the table from left to right, with rows arranged in chronological order.

\subsection{Amount}
\label{amount}
The column "Amount" ($I_{\text{sketch}}$) in Table~\ref{table:input} categorizes the number of input sketches $I_{\text{sketch}}$ into three main groups: multiple sketches, single sketches, and single sketches with additional information. The categorization is based on whether the model:
\begin{itemize}
    \item  \raisebox{-3.0pt}{\includegraphics[width=0.5cm]{images/flexiblesketch2.pdf}} works with a single sketch or strokes,
    \item  \raisebox{-3.0pt}{\includegraphics[width=0.5cm]{images/flexiblesketch1.pdf}} requires multiple sketches, and/or   
      \item \raisebox{-3.0pt}{\includegraphics[width=0.5cm]{images/flexiblesketch3.pdf}} works with a single or multiple sketches and information $i$.
    \end{itemize} 
    
\begin{figure}[H]
    \centering
\includegraphics[width=\linewidth]{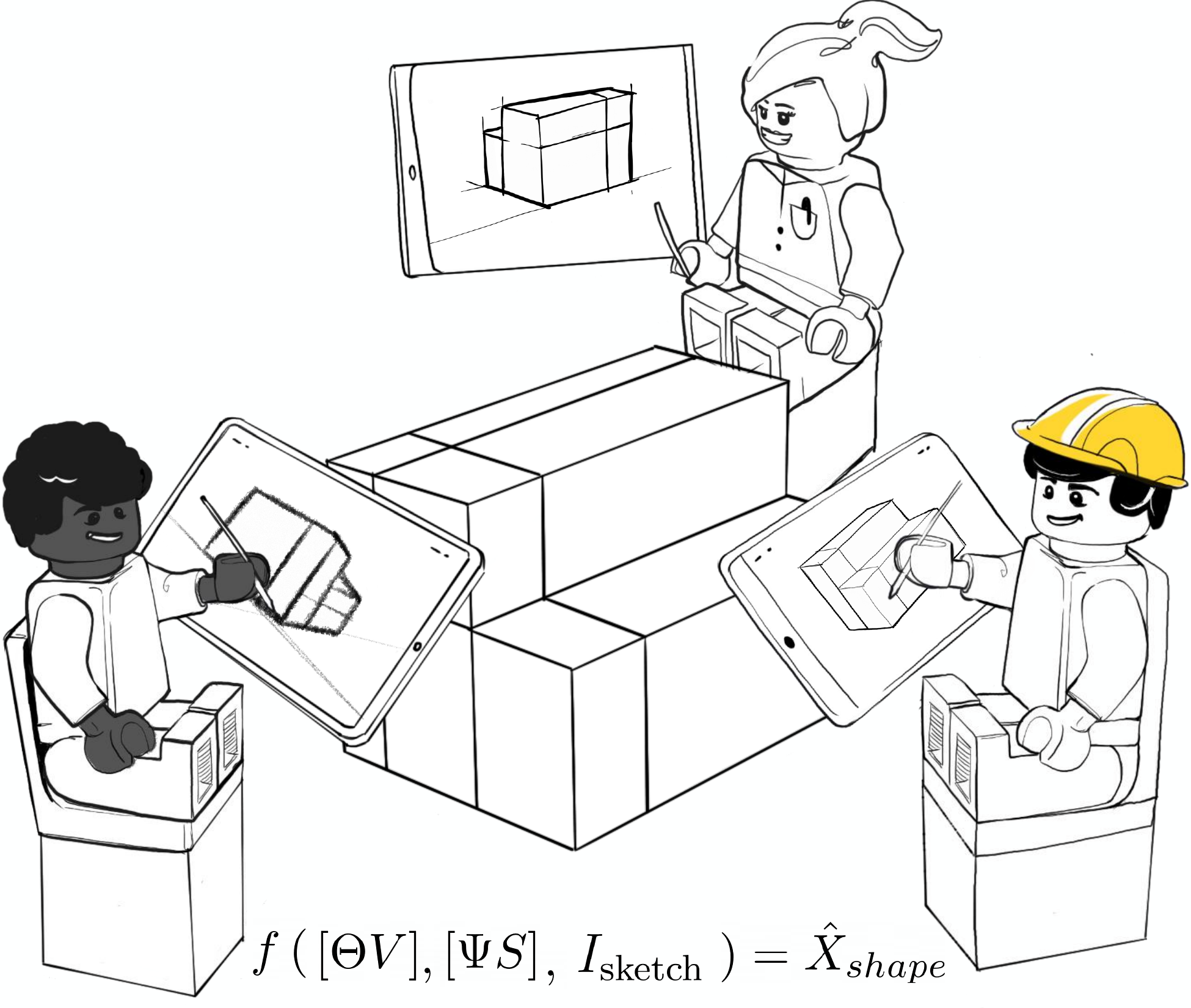}
    \caption{$\neuralsketchmethods$ are designed to accommodate a diverse array of sketching styles \cite{Xiao2022DifferSketching} and viewpoints \cite{sketchstyle}, including bird’s-eye, street-level, front, top, side, axonometric, and perspective views \cite{WherepeopledrawlinesCole2008}. These methods are robust to variations in sketch characteristics, whether the lines are wavy or straight, single or multiple, and whether the sketches are shaded or unshaded. }
    \label{fig:flexibleimage}
\end{figure}

    For information $i$, $\neuralsketchmethods$ is increasingly focusing on methods that utilize a single sketch supplemented by non-geometrical related information, such as appearance \cite{Wu2023sketch2pointcolored} and contextual information \cite{TONOVitruvio22, chen2023reality3dsketchrapid3dmodeling}, often provided through text description \cite{chen2024sketch2nerfmultiviewsketchguidedtextto3d, zheng2024sketch3d, liu2024sketchdreamsketchbasedtextto3dgeneration}. However, there are exceptions, particularly for industrial designers who need multiple drawings \cite{dualshape2024, Free2CAD, Li2020Sketch2CAD} to effectively reduce geometric-related ambiguity \cite{sketchhiddenline2024}. Our analysis primarily examines $\neuralsketchmethods$ from a single-user perspective. However, we acknowledge that more complex considerations are necessary for collaborative settings, where multiple stakeholders contribute to the design process. For instance, GroundUp \cite{unlu2024eccv} facilitates collaboration between urban planners and architects by incorporating additional top-view sketches provided by planners. This approach helps mitigate sketch ambiguity and enhances the accuracy of city-scale 3D generation, demonstrating the potential benefits of integrating multi-user inputs in $\neuralsketchmethods$ systems as shown in Figure \ref{fig:flexibleimage}. Similar considerations are taken when considering the methods' ability to operates with different type of viewpoint (Section \ref{view}) or style (Section \ref{style}).

\subsection{View} 
\label{view}
The column "View" ($\Theta V$) in Table~\ref{table:input} examines if the type of viewpoint uses:

\begin{itemize}
    \item \raisebox{-3.0pt}{\includegraphics[width=0.5cm]{images/flexibleview1.pdf}} fixed viewpoints,
    \item \raisebox{-3.0pt}{\includegraphics[width=0.5cm]{images/flexibleview2.pdf}} learned camera parameters, and/or
    \item \raisebox{-3.0pt}{\includegraphics[width=0.5cm]{images/flexibleview3.pdf}} flexible to the viewpoint, by leveraging additional information.
\end{itemize}
 Early $\neuralsketchmethods$ addressed the challenge of view ambiguity in sketches by relying on strict view assumptions rather than developing methods capable of accommodating flexible viewpoints. These assumptions include fixed viewpoints, such as axonometric views \cite{Free2CAD}, frontal views \cite{3dshapereconstructionmvcnn}, or other predefined camera positions, to simplify the mapping between the 2D sketch and the resulting 3D geometry. For example, Free2CAD \cite{Free2CAD} assumed an axonometric perspective, sidestepping the inherent ambiguity of user-provided sketches (see Figure \ref{hiddenlines}).

In more recent work, researchers have shifted toward actively tackling sketch ambiguity by learning view-related parameters and developing view-aware methods \cite{sketch2model}. View-aware approaches \cite{sketch2model, zhou2023gasketchingshapemodelingmultiview, zheng2023lasdiffusion} incorporate explicit camera parameters or geometric cues to infer 3D shapes more accurately, acquiring knowledge of the sketch's viewpoint. These approaches leverage camera information, such as position, field of view, and perspective, for the 3D generation. The 3D clues derived by specific camera information are added to the sketch during the generation process. For example, depth-guided warping \cite{Fehn2004DIBRdepthimagebasedrendering, somraj2020posewarping} moves the sketch \cite{liu2024sketchdreamsketchbasedtextto3dgeneration} to a depth-based space (2.5D) where it can be quickly wrapped to generate additional views and geometric information removing sketch's ambiguity and improving faithfulness. SketchSampler \cite{sketchsampler2024_version2} instead goes directly to 3D; its sketch translator module extracts spatial information and generates a 3D point cloud that conforms to the shape depicted in the sketch. LAS-diffusion \cite{zheng2023lasdiffusion} uses a view-aware local attention mechanism to match path images to 3D volume features as displayed in Figure \ref{fig:viewawareattention}. Furthermore, recently few viewpoint independent methods \cite{DemSketchto3DShapeRetrievalPivoting23, bandyopadhyay2024doodle} emerged. They handle sketches independently of specific viewpoints, using features learned from foundation models and implicit representations to resolve ambiguities. Both approaches remain sensitive to the sketch’s position on the canvas and require a proper alignment for optimal translation. For example, Zhong et al. \cite{zhong2020towards} addressed this centering issue by employing spatial transformer networks to automatically center and align sketches. The spatial transformer network ensures consistent and accurate model predictions irrespective of the sketch’s initial placement. This evolution reflects a transition from rigid assumptions to more flexible and robust techniques for interpreting sketches. In fact, viewpoint independent methods, instead, are methods in which the network does not explicitly consider the camera parameters and positions, presenting more flexible methods \cite{bandyopadhyay2024doodle} that leverage textual information \cite{dreamcontrol2024tianyu}.

\subsection{Style}
\label{style}
\begin{figure}[ht!]
    \centering
    \includegraphics[width=\columnwidth]{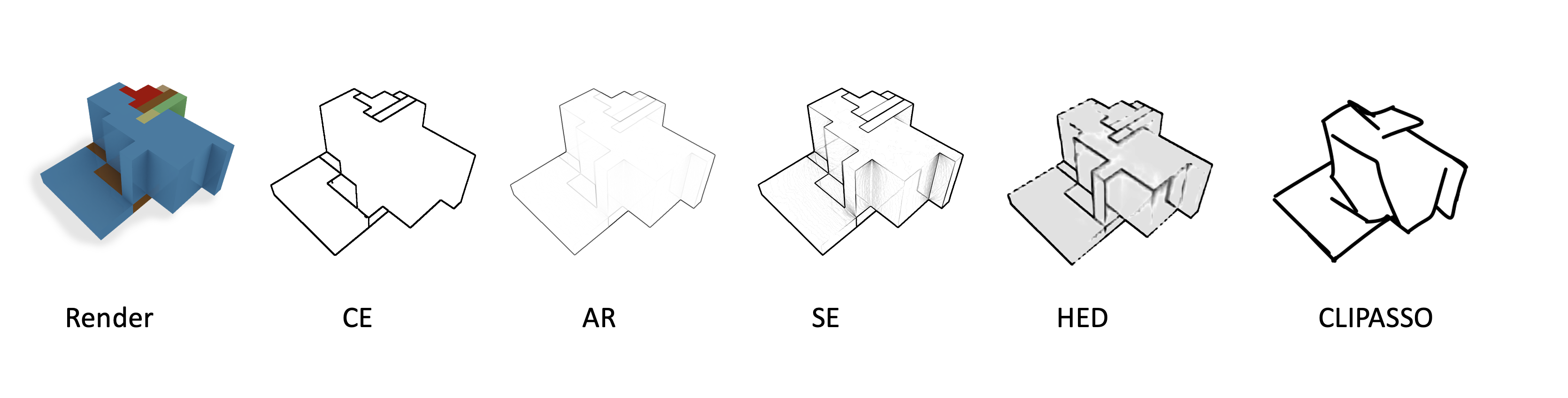}  
    \caption{BuildingGAN \cite{buildingan} with different sketching styles. CLIPASSO \cite{vinker2022clipasso}, Apparent Ridges \cite{Apparentridges07} (AE), Canny Edge \cite{canny1996edgedetection} (CE), Hollistically Nested-Edge  \cite{xie2015holistically} (HED), Sobel Edge \cite{sobel_operator} (SE). These have been generated using OpenCV, and filters have been applied to the initial render. }
    \label{fig:stylesketch}
\end{figure}

The column "Style" ($\Psi S$) in Table~\ref{table:input} focuses on the sketching style, considering the styles used to train the AI models and those that yield the best performance during inference. The subcategories are divided based on when methods: \begin{itemize}
    \item \raisebox{-3.0pt}{\includegraphics[width=0.5cm]{images/flexiblestyle1.pdf}} use a fixed sketching style during training,
    \item \raisebox{-3.0pt}{\includegraphics[width=0.5cm]{images/flexiblestyle2.pdf}} account for style, with adapters, and/or
    \item \raisebox{-3.0pt}{\includegraphics[width=0.5cm]{images/flexiblestyle3.pdf}} are flexible to the style, by leveraging additional information.
\end{itemize}
 As observed from Table \ref{table:input}, handling sketch style remains a challenging aspect of $\neuralsketchmethods$. Achieving domain-free adaptation to sketch styles, meaning the model must be invariant to distribution shifts caused by variations in sketching styles, as discussed in Section~\ref{input} (denoted as $\Psi S$) requires large and diverse datasets of real sketches. However, collecting such datasets is currently prohibitive \cite{WherepeopledrawlinesCole2008}. Among various sketching styles, doodles offer an effective medium for conveying abstract semantic information through visual representation \cite{PragmaticInferenceVisualAbstractionFan2020, vinker2023clipascenescenesketchingdifferent, SwiftSketch2025Yael} . Doodles represent an abstract style of sketches, typically created by humans in under 20 seconds (as exemplified in the QuickDraw dataset \cite{sketchrnn}). To replicate this level of semantic abstraction digitally, CLIPasso \cite{vinker2022clipasso} pioneered a method for synthetically generating doodle-like sketches using only a few strokes. While CLIPasso has proven effective for enhancing abstraction modeling in synthetic edgemaps \cite{bandyopadhyay2024doodle}, whether it serves as an adequate proxy for authentic human drawing behavior remains an open research question. While the term "sketch" broadly encompasses both synthetic and natural representations, real sketches are typically hand-drawn on digital or physical surfaces (e.g., paper). In contrast, synthetic sketches are algorithmically generated and can take on various forms, as illustrated in Figure~\ref{fig:stylesketch}. A further distinction is related to the sketch input format to the neural network which influences the encoder design choice. There are mainly two formats: \textbf{bitmap} $\boldsymbol{I}_s$ and \textbf{vector} inputs $\boldsymbol{I}_{\vec{s}}$. Each category includes both real and synthetic sketch representations. A real sketch refers to one created by a human \cite{zhong2020towards}, while a synthetic sketch is generated by a machine using techniques designed to mimic a sketch-like form \cite{canny1996edgedetection, vinker2022clipasso}. These sketches, paired with 3D shapes, are used to train $\neuralsketchmethods$ models (see Dataset Table~\ref{datasets_table}), by learning to associate a user’s sketch with specific 3D representations. Since sketches can be drawn by different users and from various perspectives, $\neuralsketchmethods$ must be flexible to accommodate this diversity.

During \textit{inference}, input sketches can be drawn by both \textit{novice} and \textit{professional} designers \cite{zhong2020towards}. Independent of their style, these sketches can be captured in bitmap format if an image is taken from the physical drawing, or in vector format \cite{bitmaporvector2019} if the sketch is performed directly on a digital interface. However, when differentiating between novice and professional sketches, novice sketches tend to be less structured and more abstract, while expert designers often use \textit{scaffolds} \cite{OpenSketch19} to guide their drawings, resulting in more precise and detailed designs \cite{How2SketchHennessey16}. This variation requires $\neuralsketchmethods$ to be flexible enough to handle both simple and complex inputs.

\begin{figure}[ht!]
    \centering
    \includegraphics[width=\columnwidth]{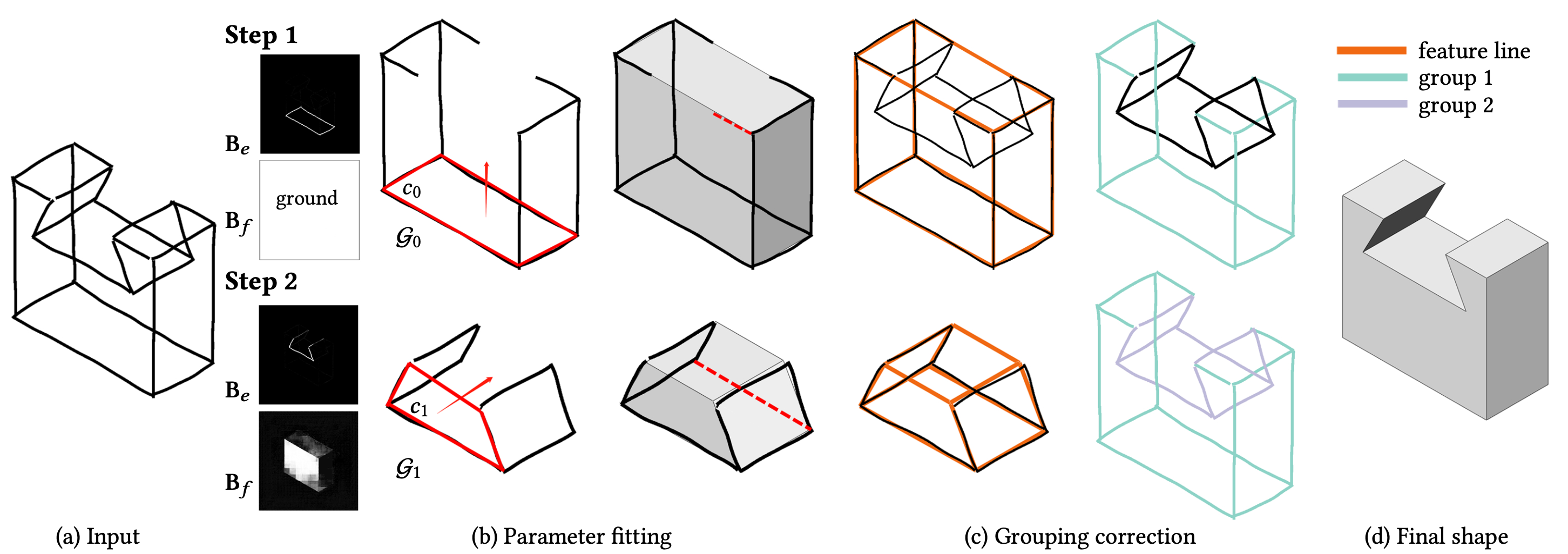}  
    \caption{Free2CAD \cite{Free2CAD} used transformers to group sketch input to specific CAD sequences. The geometry is divided in different, but they do not carry any semantic meaning, they are related to the CAD sequence of commands \raisebox{-3.0pt}{\includegraphics[width=0.4cm]{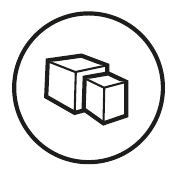}}. }
    \label{Free2CAD}
\end{figure}

During \textit{training}, collecting large datasets of paired 3D shapes $X_{3D}$ and corresponding hand-drawn sketches $I_{sketches}$ is prohibitive. To address this limitation, the research community has developed synthetic proxies, primarily through non-photorealistic rendering (NPR), applicable as a 2D filter or 3D renderer. \\ Filter-based NPR methods translate 2D images into sketch-like formats, common techniques include Line Rendering \cite{linerendering1990saito}, Apparent Ridges \cite{Apparentridges07}, Canny Edge \cite{canny1996edgedetection}, Salient Outline \cite{SalientBuildingOutline19}, Hollistically Nested-Edge - HED \cite{xie2015holistically}, Sobel Edge \cite{sobel_operator}, Difference of Gaussians (XDoG) \cite{xdog2012}, Sketch Simplification via GANs \cite{simoserra2016gansketchlearning}, PaintsUNDO, and others \cite{liao2024freehand, lee2023learning}. \textit{Synthetic Sketch 3D renderer}: these NPR techniques rely on 3D representation to produce more geometrically plausible synthetic sketches, such as Suggestive Contours \cite{decarlo2003suggestive, suggestivecontours} or Sketch Simplification \cite{DeepSketch2FaceHan2017}. For \textit{vector formats}, methods like CLIPasso \cite{vinker2022clipasso} represent sketches as sequences of user input strokes \cite{sketchrnn,Li2020Sketch2CAD}, which can be grouped into primitives, as demonstrated in Free2CAD \cite{Free2CAD} showed in Fig. \ref{Free2CAD}. CLIPasso has been used \cite{bandyopadhyay2024doodle} to pair these more abstract synthetic sketches and 3D shapes \cite{DemSketchto3DShapeRetrievalPivoting23}, mitigating the domain shift between synthetic sketches and real sketches.

After analyzing various datasets based on their respective sketch styles (refer to Table~\ref{datasets_table}), we observed that most synthetic datasets predominantly rely on widely-used sketch styles, such as Canny Edges or Suggestive Contours. This raises an important question: “Do these synthetic sketch styles accurately represent the unique styles used in $X$?” Here, $X$ could refer to a specific design discipline (e.g., architecture, industrial design) or a particular 3D object category (e.g., chairs, buildings, or trees).

While some researchers have attempted to address this question \cite{Xiao2022DifferSketching}, the growing demand for more refined synthetic representations \cite{mechanicalpartspurdue, liao2024freehand} underscores the importance of developing domain-specific sketching styles. These styles must better reflect the unique characteristics and requirements of particular fields, emphasizing the need for tailored datasets that capture the nuances of sketch-based design within specific contexts.

\label{datasets}
\textbf{Datasets:} For a broader survey on sketch datasets, we direct the reader to the work of \cite{sketchsurvey}. Table~\ref{datasets_table} summarizes key datasets and their characteristics, focusing on 3D objects used in $\neuralsketchmethods$. Many of these methods require generating custom datasets to address the challenge of domain shift between images and sketches \cite{bandyopadhyay2024doodle}. This shift occurs because models trained on photographs often perform poorly with sketch inputs without proper fine-tuning; a synthetic-trained mapper does not generalize to human drawings that are vague and ambiguous, lacking geometric and perspective clues \cite{liao2024freehand, freehandreconstruction}. For instance, Sketch2Model \cite{sketch2model} collected 1,300 sketches of ShapeNet objects from novices and professionals. Vitruvio \cite{TONOVitruvio22} not only generated 1,000 3D building \cite{realcity3d, YangSketchto3DHouses23, BuildingNet} shapes as occupancy functions \cite{Occnet2019} (Section~\ref{deepgenmodelimplicit}) but rendered them producing a total of 24,000 synthetic sketches using suggestive contours rendered in Blender. Sketch2CAD \cite{Li2020Sketch2CAD} generated 50,000 3D shapes, with different permutations of CAD modeling operations \cite{jignasu2024slice100kmultimodaldatasetextrusionbased, fusion360gallery, Sketch2Seq2025Sun, man2025videocaddatasetmodellearning}, and corresponding sketches to address the domain shift issue and the lack of a CAD dataset. GeoCode \cite{pearl2022geocodeinterpretableshapeprograms} used Blender Geometry Nodes to create models for chairs, vases, and tables as the dataset, with 59, 39, and 36 human-interpretable parameters as input, respectively, and paired them with NPR style. Sketch3D \cite{zheng2024sketch3d, chen2024sketch2nerfmultiviewsketchguidedtextto3d, liu2024sketchdreamsketchbasedtextto3dgeneration, oh2024controldreamerblendinggeometrystyle} used larger datasets \cite{deitke2022objaverseuniverseannotated3d}, and re-rendering all these 3D shapes in Suggestive Contours style become computational expensive, therefore other 2D-filter based methods have been used.
While previous methods primarily relied on synthetic sketch styles, Deepsketch \cite{zhong2020towards} was the first to incorporate 1,500 professional hand-drawn sketches of ShapeNet objects \cite{chang2015shapenet}, highlighting the importance of accounting for these nuances in sketch-based modeling \cite{tracingvsfreehand21}. This shift underscores the need to differentiate between freehand sketches and those generated through synthetic means to better capture the intricacies of human input.

\begin{table}[htbp]
\centering
\tiny  
\setlength{\tabcolsep}{2pt}  

\resizebox{\columnwidth}{!}{%
\begin{tabular}{p{2.3cm}|p{2cm}p{1cm}p{1cm}p{1.3cm}p{1.2cm}}
\toprule
\rowcolor[gray]{0.9} \textbf{Dataset} &
  \textbf{Category} &
  \textbf{Shapes} &
  \textbf{Sketches} &
  \textbf{Style} &
  \textbf{Views} \\ \midrule
Nishida \cite{Nishida2016} &
  Param. Prim. &
  4 &
  47,000 &
  SC &
  1 \\ \midrule
ShapeCOSEG \cite{Delanoy20203DJohanna} &
  Chair, Vase &
  700 &
  5,600 &
  SC &
  8 \\ \midrule
OpenSketch \cite{OpenSketch19}&
  Product Design &
  12 &
  400 &
  H &
  3 \\ \midrule
ShapeNet &
  13 Categories &
  43,783 &
  39,423 &
  Canny/H &
  48 \\ \midrule
Manhattan 1k \cite{TONOVitruvio22} &
  Buildings &
  1,000 &
  24,000 &
  SS &
  24 \\ \midrule
ShapeNet-S3D \cite{zheng2024sketch3d} &
  ShapeNet+Text &
  11,000 &
  220,000 &
  Canny &
  20 \\ \midrule
Objaverse \cite{deitke2022objaverseuniverseannotated3d}&
  3D Objects &
  400,000 &
  - &
  Canny &
  30 \\ \midrule
OmObject3D \cite{wu2023omniobject3d}&
 20 Categories &
  - &
  -  &
  HED &
  -- \\ \midrule



  Sketch3D Hand \cite{sketch2vox} &
  9 Categories &
  - &
  12,877 &
  - &
  - \\ \midrule

  SpeedTracer \cite{tracingvsfreehand21} &
   Various &
   63+ &
   1,498 &
   H & 
   Various \\ \midrule

  DifferSketching \cite{Xiao2022DifferSketching} &
  9 Categories &
  136 &
  3,620 &
  H &
  2-3 \\

\bottomrule
\end{tabular}%
}
\caption{Overview of datasets containing paired sketches and 3D shapes. While ShapeNet \cite{chang2015shapenet} serves as a foundation for many methods, each approach \cite{freehandreconstruction, sketch2model, zhong2020towards, deepsketchmodeling, sketch2mesh, bandyopadhyay2024doodle} employs a specific subset as detailed in the Supplementary materials. In this table, "H" denotes human-created sketches, "SC" refers to suggestive contours style, "SS" indicates unspecified synthetic sketches, and "HED" represents Holistically Nested-Edge style as shown in Figure \ref{fig:stylesketch}. "Param. Prim." abbreviates Parametric Primitive. Nishida \cite{Nishida2016} employs procedural techniques to generate multiple variations from four base 3D objects using underlying grammar rules.}
\label{datasets_table}
\end{table}

\subsection{Input Summary}
\label{conclusionflexibleinputs}

Table~\ref{table:input} reveals that multimodal inputs, particularly combining sketches with text, effectively resolve style and viewpoint ambiguities. Text prompts provide crucial context like \textit{"a chair"} for semantic clarity, \textit{"drawn from a frontal view"} for viewpoint specification, or \textit{"convert this doodle done by a 5-year-old into a chair"} for style information, enhancing neural sketch methods' expressiveness and accuracy. Recent research trends toward advanced techniques including deformation-aware, 3D-aware, and physics-informed losses \cite{uy2021jointlearning3dshape, zhong2020towards}, plus equivariant methods from geometric deep learning \cite{canonicalcapsules, vectorneurons}, promising improved robustness for sketch-based 3D modeling. 


\section{Generative AI Model}
\label{aimodel}

\begin{figure*}[ht]
    \centering
    \includegraphics[width=\textwidth]{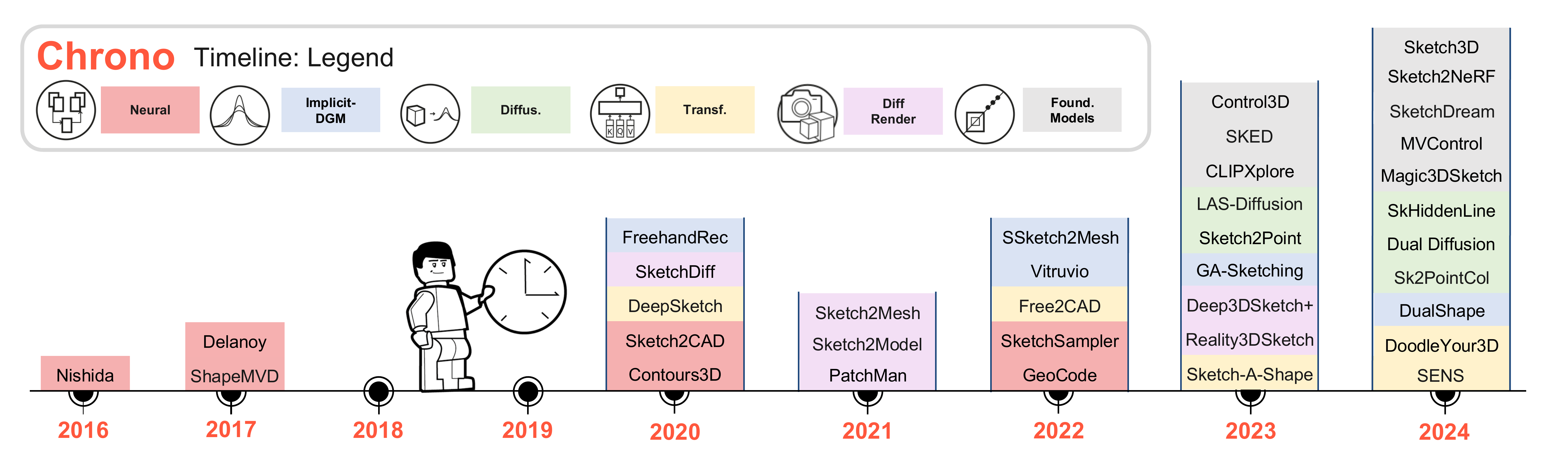}
    \caption{\textbf{Sketch-Modeling Deep Learning Based Methods}. This timeline illustrates the evolution of $\neuralsketchmethods$ methods, highlighting key innovations and breakthroughs in the field. Methods are organized chronologically and color-coded by year to facilitate cross-referencing with Table \ref{tab:model table}, enabling identification of specific research patterns and emerging directions.} 
    \label{fig:chrono}
\end{figure*}

In this section, we provide a chronological overview of $\neuralsketchmethods$ developments driven by key innovations in neural network architecture (Figure~\ref{fig:chrono}). Since these methods often combine multiple innovations simultaneously, such as transformers for structural understanding, diffusion models for geometry generation, and differentiable rendering for 2D-3D alignment, we categorize them based on their primary underlying innovation to simplify our analysis. These fundamental innovations integrate sketch-modeling with these six techniques: neural models (see Section~\ref{regression} \raisebox{-0.1\baselineskip}{\includegraphics[width=0.35cm]{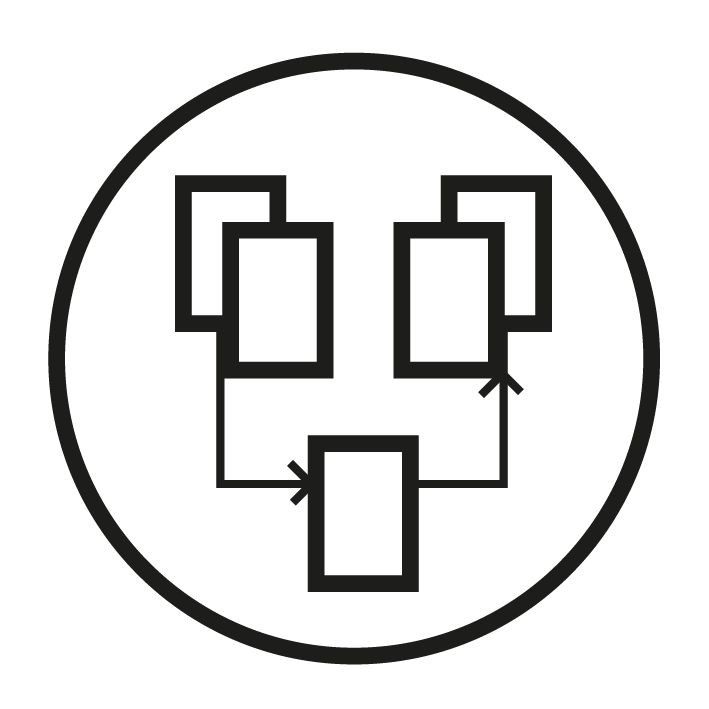}}), deep generative models and novel implicit representation (see Section~\ref{deepgenmodelimplicit} \raisebox{-0.1\baselineskip}{\includegraphics[width=0.35cm]{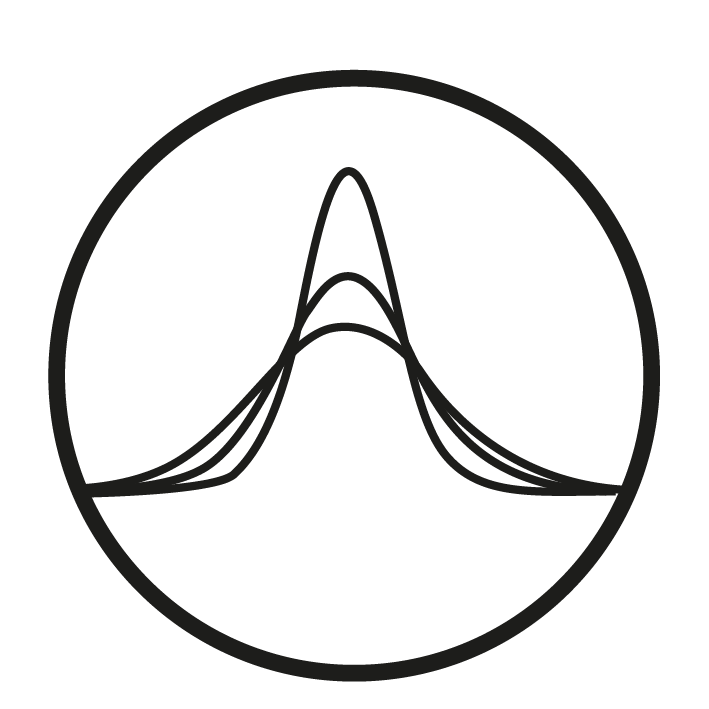}}), diffusion models (see Section~\ref{diffusionmodel} \raisebox{-0.1\baselineskip}{\includegraphics[width=0.35cm]{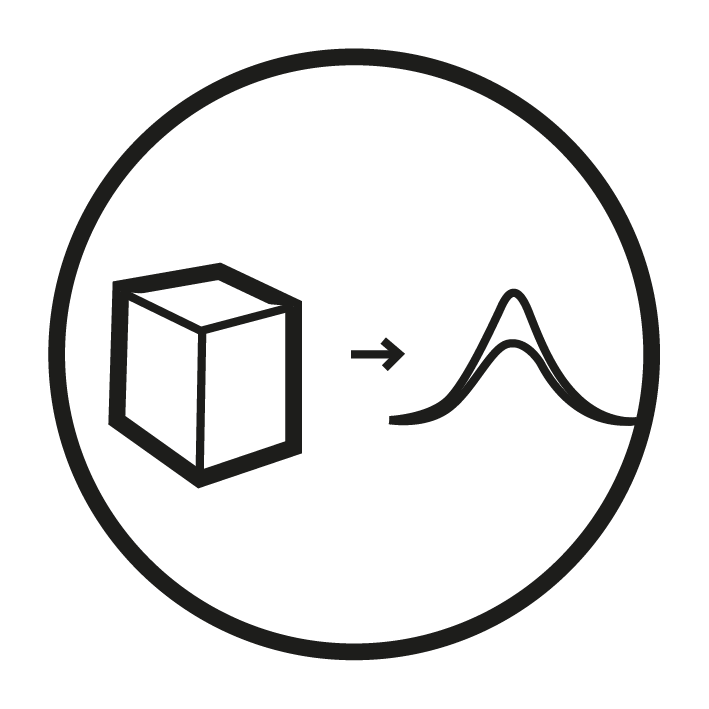}}), differentiable renderers (see Section \ref{differentiablerendering} \raisebox{-0.1\baselineskip}{\includegraphics[width=0.35cm]{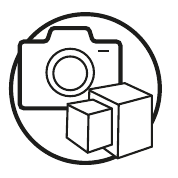}}), transformers (See Section~\ref{transformers} \raisebox{-0.1\baselineskip}{\includegraphics[width=0.35cm]{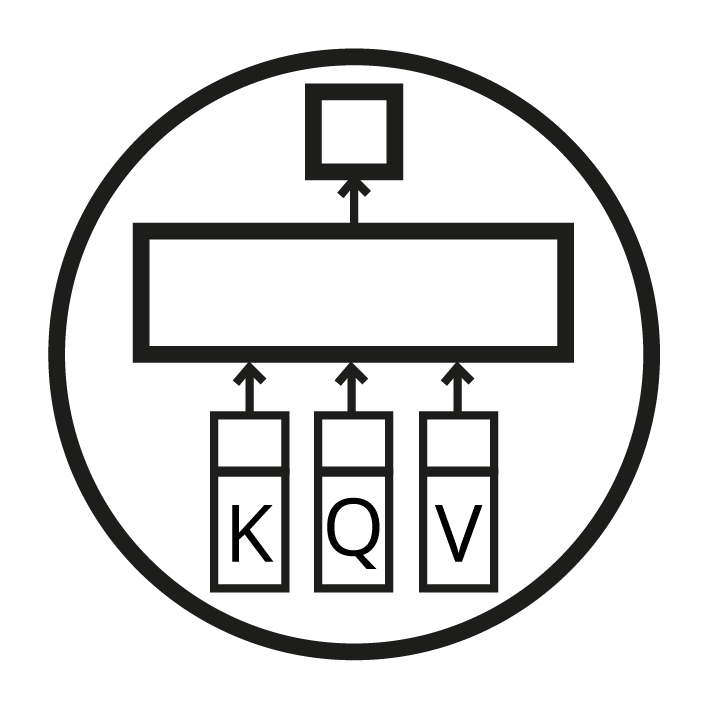}}), and optimization and pre-trained-based (CLIP) models (see Section~\ref{pretrainoptimization} \raisebox{-0.1\baselineskip}{\includegraphics[width=0.35cm]{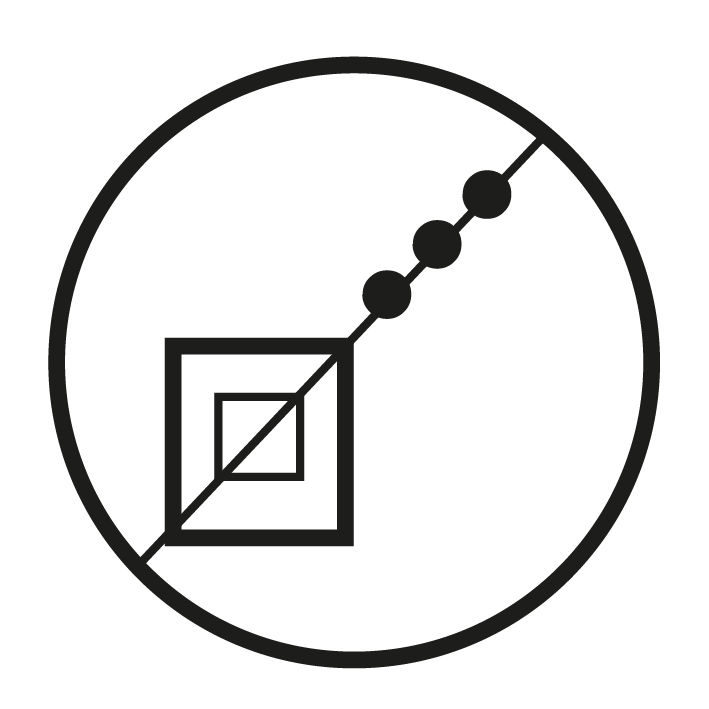}}). Since some of the $\neuralsketchmethods$ adopt several of these innovations, we provide a comprehensive summary of these methods and overall table (see Table \ref{table:neural_sketch_methods}) in the Supplementary \ref{architectures}.


\begin{table}[htbp]
\centering
\setlength{\tabcolsep}{1pt} 
\renewcommand{\arraystretch}{1} 
\small
\begin{tabular}{p{3cm}|p{0.5cm}p{0.5cm}p{0.5cm}p{0.5cm}p{0.5cm}p{0.5cm}}
    \toprule
    \textbf{Paper} & \multicolumn{6}{c}{\textbf{Models}}\\
    &
    \raisebox{-0.3\baselineskip}{\includegraphics[width=0.45cm]{images/UNetCNNRegression.png}} & 
    \raisebox{-0.3\baselineskip}{\includegraphics[width=0.45cm]{images/ImplicitDeepGenerative.png}} & 
    \raisebox{-0.3\baselineskip}{\includegraphics[width=0.45cm]{images/Diffusion.png}} & 
    \raisebox{-0.3\baselineskip}{\includegraphics[width=0.45cm]{images/Transformers.png}} & 
    \raisebox{-0.3\baselineskip}{\includegraphics[width=0.45cm]{images/flexibleview2bw.pdf}} & 
    \raisebox{-0.3\baselineskip}{\includegraphics[width=0.45cm]{images/Optimization.png}}   \\ 
    \midrule
Nishida et al. \cite{Nishida2016} & \cellcolor{pinkish} &  &  &  &  &  \\ 
Delanoy et al. \cite{Delanoy20203DJohanna} & \cellcolor{pinkish} &  &  &  &  &  \\
ShapeMVD \cite{3dshapereconstructionmvcnn} & \cellcolor{pinkish} &  &  &  &  &  \\
Contour3D \cite{Contourbased3DModelingAobo20} & \cellcolor{pinkish} &  &  &  &  &  \\
DeepSketch \cite{zhong2020towards} & \cellcolor{gray!08} &  &  & \cellcolor{lightyellow} &  &  \\
Sketch2CAD \cite{Li2020Sketch2CAD} & \cellcolor{pinkish} &  &  &  &  &  \\
SketchDiff \cite{sketchmodelingdiffrenderer} &  &  &  &  & \cellcolor{lightpurple} &  \\
FreeHandRec \cite{freehandreconstruction} &  & \cellcolor{blueish} & &  &  &  \\
Sketch2Model \cite{sketch2model} & \cellcolor{gray!08} & \cellcolor{gray!08}  &  & & \cellcolor{lightpurple} &  \\
Sketch2Mesh \cite{sketch2mesh} & \cellcolor{gray!08} &  & & & \cellcolor{lightpurple} & \\
Free2CAD \cite{Free2CAD} & \cellcolor{gray!08}  &  &  & \cellcolor{lightyellow} &  & \\
SS2Mesh \cite{bhardwaj2022singlesketch2meshgenerating3d} & \cellcolor{gray!08}  & \cellcolor{blueish} &  &  &  & \\
GeoCode \cite{pearl2022geocodeinterpretableshapeprograms} & \cellcolor{pinkish} &  &  &  &  & \\
SketchSampler \cite{sketchsampler2022eccv} & \cellcolor{pinkish} &  &  &  &  & \\
LAS-Diffusion \cite{zheng2023lasdiffusion} &  & \cellcolor{gray!08} & \cellcolor{lightgreen} &  &  & \\
Sketch-A-Shape \cite{sanghi2023sketchashapezeroshotsketchto3dshape} &  & \cellcolor{blueish} &  &  &  & \\
SKED \cite{Mikaeili_2023_sked} &  &  &  &  &  & \cellcolor{lightgray}\\
CLIPXPlore \cite{CLIPXPlore2023sketchshape} &  &  & \cellcolor{gray!08} & \cellcolor{gray!08}  &  & \cellcolor{lightgray} \\
D3DSketch+ \cite{chen2023deep3dsketchrapid3dmodeling} &  &  &  & \cellcolor{lightyellow} &  &  \\
Control3D \cite{Control3D2023} &  & \cellcolor{gray!08} & \cellcolor{gray!08} &  & \cellcolor{gray!08} & \cellcolor{lightgray}  \\
Re3DSketch \cite{chen2023reality3dsketchrapid3dmodeling} &  & \cellcolor{gray!08} &  &  & \cellcolor{lightpurple} &  \\
Sketch2Point \cite{diffrefforsketchtopointmodeling2023Di} &  &  &\cellcolor{lightgreen} &  &  &  \\
Sketch2Vox \cite{sketch2vox} &  &  & & \cellcolor{lightyellow} &  &  \\
GA-Sketching \cite{zhou2023gasketchingshapemodelingmultiview} &  & \cellcolor{blueish} &  &  & \cellcolor{gray!08} & \\
S2PointCol \cite{Wu2023sketch2pointcolored} &  & \cellcolor{gray!08} & \cellcolor{lightgreen} &  &  & \\
SketchDream \cite{liu2024sketchdreamsketchbasedtextto3dgeneration} &  & \cellcolor{gray!08} & \cellcolor{gray!08} & & \cellcolor{gray!08} & \cellcolor{lightgray}  \\
SENS \cite{sens2024binningerpartawaresketchimplicit} &  & \cellcolor{gray!08} & \cellcolor{gray!08} & \cellcolor{lightyellow} &  & \\
DY3D \cite{bandyopadhyay2024doodle} &  & \cellcolor{gray!08} & \cellcolor{gray!08} & \cellcolor{lightyellow} &  & \\
Vitruvio \cite{TONOVitruvio22} & \cellcolor{gray!08} & \cellcolor{blueish} &  &  &  &  \\
MVControl \cite{li2024controllabletextto3dgenerationsurfacealigned} &  & \cellcolor{gray!08} & \cellcolor{gray!08} & \cellcolor{gray!08} & \cellcolor{gray!08} & \cellcolor{lightgray} \\
SHLine \cite{sketchhiddenline2024} &  & \cellcolor{gray!08} & \cellcolor{gray!08} & \cellcolor{lightgreen} &  & \\
M3DSketch \cite{zang2024magic3dsketchcreatecolorful3d} &  &  &  &  & & \cellcolor{lightgray}  \\
Sketch2NeRF \cite{chen2024sketch2nerfmultiviewsketchguidedtextto3d} &  & \cellcolor{gray!08} & \cellcolor{gray!08} & & \cellcolor{gray!08} & \cellcolor{lightgray} \\ 
DualShape \cite{dualshape2024} &  & \cellcolor{blueish} & &  &  &  \\
Sketch3D \cite{zheng2024sketch3d} &  & \cellcolor{gray!08} &  & &  \cellcolor{gray!08} &  \cellcolor{lightgray} \\
    \bottomrule
\end{tabular}
\caption{Categorization of $\neuralsketchmethods$'s methods alongside their corresponding model architectures, emphasizing the primary contributions. The colored cell in each row represents what we identify as the main contribution, following the color scheme in the Figures \ref{fig:overalldivisionframework} and \ref{fig:chrono}. The light gray cells indicate the other technical aspects present in each paper.} 
\label{tab:model table}
\end{table}

\subsection{Neural Models: \raisebox{-0.3\baselineskip}{\includegraphics[width=0.65cm]{images/UNetCNNRegression.png}}}
\label{regression}
Neural models in Table~\ref{tab:model table} establish direct correspondences between 2D image coordinates and 3D representations or via parametric shape modeling or via direct coordinate mapping. In parametric approaches, neural models predict specific parameters that define a 3D shape program (see previous neurosymbolic STAR \cite{ritchie2023neurosymbolicmodelscomputergraphics}), often relying on representations that consist of procedural operations, like shape grammars or parametric models \cite{procedural_model, Huang2017ShapeSynthesissketchprocedural, smirnov2021learning}. \textbf{Nishida} et al.~\cite{Nishida2016} used a cascade of CNNs to classify partial sketches into grammar snippets and estimate their parameters. This approach allows for generating a wide variety of buildings by combining different grammar rules and adjusting building parameters. \textbf{Free2CAD} \cite{Free2CAD} predicts stroke groupings that correspond to CAD operations, followed by parameter optimization for each group to match the strokes. The parameters are optimized to reproduce the strokes provided as input. Instead, \textbf{GeoCode} \cite{pearl2022geocodeinterpretableshapeprograms} predicts parameters for constructive solid geometry (CSG) \cite{CSGNet, yuan2024diffcsgdifferentiablecsgrasterization} operations and extrusion heights, aiming for interpretable shape programs for 3D reconstruction. \textbf{Sketch2CAD} \cite{Li2020Sketch2CAD}, similar to Free2CAD, utilizes regression to predict parameters for CAD operations. It focuses on sequential CAD modeling, parsing user sketches into a series of commands and their associated parameters, limiting their overall shape (see Section \ref{geometry} for more information). However other methods succeed in modeling complex topologies by providing a deformable parametric template composed of Coons NURBS patches to the decoder \cite{smirnov2021learning}. 

In coordinate mapping approaches, models learn direct transformations from image pixel locations to their corresponding 3D spatial positions. \textbf{Delanoy et al.} \cite{Delanoy20203DJohanna, DELANOY201965voxelnormal} uses an online method; an updater-CNN iteratively maps the input sketch to a voxel-based representation by utilizing fixed camera viewpoints and perspectives, thus determining which voxels are occupied or empty. Other methods use a two-step approach and multiple sketches: \textbf{ShapeMVD} \cite{3dshapereconstructionmvcnn} first predicts depth and normal maps from input sketches. Then, it fuses depth and averages it into a point cloud. Similarly, \textbf{DeepSketch} \cite{deepsketchmodeling, zhong2020towards} generates normal maps, 2.5D depth maps \cite{marrnet}, and foreground masks from sketches, and then fuses to the 3D space via mapping. While these approaches rely on 2.5D information, they perform poorly with more complex geometries.

\subsection{Deep Generative Models and/or Implicit Rep.: \raisebox{-0.3\baselineskip}{\includegraphics[width=0.65cm]{images/ImplicitDeepGenerative.png}}}
\label{deepgenmodelimplicit} 

\begin{figure}[ht!]
    \centering
    \includegraphics[width=\columnwidth]{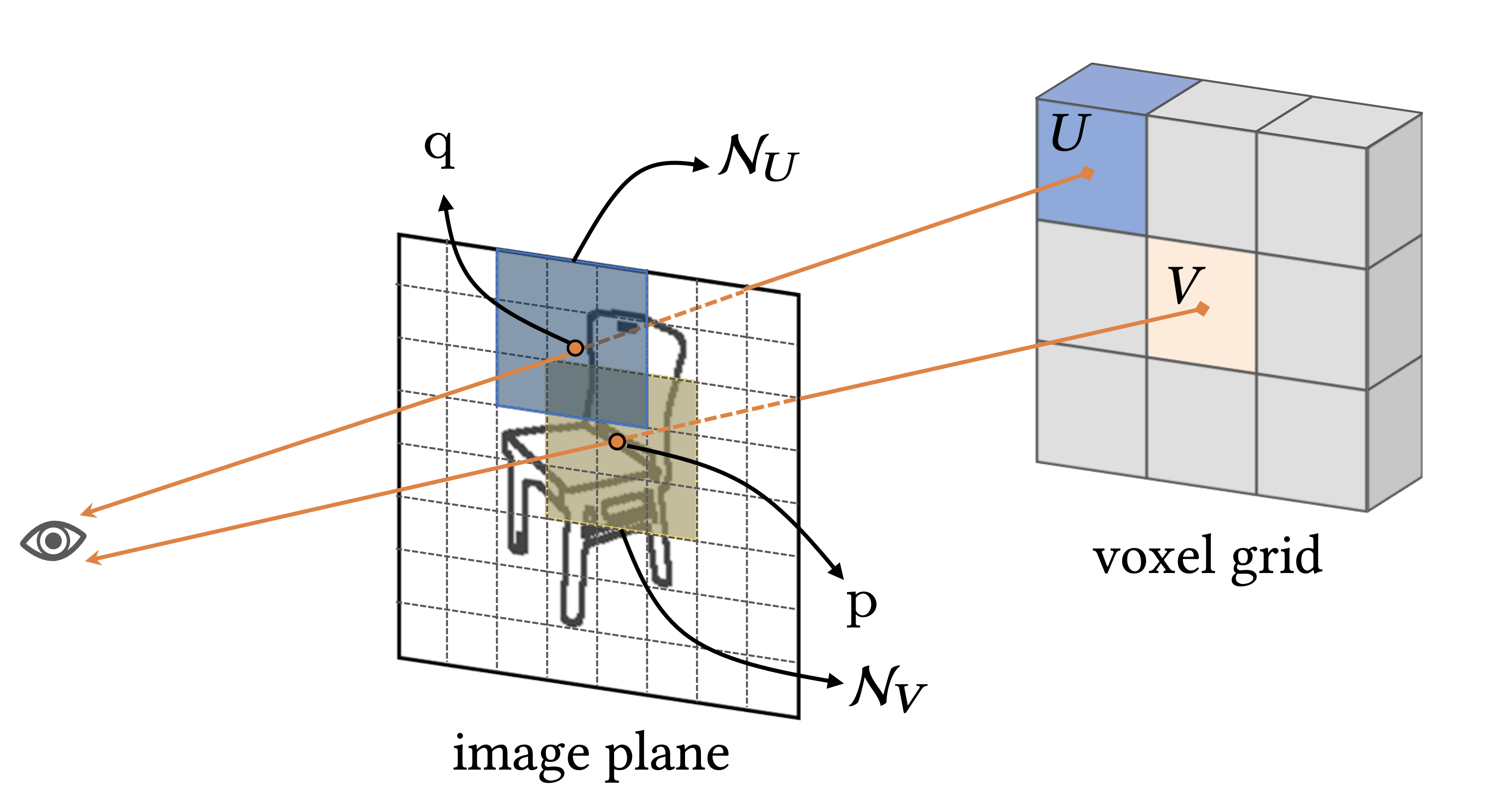}  
    \caption{Demonstration of LAS-Diffusion \cite{zheng2023lasdiffusion} approach for local view-based attention. The center of voxel $V$ is mapped to the image plane at point $p$ using a pre-determined perspective projection. The image patch features in the vicinity of point $p$ (highlighted in yellow) are used to interact with the voxel features of $V$ within the U-Net architecture through cross-attention. The same process is applied to other voxels, such as $U$. LAS-Diffusion uses 2D image patch features to guide 3D voxel feature learning methods. A voxel grid is projected to the viewer's image plane. Once this relationship is established, patch image features $f_N$ (ViT) interact with the voxel features $f_V$ via a U-Net and one-layer multi-head cross-attention \cite{Vaswani2017attentionisallyouneed}. }
    \label{fig:viewawareattention}
\end{figure}

This section focuses on methods based on deep generative models and/or implicit representations presented in Table~\ref{tab:model table}. For a more in-depth understanding of these topics, we invite the reader to review \cite{DeepGenerativeModellingreview2022, DeLuigi2023deeplearningonimplicitneuralrepresentation}. A 3D implicit representation is defined by a volumetric function $f: \mathbb{R}^3 \rightarrow \mathbb{R}$. For a point cloud, this function takes as input a point in 3D such as $ \mathbf{p}=(x, y, z) \in \mathbb{R}^3$ and passes it through a neural network to produce a scalar output in $\mathbb{R}$ written as $\quad \mathbf{p} \mapsto f(\mathbf{p})$, where $f(\mathbf{p})$ encodes information about the 3D shape at point $\mathbf{p}$. In $\neuralsketchmethods$ methods, this function can be represented by a Signed Distance Function (SDF) \cite{sketch2mesh}, Occupancy Function \cite{TONOVitruvio22}, or more broadly, Neural Implicit Representation \cite{sketch2mesh, bhardwaj2022singlesketch2meshgenerating3d, dualshape2024} composed of a neural network that is typically a multi-layer perceptron (MLP) and used as a continuous function (Deep Implicit Fields). \textbf{Sketch2Mesh} \cite{sketch2mesh} uses \cite{meshsdf} a MeshSDF encoder-decoder architecture to represent and refine a SDF to match the target external contour by using a differentiable renderer (Section \ref{differentiablerendering}). These methods tend to produce a compact implicit representation that benefits $\neuralsketchmethods$ speed \cite{sketch2mesh, TONOVitruvio22}(see the online column in the Supplementary Table~\ref{tab:model table_supplementary}). Sketch2Mesh combines view- and contour-aware methods with implicit representations, increasing the flexibility to viewpoint and sketching style. However, this flexibility comes at the cost of sensitivity, thus requiring high precision to accurately capture fine geometric details. Methods like \textbf{Sketch2Model} \cite{sketch2model} and \textbf{ShapeMVD} \cite{3dshapereconstructionmvcnn} also use implicit representations, but they do not employ deep generative models. 
Methods like \textbf{Vitruvio} \cite{TONOVitruvio22} and \textbf{DeepSketch} \cite{zhong2020towards, deepsketchmodeling} leverage deep generative models as Variational AutoEncoder (VAE) \cite{Occnet2019} and Generative Adversarial Network (GAN) \cite{marrnet} respectively. If Vitruvio directly maps a single sketch to 3D shapes, DeepSketch (3DSkVP) uses the GAN to translate multiple sketches to a 2.5D domain with normal depth and mask maps fused to generate the 3D mesh via Iterative Closest Point (ICP) and Poisson surface reconstruction \cite{poissonsurfacereconstruction}. \textbf{FreeHandRec} \cite{freehandreconstruction} uses a pix2pix-based \cite{pix2pix} sketch standardization module to reduce the style variability. Unlike Sketch2Mesh, it does not use differentiable rendering techniques to map sketches to 3D space; rather, it uses a view-aware 3D reconstruction network. FreeHandRec encodes the 3D shape with global latent space, which fails to capture fine-grained details of local parts.  To overcome this limitation \textbf{LAS-Diffusion} \cite{zheng2023lasdiffusion, sketchguided3dgenAI24} controls these local parts with local patch features provided by a view-aware locally attentional SDF (see Figure~\ref{fig:viewawareattention} and Section~\ref{diffusionmodel}).

\begin{figure}[ht!]
    \centering
    \includegraphics[width=\columnwidth]{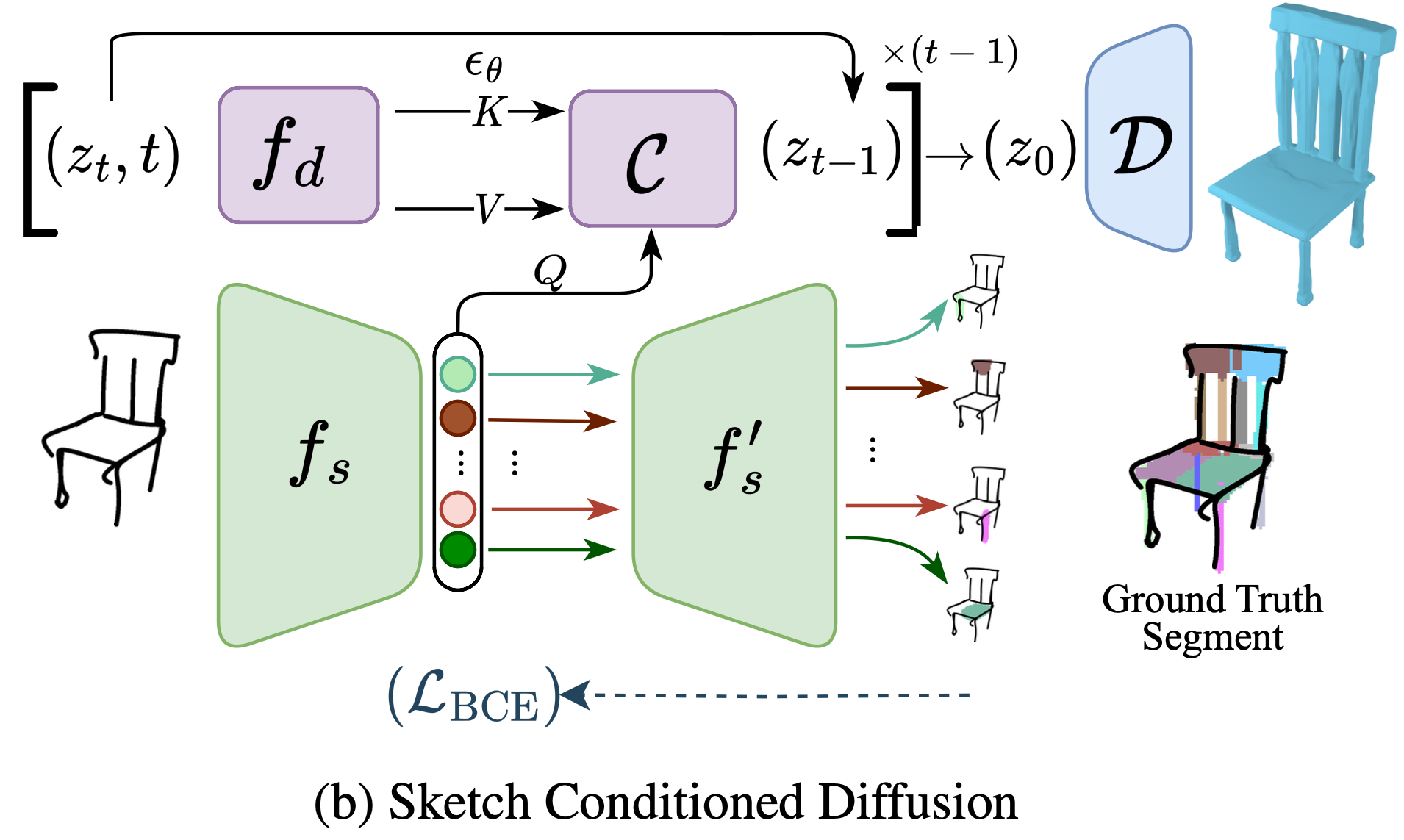} 
    \caption{Doodle Your 3D \cite{bandyopadhyay2024doodle} processes sketches by encoding them into part-disentangled representations using an encoder $f_s$, where the sketches are segmented into maps representing individual parts. A shared decoder $f_s^{\prime}$ is applied to these segmented maps. The resulting sketch representations are then passed into the attention module $\mathcal{C}$ as the Query, while the intermediate diffusion outputs from $f_d$ serve as the Key-Value pairs.}
    \label{doodleyour3d}
\end{figure}

LAS-Diffusion does not excel in hand-drawn human sketches that have highly distorted lines, cluttered linework, or inconsistent perspectives. To mitigate this issue, \textbf{Doodle Your 3D} \cite{bandyopadhyay2024doodle} leverages a latent diffusion model with a part-disentangled decoder (Figure~\ref{doodleyour3d}). It establishes correspondence among CLIPasso \cite{vinker2022clipasso} (see Section~\ref{input}) semantic edge maps, and projected 3D part regions. This part-aware Neural Implicit Shape modeling is initiated by SPAGHETTI \cite{hertz2022spaghettieditingimplicitshapes}. SPAGHETTI's decoder reconstructs a part-disentangled latent space via inversion. The alignment between the input sketch and output parts is trained with a diffusion model, enabling semantic edits that are crucial to capturing fine-grained level details.

\subsection{Diffusion Models: \raisebox{-0.3\baselineskip}{\includegraphics[width=0.65cm]{images/Diffusion.png}}}
\label{diffusionmodel}

\begin{figure*}[!htbp]
   \centering
   \includegraphics[width=2\columnwidth]{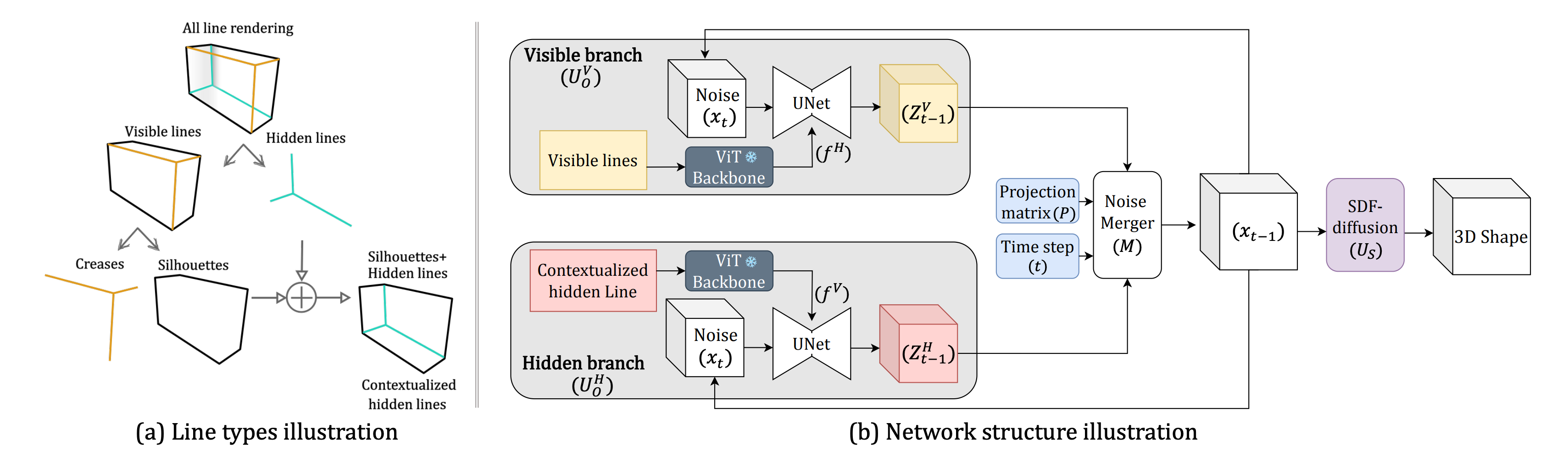}
   \caption{Fukushima et al., \cite{sketchhiddenline2024} captures also hidden lines. It captures the back information (see Figure~\ref{fig:PartialInformation}) leveraging hidden lines common in several practices \cite{Front2Back2020}.}
   \label{hiddenlines}
   \vspace{-10pt}  
\end{figure*}
\vspace{-5pt} 

In this section, we analyze papers leveraging diffusion models to implement their $\neuralsketchmethods$ methods, as displayed in Table~\ref{tab:model table}. For a more in-depth survey about diffusion models, refer to \cite{bai2023dreamdiffusiongeneratinghighqualityimages, Po2023star_diffusion_models}. Diffusion Models (DMs) are a class of deep generative models (see Section~\ref{deepgenmodelimplicit}) that learn to generate 3D shapes $\hat{x}_{shape}$ (for brevity $\hat{x}_{3D}$) from a data distribution $p(x)$ by first learning a forward diffusion process that gradually adds Gaussian noise $\epsilon$ to training data $x_{shape}$ (for brevity $x_{3D}$) over discrete time steps $t = 1, 2, ..., T$ until it becomes pure noise. The model then learns to reverse this noising process, starting from random noise and iteratively denoising to generate new 3D shapes $\hat{x}_{3D}$ \cite{denoisingdiffusionmodelddpm2020Ho}. Training typically uses an MSE loss between the actual Gaussian noise and the predicted noise $\hat{\epsilon}$, formulated as $\mathcal{L}_{MSE}=|\hat{\epsilon}_t-\epsilon_t|_2^2$. This gradual corruption of the input can be realized in the data domain, or in the latent space $z$ to reduce the dimensionality of the problem. Throughout the literature, different approaches have been adopted to consider the sketch $I_{\text{sketch}}$ (for brevity $I_s$) in the corruption process of the clean input $x_{3DS}$ as a form of conditioning \cite{ho2022classifier} (refer to \cite{cao2024controllablegenerationtexttoimagediffusion, zhan2024conditionalimagesynthesisdiffusion}). This learned noise can be represented as a function of the time $t$, the conditioning sketch $I_{s}$ and the corrupted shape $x_{3DS}$ at the time $t$ as $\epsilon_\phi(x_{3DS}; t, I_{s})$. 

\textbf{Control3D} \cite{Control3D2023} uses a 2D latent diffusion model \cite{zhang2023adding} optimized on Neural Radiance Field (NeRF) after converting the input sketch $I_s$ to a colored image $I_c$. Control3D used view-dependent prompting; they added camera specifications to the prompt $y$ transforming the noise to $\epsilon_\phi(x_{3DS}; t, y, I_{c})$. \textbf{Doodle Your 3D} \cite{bandyopadhyay2024doodle} operates directly on the 3D latent representation $z$ \cite{denoisingdiffusionmodelddpm2020Ho}. It uses a 3D part-latent diffusion model, where the implicit 3D shape representation $\hat{x}_{3DS_{implicit}}$ is divided in $m$ parts ($Z \in \mathbb{R}^{m \times d}$) aligned with the sketch-parts $E(I_{s})$ $\in \mathbb{R}^{m \times d_s}$ \cite{vinker2022clipasso, informativedrawings} via multi-head attention blocks as cross attention modules \cite{Vaswani2017attentionisallyouneed, zheng2023lasdiffusion} (see Section~\ref{transformers} and \ref{deepgenmodelimplicit}); this allows to capture part-specific 3D details that in Control3D have been overlooked due to the global latent space of the NeRF. \cite{bandyopadhyay2024doodle} establish this part-alignment for only a single category (category-specific model like \cite{sketch2model}) because each part carries specific semantic meaning, and the latent variable  $Z$  is aligned across all shapes within the chair category. Other $\neuralsketchmethods$ methods capture finer details while maintaining the ability to generalize across multiple categories. They operate directly on the global latent space ($Z \in \mathbb{R}^{d}$) of the 3D representation $x_{3DS}$. \textbf{LAS-Diffusion} \cite{zheng2023lasdiffusion} tackles the quality gap capturing fine-grained details by employing a two-stage 3D diffusion process that leverages discrete signed distance function (SDF) representation. The first stage, called occupancy diffusion, generates a coarse discrete occupancy function \cite{TONOVitruvio22, Occnet2019} to approximate the shape's shell using a 3D U-Net \cite{Delanoy20203DJohanna, DELANOY201965voxelnormal}. The forward process \cite{variationaldiffusionmodelkingma2021} is represented as: $ \mathbf{x_{3DS_t}} = \sqrt{\alpha_t}\mathbf{x_{3DS_0}} + \sqrt{1-\alpha_t} \bm{\epsilon}, $ where $\mathbf{x_{3DS}}$ is the initial 3D shape used for self-conditioning \cite{ting2022analogbits}. 
The second stage, SDF-diffusion, refines the occupied voxels from the first stage to produce a high-resolution SDF. For controllability, LAS-Diffusion uses 2D image patch features to guide 3D voxel feature learning. A voxel grid is projected to the image plane of the viewer; once this relationship is established, patch image features $f_N$ (ViT) interact with the voxel ones $f_V$ via a U-Net and one-layer multi-head cross-attention in the decoder (see usual formulation $f_V^{\text{new}} = \text{MH-Attention}(Q, K, V, \mathcal{M})$ with respectively $Q = f_V W^Q, \quad K = f_N W^K, \quad V = f_N W^V$). LAS-Diffusion achieves a higher level of control over fine-grained details of the local geometry but lacks details about the color, material, or textures. 

To capture color-related details \textbf{Sketch2PointCol} \cite{Wu2023sketch2pointcolored}, the encoder for the diffusion process uses a capsule attention mechanism to encode the sparsity of the pixels in the sketch, and then a multimodal text-sketch joint embedding to reduce the text-based ambiguities. The geometry diffusion stage employs a U-Net conditioned on sketch and text features embedding $C_g$ to generate the point cloud geometry; in this case, the noise is produced by the 3D point cloud $\left\|\epsilon-\epsilon_\theta\left(\mathbf{x_{3DS_t}}, C_g, t\right)\right\|^2$
The method also uses convolutional layers, capsule networks, and attention mechanisms. The capsule network identifies and focuses on the informative pixels within the sketch, ignoring the background. The text feature fusion uses BERT \cite{Devlin2019BERT} and multi-head attention fusion to produce the joint embedding that guides the diffusion process. However, the users with these methods lack control over the entire shape. The users draw from a frontal view and do not provide information about the rear, as shown in Figure~\ref{fig:PartialInformation}. Suppose these approaches leverage data to complete the partial information the sketch provides. In that case, SHLine\cite{sketchhiddenline2024} aligns with the users' needs, especially in automotive and industrial design industries, where 3D scaffolding is provided, and precision is paramount. \textbf{SHLine} \cite{sketchhiddenline2024} extends LAS-Diffusion to leverage hidden lines commonly found in technical drawings. SHLine distinguishes creases and silhouette lines from the hidden ones. The model has two parallel occupancy-based diffusion networks that converge to a CNN-based noise merger before entering the SDF-diffusion network. One processes visible lines, and the other processes hidden lines and silhouette lines for context. It tackles geometry and appearance separately. Both branches employ occupancy-diffusion networks, denoted as $I_{s_v}$ for visible lines and $I_{s_h}$ for hidden lines. They predict denoised occupancy grids in a unified noise volume.  This noise merger network, $M$, combines the denoised outputs from both branches, as displayed in Figure~\ref{hiddenlines}.

\subsection{Transformer-Based Models: \raisebox{-0.3\baselineskip}{\includegraphics[width=0.65cm]{images/Transformers.png}}}
\label{transformers}

In this section, we analyze papers that introduce transformers and attentions-based methods to $\neuralsketchmethods$ methods, as shown in Table~\ref{tab:model table}.
Transformers \cite{Vaswani2017attentionisallyouneed} excel at processing sequential data, making them well-suited for handling the ordered nature of sketch strokes \cite{sketchrnn}. They capture long-range dependencies within the input sequence, which is crucial for understanding the relationships between strokes in a complex sketch and for accurately grouping them into meaningful units \cite{Li2020Sketch2CAD}. For a more detailed survey on transformers, see \cite{LIN2022111transformerssurvey}. Transformers exploit self-attention mechanisms. \textbf{DeepSketch} \cite{zhong2020towards} adopted and re-purposed this self-attention mechanism to ensure
that an automatically generated attention map aligns with that of the ground-truth 3D shape. To solve previous problems, such as the one introduced at the end of Section \ref{differentiablerendering}, of shape misalignment across multiple view optimization, DeepSketch learns a global
non-linear geometric transformation between an input sketch
and its 3D shape counterpart via a spatial transform network (STN). Through their attention mechanism, transformers can selectively focus on relevant parts of the sketch, mitigating the impact of ambiguities. For instance, when encountering a sketch with unclear viewpoint information, a transformer-based model can learn to prioritize strokes that provide stronger cues about the intended 3D shape, leading to more accurate reconstructions. If they improved the performance of previous methods, they did not fully solve the ambiguity problem related to the style and viewpoint. To reduce this ambiguity, using a fixed viewpoint and a predefined sketching style proved to help. 

For example, \textbf{Free2CAD} \cite{Free2CAD} uses an axonometric fixed viewpoint to better leverage the transformer's ability to handle sequential data, introducing a novel stroke grouping task to learn CAD operations. Free2CAD encodes the strategic knowledge of CAD modeling by using the pen strokes as complete input drawings, grouping them \cite{yangsketchgnn2021}, and fitting geometric parameters. It has been trained on synthetically-generated data, but the model adapts to novice users. It takes as input the complete drawing. It mitigates the exposure bias \cite{mihaylova2019scheduledsamplingtransformers, he2021maskedautoencodersscalablevision} provided by the synthetic sequence of operation with specific transformer-based geometric fitting. To avoid strong assumptions on a fixed axonometric viewpoint, \textbf{Sketch-a-Shape} \cite{sanghi2023sketchashapezeroshotsketchto3dshape}
conditions a 3D Discrete Auto-Encoder (Vector Quantized Variational Auto-encoder SkexGen VQ-VAE \cite{xu2022skexgen}) on the CLIP features of Non Photorealist Render (NPR) from Canny edge \cite{canny1996edgedetection}. Showing the robustness of CLIP (obtained from a frozen large pre-trained vision model) \cite{vinker2022clipasso} also in 3D. The model outputs multiple shapes per sketch query thanks to the Masked Transformer, positional encoding, and cross-attention mechanism. The 3D shapes are encoded into
a sequence of discrete indices $Z$, pointing to a shape dictionary, whose distributions are then modeled later with the bi-directional masked transformer.  \cite{chang2022maskgitmaskedgenerativeimage}. However, Sketch-a-Shape does not leverage parts awareness of the different shapes.

To solve this problem, \textbf{SENS} \cite{sens2024binningerpartawaresketchimplicit} uses transformers to process visual embeddings of sketch patches and a set of part queries to predict latent part vectors, which are subsequently used to reconstruct the 3D shape. This transformer-based approach facilitates a part-aware generation process, enabling the model to handle complex shapes and avoid mere retrieval from a database of existing models. SENS generates neural implicit shapes via hand-drawing sketches. It is based on SPAGHETTI \cite{hertz2022spaghettieditingimplicitshapes} that maps input parts to the latent space. These approaches necessitate a large dataset of sketches and pair 3D data. Another SPAGHETTI-based approach has introduced a proposed solution: \textbf{Doodle Your 3D} \cite{bandyopadhyay2024doodle}. Doodle Your 3D leverages multi-head attention with a latent diffusion model for better part disentanglement in the decoder. This approach establishes a correspondence between semantic edge maps generated via CLIPasso \cite{vinker2022clipasso} and projected 3D part regions, reducing dependence on human sketch-3D shape paired datasets \cite{DemSketchto3DShapeRetrievalPivoting23} and approximating more natural human sketching patterns. The part-level modeling is initiated from SPAGHETTI \cite{hertz2022spaghettieditingimplicitshapes}; the decoder inversion reconstructs a part-disentangled latent space. The alignment between the input sketch and output parts is trained with a diffusion model, enabling semantic edits. 

\subsection{Differentiable Rendering Models: \raisebox{-0.3\baselineskip}{\includegraphics[width=0.65cm]{images/flexibleview2bw.pdf}}}
\label{differentiablerendering}

Differentiable rendering (DR) seeks to bridge the gap between 2D images and 3D scenes by incorporating the rendering process into neural network training pipelines for end-to-end learning, as grouped in Table~\ref{tab:model table}. This enables a bidirectional interaction between 2D and 3D representations, facilitated by the power of gradient-based optimization techniques, which are also applied later in this report (see Section~\ref{pretrainoptimization}). By making the rendering process differentiable, it establishes a direct connection between the pixels in the input image and the 3D parameters that define the scene \cite{ravi2020accelerating3ddeeplearningpytorch3D, jatavallabhula2019kaolinpytorchlibraryaccelerating, TensorflowGraphicsIO2019, Mitsuba3}. For an exhaustive survey about differentiable rendering, please see \cite{kato2020differentiablerenderingsurvey}. In short, DR parameters are material $\phi_m$, light $\phi_l$, geometry $\phi_s$, and camera $\phi_c$. These parameters are used as inputs to output a colored image. However, this information is not present in a single black-and-white rapid scribble. So one may wonder, how can DR be integrated and support $\neuralsketchmethods$ methods? DR needs an overall function, summarized as $\phi = {\phi_s, \phi_m, \phi_c, \phi_l}$, which generates an image defined as $I$ with color and depth information. $\neuralsketchmethods$ methods on the contrary start from the sketch $I_{\text{sketch}}$ to generate a 3D shape $\Phi_s$. If this DR works with colored 2.5D (with depth information) images, it fails with sketches due to the lack of color and depth information. To mitigate this limitation, \textbf{Sketch2Mesh} \cite{sketch2mesh} employs a hybrid approach, combining a traditional encoder/decoder architecture for initial 3D mesh generation with a differentiable rendering-based refinement step. This process helps align the projected mesh with the input sketch accurately. Sketch2Mesh uses two differentiable rendering variants: \textit{Render} and \textit{Chamfer}. \textit{Render} uses a differentiable rasterizer (SoftRas) \cite{ravi2020accelerating3ddeeplearningpytorch3D} to generate a corresponding binary mask to back-propagate the information to the 3D representation. \textit{Chamfer} directly minimize the difference between sketch and the projection of the 3D mesh's external contours, moving the optimization in the sketch contour space similar concept applied to Control3D \cite{Control3D2023}, to improve the sketch-fidelity via its sketch-consistency loss $\mathcal{L}_{\text{sketch}}$ (see Section~\ref{pretrainoptimization}). \textit{Chamfer} approach outperforms \textit{Render} for its simplicity and robustness to style changes. This robustness to the style holds only for professional sketches \cite{zhong2020towards}. 

To accommodate inputs from nonprofessional users, Smirnov et al.~\cite{smirnov2021learning} (referred to as PatchMan in the table) use sketch augmentation strategies in the dataset preparation. The authors used occluding contours and sharp edges (using the Arnold Toon Shader in Autodesk Maya). They augmented these drawings via special techniques (vectorization and augmentation by stochastic split and curve truncation). Even with this augmentation, their initial sketch still requires a good initialization from a correct perspective, which is uncommon in doodles \cite{bandyopadhyay2024doodle}. To address the aforementioned style limitations and make the method compatible with novice hand drawings or doodles, \textbf{Sketch2Model} \cite{sketch2model} disentangle the latent codes for shape and viewpoint. This disentanglement allows the model to be view-aware and explicitly leverage the viewpoint information during generation. Here, the SoftRas DR generates the silhouette from a specific viewpoint. Then a silhouette loss $L_s$ compares the generated silhouette $S_1$ to the ground truth $S_2$, summarizing $L_s = \mathcal{L}_{\text {iou }}\left(S_1, S_2\right)$. A problem arises when the method tries to leverage silhouettes from multiple views. A single silhouette is insufficient to capture all the information of a 3D object. \textbf{Deep3DSketch+} \cite{chen2023deep3dsketchrapid3dmodeling} addresses this by sampling multiple camera viewpoints and generating silhouettes from each. It introduces a structural-aware adversarial training strategy. This strategy includes a Stroke Enhancement Module (SEM) to capture the structural information and facilitate learning realistic and detailed shape structures thanks to a Shape Discriminator (SD). SD allows more consistency in the 3D representation to ensure it is used as a reference in each optimization step. Similar approaches are used in Section~\ref{pretrainoptimization} with Score Distillation Losses in 3D.

\subsection{Pre-Trained Optimization-Based Models (Found.): \raisebox{-0.3\baselineskip}{\includegraphics[width=0.65cm]{images/Optimization.png}}}
\label{pretrainoptimization}

Developments of diffusion models \cite{diffbetterthanGAN, Po2023star_diffusion_models} brought to popularity text-based 3D content generation \cite{poole2022dreamfusion, shi2024mvdreammultiviewdiffusion3d, sun2023dreamcraft3dhierarchical3dgeneration}, for more see text-to-3D surveys \cite{lee2024textto3d, liu2024comprehensivesurvey3dcontent,sun2024survey3dcontentcreationimplicit,Xia_2023surveydeepgen3dawareimagesynthesis}. In this section, we survey optimization-based generation methods that use pre-trained large language and vision models \cite{oh2024controldreamerblendinggeometrystyle} as displayed in the last column of Table~\ref{tab:model table}, represented as Foundation model \cite{foundationmodels}. These methods leverage techniques from differentiable render (see Section \ref{differentiablerendering}) to optimize both geometry and texture via specific loss function, differently from the initial image-conditioned ones based on NeRF optimizations \cite{poole2022dreamfusion, wang2023prolificdreamer} we focused on sketch-conditioned methods. For example Mikaeili et al. propose \textbf{SKED} \cite{Mikaeili_2023_sked}, introducing the ability to edit the NeRF representation via multiple 3D consistent sketch-based contour masks. SKED for editing uses (Instant-NGP) and $\mathcal{L}_{SDS}$ for the usual SDS 3D consistency loss (as highlighted in Table \ref{tab:paper_performance}). $\mathcal{L}_{pres}$ for the preservation loss, this loss uses a distance-based weighting mechanism to preserve the base 3D object's original content selectively. This approach ensures that the edits are localized to the regions specified by the user's sketches while maintaining the fidelity of the original object elsewhere. $\mathcal{L}_{sil}$ ensures that the density added during the editing process occupies the regions specified by the user's sketches. Instead of directly comparing the sketch with the edited 3D object, which is difficult due to their different representations, SKED cleverly utilizes object mask renderings. This loss penalizes the model if the rendered object mask does not closely match the provided sketch masks. Minimizing this loss encourages the generated 3D edits to fill the areas outlined in the sketches accurately. $\mathcal{L}_{sp}$ for the sparsity to minimize the entropy with the sketch masks for each view. This loss is focused on preserving the shape of the added content to ensure doesn't alter the overall initial shape. If SKED focuses on partial edits later models, such as \textbf{Control3D} \cite{Control3D2023}, generate the entire 3D objects from initial sketches. Chen et al. with Control3D introduce these sketch-conditioned capabilities with the Score Jacobian Chaining (SJC) \cite{Wang2023CVPRScoreJacobianChainingSJC} and enforce the sketch-fidelity via a novel sketch-consistency loss $\mathcal{L}_{\text{sketch}}$. This loss optimize the correspondences between the original sketch and the photo-to-sketch model $G$ \cite{Li2019photosketching} via a dot product of these two images embedding produced by the CLIP-encoder \cite{radford2021learningtransferablevisualmodels} $E$:  $\mathcal{L}_{\text{sketch}} = -E(G(x))^T E(I_s)$. Here the goal is to output photorealistic representation and the loss is specifically designed to achieve high-fidelity outputs. However, a notable limitation is the lack of explicit mechanisms to capture and preserve the user's original creative intent. While the technical focus on visual fidelity drives impressive results, the optimization process doesn't necessarily include constraints or guidance to ensure the generated content aligns with the conceptual goals or stylistic preferences the user had in mind when creating the initial sketch. 

These limitations persist in subsequent approaches, including Li et al. with \textbf{MVControl} \cite{li2024controllabletextto3dgenerationsurfacealigned} and related works \cite{chen2024sketch2nerfmultiviewsketchguidedtextto3d,wu2024recentadvances3dgaussian}. Notably, MVControl diverges from using hand-drawn sketches, instead relying on NPR edge maps derived through Canny edge filters applied to images. MVControl employs dual optimization objectives: the standard Score Distillation Sampling loss 
($\mathcal{L}_{\mathrm{SDS}}$) and a hybrid variant ($\mathcal{L}_{S D S}^{\text {hybrid}}$) that integrates both 2D ($\mathcal{L}_{S D S}^{2 D}$) and 3D ($\mathcal{L}_{S D S}^{3 D}$). Similarly, \textbf{Sketch2NeRF} \cite{chen2024sketch2nerfmultiviewsketchguidedtextto3d} uses synthetic sketches from OmniObject3D-Sketch dataset \cite{wu2023omniobject3d} and implements a camera-aware reconstruction loss. This composite loss function incorporates perceptual loss ($\mathcal{L}_{\mathrm{LPIPS}}$) \cite{Zhang2018perceptualmetric} to capture high-level structural and stylistic image features, alongside L1 loss ($\mathcal{L}_{\mathrm{L} 1}$) to preserve pixel-level sketch details. Beyond these image-space objectives, Sketch2NeRF adopts geometric regularization techniques from DreamFusion \cite{poole2022dreamfusion} that constrain the underlying 3D structure, while also implementing random viewpoint regularization to ensure consistent 3D representation. This comprehensive approach prevents overfitting to input viewpoints and produces coherent geometry from novel angles, effectively addressing common issues such as view-dependent ambiguities, floating artifacts, and near-plane geometry distortions typical of NeRF-based approaches. More advances techniques evolved with the progress made in single image-based 3D generation\cite{Zou202410324gaussiantriplanetransformersingleview3d} via Guassian Splatting \cite{Kerbl2023gaussiansplatting}. In fact, 
\textbf{Sketch3D} outputs Gaussian Splatting (survey \cite{wu2024recentadvances3dgaussian}) and SDS optimization strategies. The sketch loss $\mathcal{L}_{\text{sketch}}$ is a similarity loss ($\mathcal{L}_2$ loss) between the two CLIP-based (ResNet101) encoders with 4 layers. It uses a weighted color loss ($\mathcal{L}_{\text {Col }}$) to ensure consistent quality across viewpoints, and a structural loss $\nabla_\theta \mathcal{L}_{\mathrm{S}-\mathrm{SDS}}$) to align 3D structure with the sketch. These losses still focus on achieving photorealist results and do not include user's intent such as fabricability of the object or its costs.

\begin{figure}[ht!]
    \centering
    \includegraphics[width=\columnwidth]{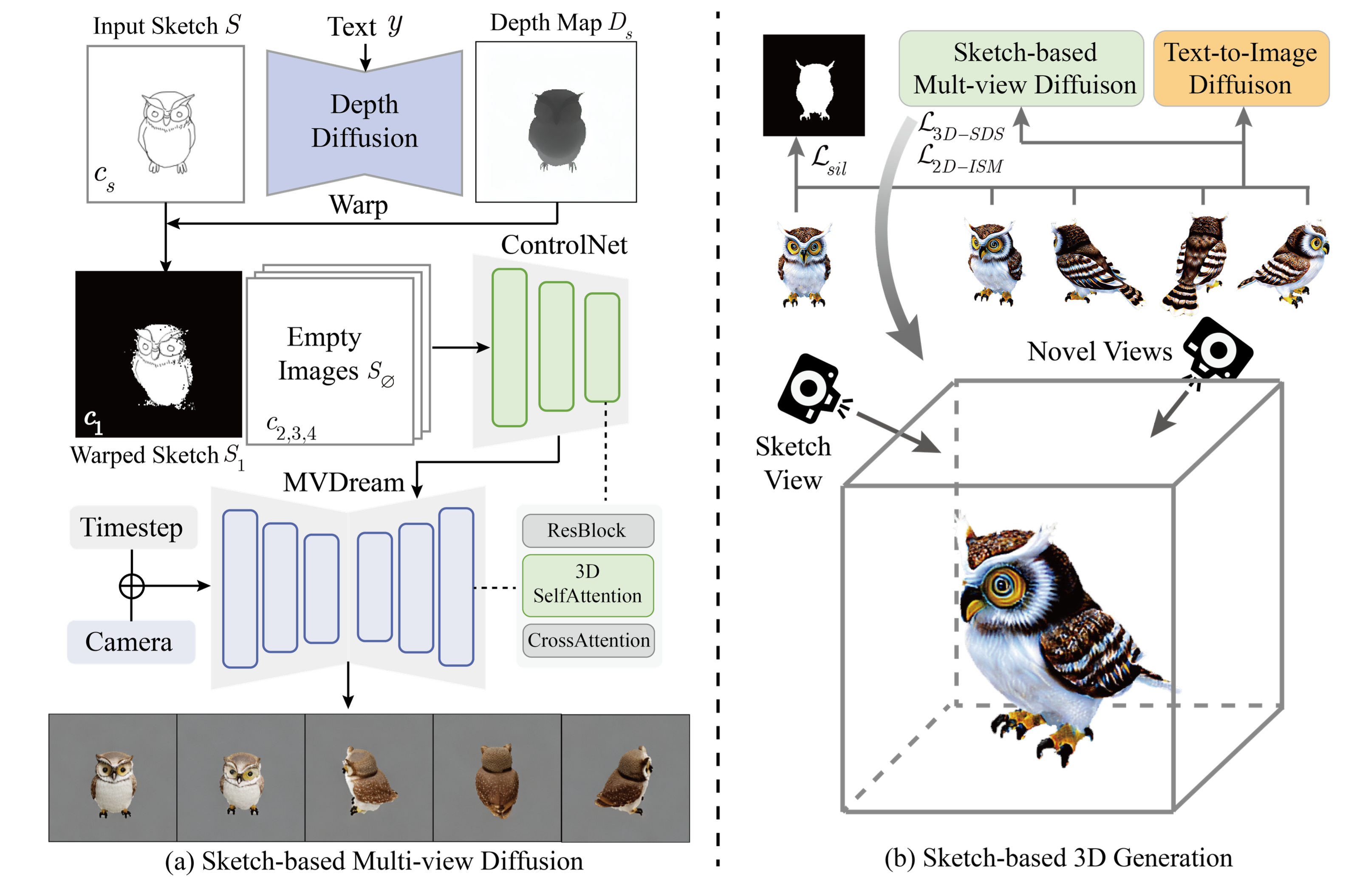}  
    \caption{SketchDream \cite{liu2024sketchdreamsketchbasedtextto3dgeneration} shows a text-based generation augmented by the sketch. The Score Distillation Sampling is facilitated by the guidance provided by the sketch.}
    \label{SketchDream}
\end{figure}


\begin{table}[ht!]
    \centering
    \renewcommand{\arraystretch}{1.3} 
    \setlength{\tabcolsep}{4pt} 

    \resizebox{\columnwidth}{!}{ 
    \begin{tabular}{lccccc}
        \toprule
        \textbf{Paper} & \textbf{Speed (min)} & \textbf{GPU} & \textbf{FT Dataset} & \textbf{Input} & \textbf{Losses} \\
        \midrule
        \cite{Mikaeili_2023_sked} & ~40* & 3090 & - & SketchMask & $\mathcal{L}_{SDS}+\mathcal{L}_{pres}+ \mathcal{L}_{sil}+\mathcal{L}_{sp}$ \\
        \cite{Control3D2023} & ~60 & V100 & - & AS** & $\mathcal{L}_{sketch}$ \\
        \cite{oh2024controldreamerblendinggeometrystyle} & - & - & - & Canny & $\mathcal{L}_{SDS}$ \\
        \cite{li2024controllabletextto3dgenerationsurfacealigned} & - & A100 & Obja/LAION & Canny  & $\mathcal{L}_{SDS}^{2D}+\mathcal{L}_{SDS}^{3D}$ \\
        \cite{chen2024sketch2nerfmultiviewsketchguidedtextto3d} & ~120 & 3090 & Omni+THum-S & - & $\mathcal{L}_{sketch}+\mathcal{L}_{reg}+\mathcal{L}_a$ \\
        \cite{liu2024sketchdreamsketchbasedtextto3dgeneration} & ~75 & A100 & Objaverse  &  -  &  $\mathcal{L}_{SDS}+\mathcal{L}_{sil}+\mathcal{L}_{ISM}$ \\
        \cite{zheng2024sketch3d} & 3 & 4090 & SS3D & Canny  & $\mathcal{L}_{\text{sketch}}+\mathcal{L}_{\text{Col}}+\mathcal{L}_{\text{SDS}}$ \\
        \bottomrule
    \end{tabular}
    } 
    \caption{Presents the evaluation performance of different methods. AS: Abstract Sketch. "*": fixed views.  **: The specific abstract sketch style is not specified. FT Datasets: dataset used to fine-tune the model. SS3D: ShapeNet-Sketch3D dataset \cite{zheng2024sketch3d}. The SketchDream loss 2D Interval Score Matching (ISM) is from LucidDreamer \cite{liang2023luciddreamerhighfidelitytextto3dgeneration}. Obja: Objaverse. Omni+THum-S: OmniObject3D-S+THuman-S.  "-": not specified. }
    \label{tab:paper_performance}
\end{table}

\subsection{Models Summary}

Examining Table~\ref{tab:model table} and Table \ref{tab:paper_performance}, we notice that while newer methods deliver superior results, they make significant trade-offs in on-device real-time interaction capabilities.  As illustrated in Figure \ref{fig:chrono}, the evolution of $\neuralsketchmethods$ methods has closely followed advancements in single-image to 3D generation techniques. This progression reveals distinct developmental phases, each with its own strengths and limitations: Neural network models initially offered precise parameter control but imposed rigid constraints on program-based output shapes, limiting creative flexibility. Deep generative models subsequently expanded output diversity but struggled to accurately capture design intent and generate high-quality 3D shapes. The introduction of diffusion models marked significant improvement in output quality, though these approaches often prioritized visual fidelity over practical design considerations. Transformer architectures brought important advances by effectively capturing relationships between sketch elements, while differentiable rendering techniques ensured view consistency across perspectives. Most recently, pre-trained optimization methods have achieved remarkable photorealistic results, albeit without adequately accounting for functional intent.
This developmental trajectory reveals a consistent limitation across all approaches: optimization objectives overwhelmingly prioritize visual appearance over user intent factors such as functionality, fabricability, and other practical considerations. The predominant loss functions ($\mathcal{L}_{\mathrm{SDS}}$, $\mathcal{L}_{\mathrm{sketch}}$, $\mathcal{L}_{\mathrm{Col}}$ ) reflect  this emphasis on visual accuracy rather than addressing practical design considerations. For example, when generating a couch from a sketch, current systems optimize primarily for visual appeal rather than the designer's intent for factors like ease of assembly, cost-effective fabrication, or ergonomic functionality. Some emergent research has begun addressing these limitations, with novel approaches becoming more body-aware for wearable objects \cite{guo2024shapecraft} and incorporating fabrication-aware constraints \cite{Fabricationawaredesign2024}. The high cost of training 3D foundation models has catalyzed a shift in $\neuralsketchmethods$ research towards leveraging multimodal inputs. By combining sketches, images, and text, a new generation of models can generate editable 3D shapes. These approaches are broadly categorized by their reliance on strong 2D priors \cite{liu2024sketchdreamsketchbasedtextto3dgeneration, zang2024magic3dsketchcreatecolorful3d}, 3D priors \cite{sanghi2024waveletlatentdiffusionwala}, or program synthesis via Large Language 3D Modelers (LL3M) \cite{lu2025ll3m}. This paradigm facilitates the use of diverse generative architectures, such as autoregressive, flow, and diffusion models, which are designed to exploit both the sequential nature of text and sketches \cite{sketchrnn} and the global properties of images. A particularly promising direction, especially for discrete representations like Computer-Aided Design (CAD), is Discrete Diffusion \cite{shi2025maskeddiffusiondiscretedata,sahoo2024simple}. These mask-based models are adept at infilling tasks without a fixed generation order, making them ideal for part-based 3D representations. They are also optimized for limited-data regimes and can operate directly on discrete data, reducing the risk of overfitting common in autoregressive models due to their Evidence Lower Bound (ELBO) type of loss. To summarize the research community continues to prioritize two critical areas: developing information-rich part-aware representations and enabling 3D foundation models to ingest multiple modalities to generate several options that accurately reflect the user's intent as we will see in the next Section.

\section{Outputs}
\label{output}
In this section, we introduce Table~\ref{awareevaluation} and structure the discussion around the outputs of $\neuralsketchmethods$. More importantly, we focus on how these outputs have been evaluated, including the metrics used to assess their quality and their alignment with user intent. Table~\ref{awareevaluation} is organized chronologically along the vertical axis, while the horizontal axis examines three key aspects: part-based semantic information (see Section \ref{partbasedsemantics}), the quantity of generated shapes (see Section \ref{optionquantity}), and the geometric characteristics of the generated shapes (see Section \ref{geometry}).

All these considerations are directly connected to how the output is evaluated via qualitative and quantitative metrics. Furthermore, Section \ref{partbasedsemantics}, \ref{optionquantity}, and \ref{geometry} present the relation between the user's intent and \textbf{metrics} used to evaluate the 3D output generated by $\neuralsketchmethods$ methods. For a more in-depth report on different 3D representations, we invite the reader to review \cite{Xia_2023surveydeepgen3dawareimagesynthesis}. The 3D output representation is critical for the design of the $\neuralsketchmethods$ model's architecture (see Supplementary \ref{supplementary} and \ref{architectures}). Therefore, to adopt end-to-end $\neuralsketchmethods$ methods, the 3D representation must be differentiable (see Supplementary for an overview of the different 3D representations). Another critical aspect concerns the information embedded in the final 3D outputs.

\begin{table}[htbp]
\centering
\setlength{\tabcolsep}{1pt} 
\renewcommand{\arraystretch}{1} 
\small
\begin{tikzpicture}
\node (table) {
\begin{tabular}{p{3cm}|p{0.5cm}p{0.5cm}p{0.5cm}|p{0.5cm}p{0.5cm}p{0.5cm}|p{0.5cm}p{0.5cm}p{0.5cm}}
    \toprule
    \rowcolor{red!5}
    \textbf{Paper} & \multicolumn{3}{c|}{\textbf{Part semantic}} & \multicolumn{3}{c|}{\textbf{Options}} & \multicolumn{3}{c}{\textbf{Geometry}} \\
    \rowcolor{red!5}
    & \raisebox{-0.3\baselineskip}{\includegraphics[width=0.45cm]{images/generativesemantic1.pdf}} & \raisebox{-0.3\baselineskip}{\includegraphics[width=0.45cm]{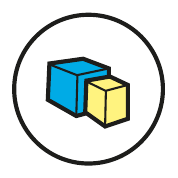}} & \raisebox{-0.3\baselineskip}{\includegraphics[width=0.45cm]{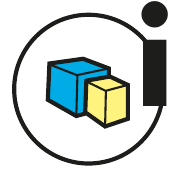}} & \raisebox{-0.3\baselineskip}{\includegraphics[width=0.45cm]{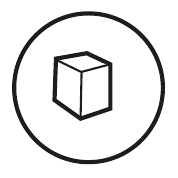}} & \raisebox{-0.3\baselineskip}{\includegraphics[width=0.45cm]{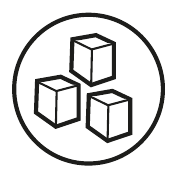}} & \raisebox{-0.3\baselineskip}{\includegraphics[width=0.45cm]{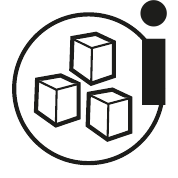}} & \raisebox{-0.3\baselineskip}{\includegraphics[width=0.45cm]{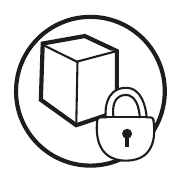}} & \raisebox{-0.3\baselineskip}{\includegraphics[width=0.45cm]{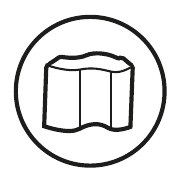}} & \raisebox{-0.3\baselineskip}{\includegraphics[width=0.45cm]{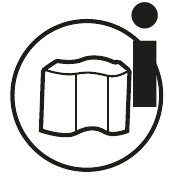}} \\
    \midrule
    Nishida et al.~\cite{Nishida2016} &  &  & \cellcolor{gray!25} &  & \cellcolor{gray!25} &  & \cellcolor{gray!25} &  &  \\
    Delanoy et al.~\cite{Delanoy20203DJohanna} &  &  & \cellcolor{gray!25} &  &  &  & \cellcolor{gray!25} &  &  \\
    ShapeMVD \cite{3dshapereconstructionmvcnn} &  &  &  &  & \cellcolor{gray!25} &  & \cellcolor{gray!25} &  &  \\
    Contour3D \cite{Contourbased3DModelingAobo20} &  &  &  & \cellcolor{gray!25} &  &  & \cellcolor{gray!25} &  &  \\
    DeepSketch \cite{zhong2020towards} & \cellcolor{gray!25} &  &  & \cellcolor{gray!25} &  &  &  & \cellcolor{gray!25} &  \\
    Sketch2CAD \cite{Li2020Sketch2CAD} &  & \cellcolor{gray!25} &  & \cellcolor{gray!25} &  &  & \cellcolor{gray!25} &  &  \\
    DiffSketch \cite{sketchmodelingdiffrenderer} &  &  &  & \cellcolor{gray!25} &  &  & \cellcolor{gray!25} &  &  \\
    Freehandrec \cite{freehandreconstruction} &  &  &  & \cellcolor{gray!25} &  &  &  & \cellcolor{gray!25} &  \\
    Sketch2Model \cite{sketch2model} &  &  &  & \cellcolor{gray!25} &  &  & \cellcolor{gray!25} &  &  \\
    Sketch2Mesh \cite{sketch2mesh} & \cellcolor{gray!25} &  &  &  & \cellcolor{gray!25} &  & \cellcolor{gray!25} &  &  \\
    Free2CAD \cite{Free2CAD} & \cellcolor{gray!25} &  &  & \cellcolor{gray!25} &  &  &  &  & \cellcolor{gray!25} \\
    SS2Mesh \cite{bhardwaj2022singlesketch2meshgenerating3d} &  &  &  & \cellcolor{gray!25} &  &  &  & \cellcolor{gray!25} &  \\
    GeoCode \cite{pearl2022geocodeinterpretableshapeprograms} &  & \cellcolor{gray!25} &  &  & \cellcolor{gray!25} &  &  & \cellcolor{gray!25} &  \\
    SketchSampler \cite{sketchsampler2022eccv} &  &  &  & \cellcolor{gray!25} &  &  &  & \cellcolor{gray!25} &  \\
    LAS-Diffusion \cite{zheng2023lasdiffusion} &  &  &  &  & \cellcolor{gray!25} &  &  & \cellcolor{gray!25} &  \\
    Sketch-A-Shape \cite{sanghi2023sketchashapezeroshotsketchto3dshape} &  &  &  &  & \cellcolor{gray!25} &  & & \cellcolor{gray!25} &  \\
    SKED \cite{Mikaeili_2023_sked} & \cellcolor{gray!25} &  &  & \cellcolor{gray!25} &  &  &  &  & \cellcolor{gray!25} \\
    CLIPXPlore \cite{CLIPXPlore2023sketchshape} & \cellcolor{gray!25} &  &  &  & \cellcolor{gray!25} &  &  &  & \cellcolor{gray!25} \\
    D3DSketch+ \cite{chen2023deep3dsketchrapid3dmodeling} &  &  &  & \cellcolor{gray!25} &  &  &  & \cellcolor{gray!25} &  \\
    Control3D \cite{Control3D2023} &  &  &  & \cellcolor{gray!25} &  &  &  & \cellcolor{gray!25} &  \\
    Re3DSketch \cite{chen2023reality3dsketchrapid3dmodeling} &  & \cellcolor{gray!25} &  &  & \cellcolor{gray!25} &  &  & \cellcolor{gray!25} &  \\
    Sketch2Point \cite{diffrefforsketchtopointmodeling2023Di} &  &  &  & \cellcolor{gray!25} &  &  &  & \cellcolor{gray!25} &  \\
    GA-Sketching \cite{zhou2023gasketchingshapemodelingmultiview} &  &  &  & \cellcolor{gray!25} &  &  &  & \cellcolor{gray!25} &  \\
    S2PointCol \cite{Wu2023sketch2pointcolored} &  &  & \cellcolor{gray!25} & \cellcolor{gray!25} &  &  &  &  & \cellcolor{gray!25} \\
    Sketch2Vox \cite{sketch2vox} & \cellcolor{gray!25} &  &  & \cellcolor{gray!25} &  &  &  & \cellcolor{gray!25} &  \\
    SketchDream \cite{liu2024sketchdreamsketchbasedtextto3dgeneration} & \cellcolor{gray!25} &  &  & \cellcolor{gray!25} &  &  &  &  & \cellcolor{gray!25} \\
    SENS \cite{sens2024binningerpartawaresketchimplicit} &  & \cellcolor{gray!25} &  & \cellcolor{gray!25} &  &  &  & \cellcolor{gray!25} &  \\
    DY3D \cite{bandyopadhyay2024doodle} &  & \cellcolor{gray!25} &  & \cellcolor{gray!25} &  &  &  & \cellcolor{gray!25} &  \\
    Vitruvio \cite{TONOVitruvio22} &  &  &  & \cellcolor{gray!25} &  &  &  &  & \cellcolor{gray!25} \\
    MVControl \cite{li2024controllabletextto3dgenerationsurfacealigned} &  &  &  & \cellcolor{gray!25} &  &  &  & \cellcolor{gray!25} &  \\
    SHLine \cite{sketchhiddenline2024} &  &  &  & \cellcolor{gray!25} &  &  & \cellcolor{gray!25} &  &  \\
    M3DSketch \cite{zang2024magic3dsketchcreatecolorful3d} & \cellcolor{gray!25}   &  &  & \cellcolor{gray!25} &  &  &  & \cellcolor{gray!25} &  \\
    Sketch2Nerf \cite{chen2024sketch2nerfmultiviewsketchguidedtextto3d} & \cellcolor{gray!25} &  &  & \cellcolor{gray!25} &  &  &  &  & \cellcolor{gray!25} \\
    DualShape \cite{dualshape2024} &  &  &  & \cellcolor{gray!25} &  &  & \cellcolor{gray!25} &  &  \\
    Sketch3D \cite{zheng2024sketch3d} & \cellcolor{gray!25} &  &  & \cellcolor{gray!25} &  &  &  & \cellcolor{gray!25} &  \\
    \bottomrule
\end{tabular}
};

\tikzset{
  pattern size/.store in=\mcSize, 
  pattern size=5pt,
  pattern thickness/.store in=\mcThickness, 
  pattern thickness=0.5pt,
  pattern radius/.store in=\mcRadius, 
  pattern radius=1pt
}
\makeatletter
\pgfdeclarepatternformonly[\mcThickness,\mcSize]{custom north west lines}
{\pgfqpoint{0pt}{0pt}}
{\pgfqpoint{\mcSize}{\mcSize}}
{\pgfqpoint{\mcSize}{\mcSize}}
{
  \pgfsetcolor{\tikz@pattern@color}
  \pgfsetlinewidth{\mcThickness}
  \pgfpathmoveto{\pgfqpoint{0pt}{0pt}}
  \pgfpathlineto{\pgfqpoint{\mcSize}{\mcSize}}
  \pgfusepath{stroke}
}
\makeatother

\begin{scope}[pattern color=red, opacity=0.22]
  \draw[fill=red!22, pattern=custom north west lines] 
    (table.north west) + (4.8cm, -9.8cm) rectangle ++(4.4cm, -13.5cm);
\end{scope}

\begin{scope}[pattern color=red, opacity=0.22]
  \draw[fill=red!22, pattern=custom north west lines] 
    (table.north west) + (6.45cm, -8.8cm) rectangle ++(5.65cm, -13.5cm);
\end{scope}

     

\end{tikzpicture}

\caption{ $\neuralsketchmethods$ methods' Output.  \raisebox{-0.3\baselineskip}{\includegraphics[width=0.45cm]{images/generativesemantic1.pdf}}: part division but without annotations, \raisebox{-0.3\baselineskip}{\includegraphics[width=0.45cm]{images/genearivesemantic2.pdf}}: with annotations. \raisebox{-0.3\baselineskip}{\includegraphics[width=0.45cm]{images/genearivesemantic3.pdf}}: with annotation and info. \raisebox{-0.3\baselineskip}{\includegraphics[width=0.45cm]{images/generativeoption1.pdf}}: single output. \raisebox{-0.3\baselineskip}{\includegraphics[width=0.45cm]{images/generativeoption2.pdf}}: multiple options. \raisebox{-0.3\baselineskip}{\includegraphics[width=0.45cm]{images/generativeoption3.pdf}}: multiple options with info. \raisebox{-0.3\baselineskip}{\includegraphics[width=0.45cm]{images/awaregeodif1.pdf}}: limited topology. \raisebox{-0.3\baselineskip}{\includegraphics[width=0.45cm]{images/awaregeodif2.pdf}}: various topology. \raisebox{-0.3\baselineskip}{\includegraphics[width=0.45cm]{images/awaregeodif3.pdf}} : Various topology with info. Red hatches show research gaps.}
\label{awareevaluation}
\end{table}

\subsection{Part-based semantics}
\label{partbasedsemantics}
 
This section categorizes $\neuralsketchmethods$ methods based on how they structure output components, as shown in Table~\ref{awareevaluation}. This subcategorization indicates the:

\begin{itemize}
    \item \raisebox{-3.0pt}{\includegraphics[width=0.5cm]{images/generativesemantic1.pdf}} geometry division,
    \item \raisebox{-3.0pt}{\includegraphics[width=0.5cm]{images/genearivesemantic2.pdf}} part-based representation, or
    \item \raisebox{-3.0pt}{\includegraphics[width=0.5cm]{images/genearivesemantic3.pdf}} part-based representation with semantic information. 
\end{itemize}

The first approach focuses on decomposing output into fundamental geometric entities. As recent works demonstrate, these decompositions manifest through various geometric forms, including topological features such as holes, and constructive geometrical operations such as extrusions.\cite{sketchhiddenline2024, Free2CAD}. Other methods employ geometric primitives, such as Gaussians \cite{xie2024sketchguidedcagebased3dgaussian, zheng2024sketch3d}, cuboids \cite{primitive3d, cuboidsprimitivestulsiani16}, superquadrics \cite{Superquadrics, SuperquadricsPaschalidou2019CVPR} or mathematical surfaces like Coons patches \cite{smirnov2021learning}. Additional geometric decomposition strategies are embraced in neural fields (NeRF) \cite{chen2024sketch2nerfmultiviewsketchguidedtextto3d, Control3D2023} and point cloud representations \cite{Wu2023sketch2pointcolored}. Notably, even when NeRF-based approaches produce visually distinct parts through color variation, these distinctions typically do not carry structural information. This consideration is expanded in Section \ref{discussion}, when we discuss editing capabilities for these parts.

The second approach incorporates higher-level structural meaning into component representations \cite{partnet, partglot, li2024pastacontrollablepartawareshape}. These methods segment shapes into functionally meaningful parts \cite{pearl2022geocodeinterpretableshapeprograms, sens2024binningerpartawaresketchimplicit}, such as the back, the seat, and the legs of a chair. This structural decomposition enhances understanding of component relationships, facilitates more intuitive editing and manipulation, improving the alignment with design intent. 

The third approach extends part-based representations by incorporating additional semantic attributes. These attributes include appearance details \cite{Wu2023sketch2pointcolored}, material properties, cost estimate, data, color information, and domain-specific metadata. This semantic enrichment enhances the usability and contextual relevance of generated models, supporting more informed design decisions. 

Additionally, empty rows in Table~\ref{awareevaluation} indicate when the shape is represented by a single unified entity \cite{TONOVitruvio22, sanghi2023sketchashapezeroshotsketchto3dshape} as a overall mesh, constructive solid geometry, or another non-decomposable structure.

\textbf{Metrics:} These approaches are evaluated with metrics that capture both geometric accuracy and semantic meaningfulness. For a meaningful part-level segmentation, ad-hoc metrics like mean Intersection over Union (mIoU) are used. In Sketch2PointColor \cite{Wu2023sketch2pointcolored}, mIoU quantify how point cloud part segmentation captures color features. Here, distinct colors represent different parts, two for table and three for an airplane, car, and chair. 
Another critical evaluation dimension is the structural stability assessment. GeoCode \cite{pearl2022geocodeinterpretableshapeprograms} implements a dual-criteria stability metric comprising structural integrity verification and physics-based simulation for practical stability assessment. This approach acknowledges that geometrically accurate reconstructions might not translate to physically stable objects, for example generated vessels with bases that are too narrow.
Similar to Sketch2PointColor \cite{Wu2023sketch2pointcolored}, NeRF-based approaches \cite{liu2024sketchdreamsketchbasedtextto3dgeneration, chen2024sketch2nerfmultiviewsketchguidedtextto3d} often employ a two-stage methodology to enable editability. This two-stage approach is evaluated with perceptual metrics. Additional specialized metrics exist for component-specific evaluation, such as the retrieval-based metrics in DualShape \cite{dualshape2024}.
To ensure a valuable output, the 3D representation and its constituent parts must align with user intent and enable a controlled and informed design process \cite{performanceinformeddesign}. User intent can vary widely; for instance, a user generating a "chair" may have specific goals for its parts. They might desire a more functional "armrest", a more cost-effective "seat" made from a different material, a more comfortable "backrest", or more structurally sound "legs". By defining these goals as the initial user intent, a part-based 3D model can provide performance feedback for each component. For example, it could calculate the maximum weight the "legs" can sustain, predict the comfort level of the "seat" with specific metrics, or estimate the aesthetic appeal for a target demographic. This performance-informed feedback empowers the user to control the design and drive further iterations effectively. See Table~\ref{tab:evaluation_metrics} for more evaluation metrics that are often used to evaluate the output as a whole.

\subsection{Amount of Output Options}
\label{optionquantity}

This section examines the number of output options generated by $\neuralsketchmethods$, as shown in Table~\ref{awareevaluation}. A fundamental challenge in $\neuralsketchmethods$ arises from the inherent ambiguity of sketch inputs, as a single sketch captures only partial information about the intended 3D representation see Figure \ref{fig:PartialInformation}. Methods have addressed this challenge through techniques that enable sampling and generating multiple 3D variations. However, the ability to produce multiple high-quality options for users to evaluate, compare, and select from remains an ongoing challenge. Significant bottlenecks persist, particularly in model architecture and computational efficiency, limiting the feasibility of real-time, diverse, and high-quality output generation.
The approaches to output generation can be categorized into three distinct categories: 

\begin{itemize}
    \item \raisebox{-3.0pt}{\includegraphics[width=0.5cm]{images/generativeoption1.pdf}} one 3D shape,
    \item\raisebox{-3.0pt}{\includegraphics[width=0.5cm]{images/generativeoption2.pdf}} multiple 3D shapes, or
    \item\raisebox{-3.0pt}{\includegraphics[width=0.5cm]{images/generativeoption3.pdf}} multiple 3D shapes with information.
\end{itemize}

\textbf{Metrics:} Recent work has begun to address the evaluation of multiple shape outputs and their diversity. For instance, Sketch-A-Shape \cite{sanghi2023sketchashapezeroshotsketchto3dshape} implements a comprehensive evaluation framework that measures accuracy across various input modalities. Their evaluation encompasses multiple datasets, including ImageNet-Sketch (IS-Acc), TU-Berlin Sketch (TU-Acc), ShapeNet-Sketch (SS-Acc), and QuickDraw (QD-Acc). Notably, Sketch-A-Shape demonstrates the capability to generate multiple shapes per sketch query, with reported metrics based on the mean performance across five sampled shapes for each sketch input.

\subsection{Geometry} 
\label{geometry}
This section categorizes output shapes based on their geometric complexity and topological properties, specifically considering the genus of the geometry (the number of "holes" or handles), as detailed in Table~\ref{awareevaluation}. We classify models according to their capability to generate:

\begin{itemize}
    \item \raisebox{-3.0pt}{\includegraphics[width=0.5cm]{images/awaregeodif1.pdf}} generates 3D shapes with \textit{limited topologies},
    \item  \raisebox{-3.0pt}{\includegraphics[width=0.5cm]{images/awaregeodif2.pdf}} generates 3D shapes with various topologies, or
    \item \raisebox{-3.0pt}{\includegraphics[width=0.5cm]{images/awaregeodif3.pdf}} generates 3D shapes with various topologies and information.
\end{itemize}

This categorization represents a progression from low-entropy to increasingly complex representations. Early approaches focused on simple models with fixed parametric ranges, requiring regression of basic geometric properties such as height, width, and depth values~\cite{Free2CAD, Nishida2016}. More sophisticated methods have evolved to generate complex parametric representations or shape programs~\cite{pearl2022geocodeinterpretableshapeprograms, smirnov2021learning}. Other methods do not depend on initial templates or programs; instead, they rely on alternative output representations such as point clouds~\cite{Wu2023sketch2pointcolored}, unsigned distance functions (uSDF)\cite{TONOVitruvio22}, signed distance functions (SDF)\cite{deepsketchmodeling}, and other geometric representations~\cite{sketch2mesh, sketch2model}. The output of these models is typically a single complex 3D shape; however, recent work has achieved part-based representations with complex part geometry in dense objects directly from sketches, as demonstrated by SENS~\cite{sens2024binningerpartawaresketchimplicit} and DoodleYour3D~\cite{bandyopadhyay2024doodle}. However, critical information regarding part materials, costs, weights, and other practical attributes remains largely unexplored in current sketch-to-3D approaches. Indeed, in Vitruvio~\cite{TONOVitruvio22}, the authors envision future models capable of generating Building Information Models (BIMs) in Universal Scene Description (USD) formats, where detailed information about each building component is preserved. This includes maintenance specifications for windows, fire safety properties, thermal transfer coefficients for walls in energy analysis, and other critical parameters necessary for shape optimization and building performance evaluation.

\textbf{Metrics:} Some  $\neuralsketchmethods$ metrics presented here answers some of the initial questions presented in the introduction \ref{introduction}. Others \cite{deepsketchmodeling, Mikaeili_2023_sked, liu2024sketchdreamsketchbasedtextto3dgeneration} evaluate how effectively the methods capture the geometric details of the sketch (see Table \ref{tab:evaluation_metrics}). How accurately must the sketch convey geometric information for the model to interpret it correctly? Finally, does the model account for the context of the sketch, including user preferences for color, texture, materials, and even broader design considerations? Does it understand the user's intent beyond the overall 3D shape? Does it integrate more nuanced elements like physics \cite{cardesignsketch2daerodynamic} for more functional design \cite{guo2024shapecraft}? All these questions need to be translated in both qualitative and quantitative metrics capable to answer them. Therefore, to provide a clearer structure, this section focuses primarily on quantitative and qualitative metrics, as they are more directly related to the overall 3D geometry and connected with the user's intent. Based on insights from previous surveys, we further categorize the \textbf{quantitative} metrics discussion into reconstruction and generation-based metrics, reflecting the distinct evaluation approaches used in existing methods \cite{Yao_2025_CVPRRecontructionVsGeneration}. After that we introduce the qualitative metrics.

\begin{table*}[htbp]
    \centering
    \small  
    \setlength{\tabcolsep}{4pt} 
    \resizebox{\textwidth}{!}{ 
    \begin{tabular}{l|ccccc|ccc}
        \toprule
        \textbf{Paper} & \multicolumn{5}{c|}{\textbf{Qualitative Metrics}} & \multicolumn{3}{c}{\textbf{Quantitative Metrics}} \\
        \cmidrule(r){2-6} \cmidrule(l){7-9}
        & \textbf{People} & \textbf{Age} & \textbf{Test} & \textbf{Topics} & \textbf{Scale} & \textbf{Reconstruction-Based} & \textbf{Generation-Based} & \textbf{Modeling-Based} \\
        \midrule

        Delanoy et al. \cite{Delanoy20203DJohanna} &  6 & - & 2s & -   & Likert & Speed, IoU & -  & -\\
        
        Sketch2CAD \cite{Li2020Sketch2CAD} &  6 & - & 3s & -   & Likert & - & -  & Operator Accuracy\\
        

        GeoCode \cite{pearl2022geocodeinterpretableshapeprograms} &  12 & - & Proc & -   & Likert & CD & -  & Robustness \\

        GA-Sketching \cite{zhou2023gasketchingshapemodelingmultiview} &  8 & 18-28 & - &  U/A & SUS, NASA-TLX & CD, IoU, NC & -  & CD-viewpoint \\

        Control3D \cite{Control3D2023} &  6 & - & v & U   & Pref. Score & - & -  & -\\
        
        CLIPXplore \cite{CLIPXPlore2023sketchshape} & 10 & - & 50v & Q/S & Likert & FID, CD, CLIP-Score & -  & - \\

        LAS-Diffusion \cite{zheng2023lasdiffusion}  & - & - & - & -  & - & FID, CD, IoU,  CLIP-Score & COV, MMD, 1-NNA &  - \\
        
        Sketch3D \cite{zheng2024sketch3d} & 9 & - & 50v & S/A &  Likert & CD-sp2p, CD-sp2t, SSIM &  & - \\

        DreamSketch \cite{liu2024sketchdreamsketchbasedtextto3dgeneration} & 41 & 18-40 & 25 & Q/S/A  & Likert & CD-mean, CD-std  & - &  - \\

        Sketch2Vox \cite{sketch2vox} &  18 & 21-26 & 36s & Q/S   & Likert & mIoU & -  & -\\

        \bottomrule

    \end{tabular}
    }
    \caption{Shows the qualitative and quantitative evaluation metrics used in different $\neuralsketchmethods$ methods. In the `Test' column, subjects evaluated the 3D shape, primarily represented as a 360° video, with the number of videos per shape indicated by the letter (v), or shape (s) generated with the related interface. In other studies, a series of images was used for evaluation, or `Proc' for procedural parameters. The qualitative metrics include: Q for Quality, S for Similarity, A for Alignment, and U for Usability.  `SUS' for System Usability Scale questionnaire and NASA-TLX for the NASA Task Load Index.  The Likert scale used for evaluation ranges from 1 to 5, except for Sketch2Vox and Delanoy et al. that adopted a 1-7 scale. Quantitative metrics reference the division based on the Table \ref{tab:model table_supplementary}, regarding reconstruction, generation, and modeling focus of these methods. Reconstruction metrics include FID (Fréchet Inception Distance), CD (Chamfer Distance), CD-sp2p (Chamfer score between 2 points), CD-sp2t (Chamfer score between shape and target), SSIM (Structural Similarity Index), NC (Normal Consistency), CD-mean (mean Chamfer Distance), CD-std (standard deviation of Chamfer Distance), CD-viewpoint (specific for different viewpoints, with different elevation and azimuth values. Generation metrics incluse COV (Coverage measured with CD and EMD that stands for Earth Moving Distance), MMD (Minimum Matching Distance), 1-NNA (1-Nearest Neighbor Accuracy).  For more details, see the Supplementary material. }
    \label{tab:evaluation_metrics}
\end{table*}

\textbf{Reconstruction-based metrics} evaluate the final model output in direct relation to the dataset used to train $\neuralsketchmethods$ models; here, we report a list of metrics with their strengths and weaknesses in correlating the user's intent with the output. For a more exhaustive and in-depth list, we recommend the reader to review \cite{SVRsurvey}. \textit{Normal consistency ($NC$)}  measures the accuracy and completeness of the shape normals \cite{zhou2023gasketchingshapemodelingmultiview, dualshape2024}. While it performs well for smooth surfaces, $NC$ is sensitive to noise and insufficient for capturing global shape fidelity and topology. To address this limitation, \textit{Earth Mover's Distance (EMD)} and \textit{Chamfer Distance (CD)} have been adopted, as they effectively capture global features and reduce noise sensitivity. They measure the distance between points in two point clouds, and use these measurements to compare the reconstructed point cloud to a ground truth or the cost of moving point density from one point cloud to another \cite{dualshape2024, deepsketchmodeling, zhong2020towards, bandyopadhyay2024doodle, freehandreconstruction, zheng2023lasdiffusion, diffrefforsketchtopointmodeling2023Di}. These metrics capture the geometric output qualities for reconstruction tasks but are sensitive to outliers. There are metrics less prone to this sensitivity \cite{zhong2020towards} such as F-Score, Intersection over Union (IoU), Accuracy, and others. Specifically, F-Score evaluates the overlap between predicted and ground-truth point clouds based on precision and recall. \textit{IoU} is used in retrieval tasks and provides an even more global evaluation. IoU measures how well the defined volumes overlap but overlooks fine details. IoU shows the model's accuracy in retrieving shapes from the dataset. While these metrics emphasize the accurate geometric reconstruction of shapes, they often fail to capture the subtleties of how humans perceive visual quality. Furthermore, these metrics primarily evaluate the generated output in isolation, without establishing a direct connection to the input sketch, assessing output quality by comparing it to similar shapes within the dataset, rather than considering its alignment with the user's original intent. As a result, they are more suited for retrieval-based tasks rather than capturing the nuances of sketch-to-3D generation, where user intent plays a more central role. To address disconnection with user's intent, novel metrics evaluate the similarity also in the image space: perceptual metrics shift the focus from strict geometric accuracy to more subjective qualities, such as realism and visual coherence. These perceptual metrics evaluate the model's output regarding shape fidelity and how well the generated images align with human perception. For example, \textit{Structural Similarity (SSIM)} \cite{Mikaeili_2023_sked, zheng2024sketch3d} measures the similarity between two images, considering luminance, contrast, and structure.  To add additional information and remove further ambiguity sketches have been paired with textual descriptions, and with the advent of CLIP \cite{radford2021learningtransferablevisualmodels, poole2022dreamfusion, dreamfields_zeroshot}, novel metrics now also focus on the alignment of generated images with text inputs \cite{sketchtripletWu2025}, enhancing the evaluation of models in tasks that involve both visual and textual data \cite{CLIPXPlore2023sketchshape, Mikaeili_2023_sked, liu2024sketchdreamsketchbasedtextto3dgeneration, chen2024sketch2nerfmultiviewsketchguidedtextto3d, xie2024sketchguidedcagebased3dgaussian}. \textit{CLIP-score} measures fidelity to the user's input by combining CLIP-T, CLIP-I, and CLIP-S, which assess the similarity between generated shapes and a text prompt in the latent space of the CLIP embedding. CLIP-I measures the similarity between generated shapes and a reference image. The CLIP-Similarity (CLIP-S) measures the consistency of generated shapes with given conditions in the CLIP space. The text-guided mode assesses the similarity between a rendered image of the shape and the target text. The sketch-guided mode compares the rendered sketch of the generated shape with the input sketch \cite{Mikaeili_2023_sked, zheng2024sketch3d, CLIPXPlore2023sketchshape, zheng2023lasdiffusion, chen2024sketch2nerfmultiviewsketchguidedtextto3d}. \textit{Fr\'{e}chet Inception Distance (FID)} \cite{li2024controllabletextto3dgenerationsurfacealigned, CLIPXPlore2023sketchshape, zheng2024sketch3d} evaluates the distance between the distributions of generated shapes and real shapes in a feature space learned by a pre-trained Inception network. 

As established in the literature, the reconstruction-based metrics primarily quantify text alignment with user intent, aiming to recreate shapes that users have previously encountered. However, these metrics do not directly assess the ability to generate entirely novel and original shapes that fall outside the distribution of the training dataset. To address this, \textbf{generation-based metrics} have been introduced, specifically designed to evaluate the diversity of newly generated shapes \cite{Wu2023sketch2pointcolored, bandyopadhyay2024doodle}. These metrics measure how effectively deep generative models sample from learned distributions, capturing variations beyond those present in the original dataset. Hence, it is difficult to compare these distributions, the original dataset distribution, and the generated sample distributions. These distributions should be similar, representing that the generated 3D shapes preserve similar characteristics, but the individual shapes should be slightly different in their details. This requirement is also a limitation of current models, which still do not generalize for out-of-distribution tasks and for this specific evaluation need access to the original training dataset. 
These $\neuralsketchmethods$ face challenges in generating entirely novel 3D shapes, which is closely tied to the inherent challenge of quantifying such novelty. To properly measure the similarity between these two distributions, the $\neuralsketchmethods$ should sample an amount of 3D shapes comparable with the number of 3D shapes present in the test set. Coverage (COV) measures the percentage of test samples covered by generated samples. A test sample is covered if it is the nearest neighbor of a generated sample \cite{hertz2022spaghettieditingimplicitshapes, zheng2023lasdiffusion}. Minimum Matching Distance (MMD) measures the distance between test samples and generated samples \cite{nam20223dldm}. 1-Nearest Neighbor Accuracy (1-NNA) measures how well test samples and generated samples are mixed by penalizing samples from the test set and generated set that have their nearest neighbor in the same set. COV, MMD, 1-NNA cannot be used without a proper test set, as in the case of pre-trained foundation models. 

The main limitation of quantitative metrics is that they are disconnected with the user's intent, and often they do not consider the human interaction with the sketch-based interface, lacking a deep connection between input sketch and output shape. To address this limitation surveys and user studies are used to evaluate $\neuralsketchmethods$ methods \textbf{qualitatively} (see Table~\ref{tab:evaluation_metrics}). They often rely on common perception studies from the Human-Computer Interaction (HCI) community \cite{waysofknowing2014}. For example, DreamSketch \cite{liu2024sketchdreamsketchbasedtextto3dgeneration} uses the one-way ANOVA test and box plots and paired T-test to evaluate its model's performance. Participants used a 1\textendash5 Likert scale to evaluate text faithfulness, sketch faithfulness, geometry quality, texture quality, and overall quality. The participants reviewed 20 cases; each was created by pairing a 3D shape with the sketch and text inputs used to generate it, a similar pairing is performed in Sketch2Vox \cite{sketch2vox}. To evaluate the 3D shapes, some $\neuralsketchmethods$ methods allow a free rotation of the 3D digital object, while other methods present videos \cite{Control3D2023} (the video is captured with a full rotation in azimuth and a fixed elevation of 15 degrees). Evaluating the final 3D shape based on someone else's sketch and text fails to capture the original user's intent and the abstract vision they had in mind while creating the object. This disconnect can obscure how accurately the generated 3D model reflects the initial creative process. Additionally, most of these methods conducted qualitative evaluations on shapes with a similar genus, which may not fully showcase the model's awareness capabilities for topology diversification. DeepSketch \cite{deepsketchmodeling} aimed to evaluate the shape diversity that could be outputted from their model. They selected questions and evaluation metrics guided by these principles: easy to sketch, generality, view differentiability, shape genius higher than 1, and sizable inter-category variance. They generated three shape categories from ShapeNet \cite{chang2015shapenet} with distinctive styles (naive, stylized, and style-unified). They considered professional drawings. However not all $\neuralsketchmethods$ users are professional designers, therefore, how can $\neuralsketchmethods$ be adapted to novice designers. For example, one method that considers amateur or novice sketching style is Doodle your 3D \cite{bandyopadhyay2024doodle}, the authors asked 30 users to draw ten sketches each on the demo canvas (Gradio) and rate the generated shapes based on how well they matched their expectations. This approach addresses a key limitation in the evaluation method of DreamSketch \cite{liu2024sketchdreamsketchbasedtextto3dgeneration} showed in Fig. \ref{SketchDream}, where the participants in the evaluation of the 3D outputs were different from the users providing the initial sketch, making it impossible to track intent accurately. To improve future studies, we recommend that researchers have participants use their novel $\neuralsketchmethods$ methods for generating and evaluating these shapes, ensuring a more direct alignment between input and output. This intent-aligned evaluation captures if the user's concepts and ideas are faithfully translated into 3D shapes. However, the final 3D representation should also be assessed in terms of its editing capabilities, which we do not cover in this report. For instance, Sketch2CAD \cite{Li2020Sketch2CAD} enrolled six novices in their user studies. These participants were asked to evaluate whether the $\neuralsketchmethods$ method accurately translated their intended CAD operations and parameters properly, given the iterative nature of the Sketch2CAD method (see the Supplementary materials \ref{questionnaire} for more). An intent-aligned evaluation example is provided by DualShape \cite{dualshape2024}. DualShape's authors asked, "Did the generated car shell models meet your expectations?" While a simple Yes or No answer may not fully evaluate the method, it is a step toward better understanding and addressing user needs.

\subsection{Output Summary}
\label{conclusionawareouput}

This section presents three key insights derived from analyzing the output representations in Table~\ref{awareevaluation} and evaluation metrics in Table~\ref{tab:evaluation_metrics} across $\neuralsketchmethods$
 methods. Most approaches generate holistic 3D shapes without explicit part decomposition, limiting their utility for downstream editing tasks. Part-aware representations require models to distinguish individual components through one of several strategies: (1) part-level latent codes \cite{Koo_2023_ICCV_SALAD} {\includegraphics[width=0.3cm]{images/Diffusion.png}}, where each part occupies a distinct region in latent space \cite{sens2024binningerpartawaresketchimplicit, hertz2022spaghettieditingimplicitshapes, lee2025pastapartawaresketchto3dshape}; (2) neurosymbolic decomposition \cite{pearl2022geocodeinterpretableshapeprograms, ritchie2023neurosymbolicmodelscomputergraphics}, which represents shapes as executable programs \cite{DAGamendmentinversecontrol21}; or (3) hierarchical attention mechanisms {\includegraphics[width=0.3cm]{images/Transformers.png}} that parse part boundaries. However, these part-aware architectures introduce substantial computational overhead, explaining their limited adoption despite clear advantages for interactive modeling workflows.

Second, current neural sketch methods rarely support multi-candidate generation, hindering iterative design workflows. Only two methods \cite{sanghi2023sketchashapezeroshotsketchto3dshape, sketch2mesh} generate multiple output variations from a single sketch, and none provide auxiliary information such as material cost, and engineering performance. This contrasts with procedural generative design pipelines \cite{proceduralmodeling2024siggraph}, which routinely produce several design alternatives in parallel for comparative evaluation. Procedural approaches achieve this scalability by leveraging computationally efficient parametric models {\includegraphics[width=0.3cm]{images/UNetCNNRegression.png}} that can generate variants through parameter sampling rather than full neural inference. 

Finally, methods are shifting toward texture-enriched representations, but evaluation frameworks lag behind. Recent approaches generate geometry alongside surface appearance \cite{liu2024sketchdreamsketchbasedtextto3dgeneration}, yet existing metrics only assess geometric accuracy. This creates a gap in evaluating whether generated textures align with user intent. Texture synthesis fundamentally requires optimization-based architectures {\includegraphics[width=0.3cm]{images/flexibleview2bw.pdf}} {\includegraphics[width=0.3cm]{images/Optimization.png}} with differentiable rendering pipelines that iteratively refine appearance through gradient descent, enabling explicit control over material parameters. As discussed in Section \ref{discussion}, scenarios like game development, and architecture need perceptual evaluations measured via user preference studies, to properly validate the texture and material information outputted.

\section{Ethical Considerations}
\label{ethics}
Deep sketch-based 3D modeling necessitates a human-centric paradigm: creation begins with a user's sketch and additional information, making full automation neither feasible nor desirable. Key ethical challenges lie in defining the scope of human control: What constitutes a "sufficiently detailed" sketch? What additional information should be provided? And for which audiences and application domains?
Historically, the relationship between sketch and 3D shapes has followed a specific direction: observing three-dimensional reality and translating it into two-dimensional sketches \cite{ArtistasneuroscientistCavanagh2005} (Table \ref{tab:model table_supplementary}, where initial approaches focused on reconstruction). Producing an effective sketch demands mastery of proportion, perspective, and observational acuity, skills codified by figures such as Brunelleschi, Leon Battista Alberti, and Leonardo da Vinci \cite{ArtistasneuroscientistCavanagh2005,DrawtolearnFan2015, learningtodrawtosee2025}. Sketching, like writing, is not merely a communicative act but also a cognitive one \cite{DrawingasversatilecognitivetoolFan2023}; it shapes perception and thought \cite{kosmyna2025brainchatgptaccumulationcognitive, agrawalarenderingeffectiveroutemaps2001, PragmaticInferenceVisualAbstractionFan2020}.
Now, with these novel deep sketch-based 3D modeling approaches, sketches become the basis for generating 3D forms. This inversion raises ethical questions: Should these tools simply output 3D shapes, or should they also foster user learning, helping users refine sketching skills to better communicate ideas and participate in collaborative creation? Should we study and analyze the consequences of these novel tools and how they affect our creative processes, how they augment us?
Implementing a human-centered approach while keeping these questions and considerations in mind could support the development of novel metrics more aligned with keeping humans in control and defining the level of information required as input and the specific information to provide as output.
As these tools become widespread, should $\neuralsketchmethods$'s methods augment humans, reinforcing their sense of community, feeling of belonging, and creative empowerment? If so, how \cite{wang2024phidiasgenerativemodelcreating, Somepalli_2023_CVPR, wang2024attributebyunlearning, OwnDiffusion23, usCopyrightOffice2024}?

\section{Discussion and Future Directions}
\label{discussion}

This section demonstrates how the $\designspace$ design space can guide research in $\neuralsketchmethods$ by presenting concrete application scenarios. Each case study illustrates an integrated workflow from input to output, with entry points at different stages of the IMO-based design space. 



First, consider the challenge of generating production-ready 3D assets from concept sketches in game development scenarios. Researchers can begin by examining the \textbf{Model} dimension, noting the historical shift \cite{sketchmodeling2009, dingsketchmodelingsurvey2016, bonnici2019} to data-driven approaches enabled by breakthroughs in computer vision and graphics (Figure~\ref{fig:chrono}). Using $\designspace$ to constrain the design space (fixing Input to ``multiple sketches and annotations'' typical of professional storyboards and Output to ``high-fidelity 3D assets with various topologies and material information'') reveals a clear architectural trend. Our comprehensive comparison table (Supplementary Material) shows that recent methods predominantly employ pre-trained foundation models with optimization-based generation \cite{zheng2024sketch3d, chen2024sketch2nerfmultiviewsketchguidedtextto3d, zhou2025sketch2symmsymmetryawaresketchtoshapegeneration}, producing Neural Radiance Fields or Gaussian Splatting representations. This landscape analysis directly informs evaluation strategy. For game development, the critical trade-off lies between visual quality and runtime performance: lower polygon meshes enable fluid gameplay but may sacrifice fidelity. Proper assessment therefore requires both quantitative geometry metrics and user-centered perceptual studies that measure whether quality degradation impacts the player experience.

Second, in preliminary design phases, generating multiple plausible options is essential for productive client-designer dialogue. Analyzing the \textbf{Output} categories in Table \ref{awareevaluation} through $\designspace$ reveals a critical gap: while some methods generate multiple geometric variations \cite{sanghi2023sketchashapezeroshotsketchto3dshape} and others provide part-level semantics \cite{Wu2023sketch2pointcolored}, no existing method efficiently produces multiple options with associated metadata (e.g., cost estimates, performance metrics) in real-time (as discussed in Section \ref{conclusionawareouput}). This gap defines a clear research opportunity. Consider an architect sketching an office building layout. An ideal $\neuralsketchmethods$ system would generate multiple design alternatives, each annotated with energy performance, construction cost, material specifications, and buildability constraints. To address this requirement, researchers can use $\designspace$ to specify the target output: \textit{multiple semantic options} represented as lightweight graph-based room layouts with cost metadata. Consulting the framework's model taxonomy identifies discrete diffusion models as strong candidates, given their success in graph-based architectural generation \cite{buildingan, BuildingNet} and discrete representation spaces \cite{inoue2023layoutdmdiscrete, shi2025maskeddiffusiondiscretedata}. With this output configuration fixed, researchers can identify appropriate baselines sharing similar input modalities and establish evaluation metrics. These metrics must assess both diversity across generated options and fidelity to the original design intent \cite{vinker2024sketchagentlanguagedrivensequentialsketch}, alongside domain-specific measures like spatial efficiency and regulatory compliance.

Finally, researchers can also begin from the \textbf{Input} dimension in Table \ref{table:input} by targeting a specific audience \cite{Smartcanvas2016Zheng, secondskin2015layered3D, insituPaczkowski2011sketchcontext}. For example, let's consider democratizing industrial design for novice users, such as users designing their custom footwear. The research problem begins with accommodating \textit{flexible sketch styles} and \textit{multiple views} to resolve geometric ambiguity. Consulting Table~\ref{table:input} reveals that methods supporting both ``Multiple Sketches'' and ``Flexible Style'' are sparse or absent, immediately isolating a high-value research direction. Even without direct competitors, researchers can identify partial baselines sharing one of these traits for comparison. This input constraint propagates downstream: outputs must be fabrication-ready CAD formats \cite{Free2CAD, pearl2022geocodeinterpretableshapeprograms} or Signed Distance Functions better suited for Finite Element Analysis \cite{li2025meshpad, Li2020Sketch2CAD, sketch2mesh, dualshape2024, sens2024binningerpartawaresketchimplicit}, excluding implicit representations like NeRF that resist physical simulation. The model architecture should therefore employ differentiable rendering with optimization losses tuned for manufacturing constraints such as material cost and structural durability.

The diverse requirements across these scenarios, ranging from environmental performance and fabrication constraints to player experience  \cite{dualshape2024, xu2024sketch2sceneautomaticgenerationinteractive, unlu2024eccv}, underscore the necessity for comprehensive \textbf{human-centered metrics}. The evaluation of $\neuralsketchmethods$ presents unique challenges: while existing approaches prioritize reconstruction fidelity, they often fail to capture the critical human factors that determine interface success, such as interaction fluidity and cognitive load. As input technologies evolve from standard tablets to novel brain-computer interfaces—for instance, integrating EEG data \cite{bai2023dreamdiffusiongeneratinghighqualityimages} in evaluation protocols could be considered. Future research should aim to standardize these user experience metrics, establishing benchmarks that balance technical geometric precision with user satisfaction and intent alignment across varying levels of expertise.

\section{Conclusion}
We survey previous work in \textbf{Deep Sketch-Based 3D Modeling} ($\neuralsketchmethods$) and propose $\designspace$, a design space designed around the input-model-output framework. Through $\designspace$ we highlight limitations and find venues for future research. $\neuralsketchmethods$ techniques combine sketch modeling with deep learning to generate 3D representations for design applications. We emphasize the need for a controlled and informed design where the user's intent can be better-captured thanks to novel evaluation metrics. Furthermore, $\designspace$ facilitates the evaluation and categorization of $\neuralsketchmethods$ methods, highlighting their trends: higher level of 3D control and information-richer outputs. Finally, $\designspace$ helps researchers to identify limitations of previous methods, thus offering future research directions, and supports industry-focused readers in selecting the most suitable method and metrics for their use cases.

\section{Acknowledgment}
We acknowledge Karen Liu, Yael Vinker, Judith Fan, Alexandra Bonnicci, Dima Smirnov, Elena Colombini and Paul Guerrero for their feedback. Furthermore, the authors thank  Hannah Luxenberg-Tono, Alberto Tauiti, Simge Girgin, Eleni Alexandraki, Luc Houriez, Allie Cemalovic, Bochen Zhang, Alissa Cooperman, Andrej Krevl, Simi Aluko, Robyn Brinks Lockwood, Lisa Modifica, Veronica Augustina Bot, Samantha Bennett, Tara Srirangarajan, Collin Anthony Chen, and Yulia Gryaditskaya for their inspiration, suggestions, reviews, and support throughout the publication. The work is supported by CIFE Seed Grants, the Wu Tsai Neurosciences Institute and the Koret Human Neurosciences Community Lab (HNCL), Amazon (AWS), NVIDIA, Adobe, Google, the McCoy Family Center for Ethics in Society, and Stanford HAI.




\bibliographystyle{eg-alpha-doi}
\bibliography{egbibsample}
\newpage
\newpage
\clearpage

\section{Supplementary}
\label{supplementary}

In the supplementary, we summarize the different datasets used in the literature for $\neuralsketchmethods$ in Table \ref{datasets_table2} for our $\designspace$ design space. 

\textbf{Disclosure Statement}: None declared

\begin{table*}[htbp]
\centering
\setlength{\tabcolsep}{1pt} 
\renewcommand{\arraystretch}{1.1} 
\small
\resizebox{\textwidth}{!}{
\begin{tabular}{p{3cm}|p{0.5cm}p{0.5cm}p{0.5cm}|p{0.5cm}p{0.5cm}p{0.5cm}|p{0.5cm}p{0.5cm}p{0.5cm}|p{0.5cm}p{0.5cm}p{0.5cm}p{0.5cm}p{0.5cm}p{0.5cm}|p{0.5cm}p{0.5cm}p{0.5cm}|p{0.5cm}p{0.5cm}p{0.5cm}|p{0.5cm}p{0.5cm}p{0.5cm}}
    \toprule
    \multirow{2}{*}{\textbf{Paper}} & 
    \multicolumn{9}{c|}{\cellcolor{green!5}\textbf{Input}} & 
    \multicolumn{6}{c|}{\textbf{Models}} & 
    \multicolumn{9}{c}{\cellcolor{red!5}\textbf{Output}} \\
    
    & \multicolumn{3}{c|}{\cellcolor{green!5}\textbf{Amount}} & \multicolumn{3}{c|}{\cellcolor{green!5}\textbf{View}} & \multicolumn{3}{c|}{\cellcolor{green!5}\textbf{Style}} & 
    \multicolumn{6}{c|}{} & 
    \multicolumn{3}{c|}{\cellcolor{red!5}\textbf{Part Semantic}} & \multicolumn{3}{c|}{\cellcolor{red!5}\textbf{Options}} & \multicolumn{3}{c}{\cellcolor{red!5}\textbf{Geometry}} \\
    
    & 
    \raisebox{-0.3\baselineskip}{\includegraphics[width=0.45cm]{images/flexiblesketch2.pdf}} & \raisebox{-0.3\baselineskip}{\includegraphics[width=0.45cm]{images/flexiblesketch1.pdf}} & \raisebox{-0.3\baselineskip}{\includegraphics[width=0.45cm]{images/flexiblesketch3.pdf}} & 
    \raisebox{-0.3\baselineskip}{\includegraphics[width=0.45cm]{images/flexibleview1.pdf}} & \raisebox{-0.3\baselineskip}{\includegraphics[width=0.45cm]{images/flexibleview2.pdf}} & \raisebox{-0.3\baselineskip}{\includegraphics[width=0.45cm]{images/flexibleview3.pdf}} & 
    \raisebox{-0.3\baselineskip}{\includegraphics[width=0.45cm]{images/flexiblestyle1.pdf}} & \raisebox{-0.3\baselineskip}{\includegraphics[width=0.45cm]{images/flexiblestyle2.pdf}} & \raisebox{-0.3\baselineskip}{\includegraphics[width=0.45cm]{images/flexiblestyle3.pdf}} & 
    \raisebox{-0.3\baselineskip}{\includegraphics[width=0.45cm]{images/UNetCNNRegression.png}} & \raisebox{-0.3\baselineskip}{\includegraphics[width=0.45cm]{images/ImplicitDeepGenerative.png}} & \raisebox{-0.3\baselineskip}{\includegraphics[width=0.45cm]{images/Diffusion.png}} & \raisebox{-0.3\baselineskip}{\includegraphics[width=0.45cm]{images/Transformers.png}} & \raisebox{-0.3\baselineskip}{\includegraphics[width=0.45cm]{images/flexibleview2bw.pdf}} & \raisebox{-0.3\baselineskip}{\includegraphics[width=0.45cm]{images/Optimization.png}} & 
    \raisebox{-0.3\baselineskip}{\includegraphics[width=0.45cm]{images/generativesemantic1.pdf}} & \raisebox{-0.3\baselineskip}{\includegraphics[width=0.45cm]{images/genearivesemantic2.pdf}} & \raisebox{-0.3\baselineskip}{\includegraphics[width=0.45cm]{images/genearivesemantic3.pdf}} & 
    \raisebox{-0.3\baselineskip}{\includegraphics[width=0.45cm]{images/generativeoption1.pdf}} & \raisebox{-0.3\baselineskip}{\includegraphics[width=0.45cm]{images/generativeoption2.pdf}} & \raisebox{-0.3\baselineskip}{\includegraphics[width=0.45cm]{images/generativeoption3.pdf}} & 
    \raisebox{-0.3\baselineskip}{\includegraphics[width=0.45cm]{images/awaregeodif1.pdf}} & \raisebox{-0.3\baselineskip}{\includegraphics[width=0.45cm]{images/awaregeodif2.pdf}} & \raisebox{-0.3\baselineskip}{\includegraphics[width=0.45cm]{images/awaregeodif3.pdf}} \\
    \midrule
    
    Nishida et al. \cite{Nishida2016} & & & \cellcolor{gray!25} & & & & \cellcolor{gray!25} & & & \cellcolor{gray!25} & & & & & & & & \cellcolor{gray!25} & & \cellcolor{gray!25} & & \cellcolor{gray!25} & & \\
    Delanoy et al. \cite{Delanoy20203DJohanna} & \cellcolor{gray!25} & & & \cellcolor{gray!25} & & & & & & \cellcolor{gray!25} & & & & & & & & \cellcolor{gray!25} & & & & \cellcolor{gray!25} & & \\
    ShapeMVD \cite{3dshapereconstructionmvcnn} & \cellcolor{gray!25} & \cellcolor{gray!25} & & \cellcolor{gray!25} & & & \cellcolor{gray!25} & & & \cellcolor{gray!25} & & & & & & & & & & \cellcolor{gray!25} & & \cellcolor{gray!25} & & \\
    Contour3D \cite{Contourbased3DModelingAobo20} & & \cellcolor{gray!25} & & \cellcolor{gray!25} & & & \cellcolor{gray!25} & & & \cellcolor{gray!25} & & & & & & & & & \cellcolor{gray!25} & & & \cellcolor{gray!25} & & \\
    DeepSketch \cite{zhong2020towards} & \cellcolor{gray!25} & & & & \cellcolor{gray!25} & & & \cellcolor{gray!25} & & \cellcolor{gray!25} & & & \cellcolor{gray!25} & & & \cellcolor{gray!25} & & & \cellcolor{gray!25} & & & & \cellcolor{gray!25} & \\
    Sketch2CAD \cite{Li2020Sketch2CAD} & \cellcolor{gray!25} & \cellcolor{gray!25} & & \cellcolor{gray!25} & & & \cellcolor{gray!25} & & & \cellcolor{gray!25} & & & & & & & \cellcolor{gray!25} & & \cellcolor{gray!25} & & & \cellcolor{gray!25} & & \\
    SketchDiff \cite{sketchmodelingdiffrenderer} & & & \cellcolor{gray!25} & \cellcolor{gray!25} & & & \cellcolor{gray!25} & & & & & & & \cellcolor{gray!25} & & & & & \cellcolor{gray!25} & & & \cellcolor{gray!25} & & \\
    FreeHandRec \cite{freehandreconstruction} & & & \cellcolor{gray!25} & & \cellcolor{gray!25} & & & & \cellcolor{gray!25} & & \cellcolor{gray!25} & & & & & & & & \cellcolor{gray!25} & & & & \cellcolor{gray!25} & \\
    Sketch2Model \cite{sketch2model} & & & \cellcolor{gray!25} & & \cellcolor{gray!25} & & & \cellcolor{gray!25} & & \cellcolor{gray!25} & \cellcolor{gray!25} & & & \cellcolor{gray!25} & & & & & \cellcolor{gray!25} & & & \cellcolor{gray!25} & & \\
    Sketch2Mesh \cite{sketch2mesh} & & & \cellcolor{gray!25} & & \cellcolor{gray!25} & & \cellcolor{gray!25} & & & \cellcolor{gray!25} & & & & \cellcolor{gray!25} & & \cellcolor{gray!25} & & & & \cellcolor{gray!25} & & \cellcolor{gray!25} & & \\
    Free2CAD \cite{Free2CAD} & \cellcolor{gray!25} & \cellcolor{gray!25} & & \cellcolor{gray!25} & & & \cellcolor{gray!25} & & & \cellcolor{gray!25} & & & \cellcolor{gray!25} & & & \cellcolor{gray!25} & & & \cellcolor{gray!25} & & & & & \cellcolor{gray!25} \\
    SS2Mesh \cite{bhardwaj2022singlesketch2meshgenerating3d} & & & \cellcolor{gray!25} & \cellcolor{gray!25} & & & & \cellcolor{gray!25} & & \cellcolor{gray!25} & \cellcolor{gray!25} & & & & & & & & \cellcolor{gray!25} & & & & \cellcolor{gray!25} & \\
    GeoCode \cite{pearl2022geocodeinterpretableshapeprograms} & \cellcolor{gray!25} & & & \cellcolor{gray!25} & & & \cellcolor{gray!25} & & & \cellcolor{gray!25} & & & & & & & \cellcolor{gray!25} & & & \cellcolor{gray!25} & & & \cellcolor{gray!25} & \\
    SketchSampler \cite{sketchsampler2022eccv} & \cellcolor{gray!25} & & & \cellcolor{gray!25} & & & & & \cellcolor{gray!25} & \cellcolor{gray!25} & & & & & & & & & \cellcolor{gray!25} & & & & \cellcolor{gray!25} & \\
    LAS-Diffusion \cite{zheng2023lasdiffusion} & & & \cellcolor{gray!25} & & \cellcolor{gray!25} & & & \cellcolor{gray!25} & & & \cellcolor{gray!25} & \cellcolor{gray!25} & & & & & & & & \cellcolor{gray!25} & & & \cellcolor{gray!25} & \\
    Sketch-A-Shape \cite{sanghi2023sketchashapezeroshotsketchto3dshape} & & & \cellcolor{gray!25} & & & \cellcolor{gray!25} & & & \cellcolor{gray!25} & & \cellcolor{gray!25} & & & & & & & & & \cellcolor{gray!25} & & & \cellcolor{gray!25} & \\
    SKED \cite{Mikaeili_2023_sked} & & \cellcolor{gray!25} & & \cellcolor{gray!25} & & & & & \cellcolor{gray!25} & & & & & & \cellcolor{gray!25} & \cellcolor{gray!25} & & & \cellcolor{gray!25} & & & & & \cellcolor{gray!25} \\
    CLIPXPlore \cite{CLIPXPlore2023sketchshape} & & & \cellcolor{gray!25} & & & \cellcolor{gray!25} & & \cellcolor{gray!25} & & & & \cellcolor{gray!25} & \cellcolor{gray!25} & & \cellcolor{gray!25} & \cellcolor{gray!25} & & & & \cellcolor{gray!25} & & & & \cellcolor{gray!25} \\
    D3DSketch+ \cite{chen2023deep3dsketchrapid3dmodeling} & & & \cellcolor{gray!25} & \cellcolor{gray!25} & & & \cellcolor{gray!25} & & & & & & \cellcolor{gray!25} & & & & & & \cellcolor{gray!25} & & & & \cellcolor{gray!25} & \\
    Control3D \cite{Control3D2023} & & & \cellcolor{gray!25} & & & \cellcolor{gray!25} & & & \cellcolor{gray!25} & & \cellcolor{gray!25} & \cellcolor{gray!25} & & \cellcolor{gray!25} & \cellcolor{gray!25} & & & & \cellcolor{gray!25} & & & & \cellcolor{gray!25} & \\
    Re3DSketch \cite{chen2023reality3dsketchrapid3dmodeling} & & & \cellcolor{gray!25} & & \cellcolor{gray!25} & & \cellcolor{gray!25} & & & & \cellcolor{gray!25} & & & \cellcolor{gray!25} & & & \cellcolor{gray!25} & & & \cellcolor{gray!25} & & & \cellcolor{gray!25} & \\
    Sketch2Point \cite{diffrefforsketchtopointmodeling2023Di} & & & \cellcolor{gray!25} & & & \cellcolor{gray!25} & & \cellcolor{gray!25} & & & & \cellcolor{gray!25} & & & & & & & \cellcolor{gray!25} & & & & \cellcolor{gray!25} & \\
    GA-Sketching \cite{zhou2023gasketchingshapemodelingmultiview} & & & \cellcolor{gray!25} & & & \cellcolor{gray!25} & & \cellcolor{gray!25} & & & \cellcolor{gray!25} & & & \cellcolor{gray!25} & & & & & \cellcolor{gray!25} & & & & \cellcolor{gray!25} & \\
    S2PointCol \cite{Wu2023sketch2pointcolored} & & & \cellcolor{gray!25} & & & \cellcolor{gray!25} & & & \cellcolor{gray!25} & & \cellcolor{gray!25} & \cellcolor{gray!25} & & & & & & \cellcolor{gray!25} & \cellcolor{gray!25} & & & & & \cellcolor{gray!25} \\
    Sketch2Vox \cite{sketch2vox} & \cellcolor{gray!25} & & & & \cellcolor{gray!25} & & & \cellcolor{gray!25} & & & & & \cellcolor{gray!25} & & & \cellcolor{gray!25} & & & \cellcolor{gray!25} & & & & \cellcolor{gray!25} & \\
    SketchDream \cite{liu2024sketchdreamsketchbasedtextto3dgeneration} & & & \cellcolor{gray!25} & & \cellcolor{gray!25} & & \cellcolor{gray!25} & & & & \cellcolor{gray!25} & \cellcolor{gray!25} & & \cellcolor{gray!25} & \cellcolor{gray!25} & \cellcolor{gray!25} & & & & \cellcolor{gray!25} & & & & \cellcolor{gray!25} \\
    SENS \cite{sens2024binningerpartawaresketchimplicit} & & & \cellcolor{gray!25} & & & \cellcolor{gray!25} & & & \cellcolor{gray!25} & & \cellcolor{gray!25} & \cellcolor{gray!25} & \cellcolor{gray!25} & & & & \cellcolor{gray!25} & & & \cellcolor{gray!25} & & & \cellcolor{gray!25} & \\
    DY3D \cite{bandyopadhyay2024doodle} & & & \cellcolor{gray!25} & & & \cellcolor{gray!25} & & & \cellcolor{gray!25} & & \cellcolor{gray!25} & \cellcolor{gray!25} & \cellcolor{gray!25} & & & & \cellcolor{gray!25} & & & \cellcolor{gray!25} & & & \cellcolor{gray!25} & \\
    Vitruvio \cite{TONOVitruvio22} & \cellcolor{gray!25} & & & & \cellcolor{gray!25} & & & \cellcolor{gray!25} & & \cellcolor{gray!25} & \cellcolor{gray!25} & & & & & & & & \cellcolor{gray!25} & & & & & \cellcolor{gray!25} \\
    MVControl \cite{li2024controllabletextto3dgenerationsurfacealigned} & & & \cellcolor{gray!25} & & \cellcolor{gray!25} & & & \cellcolor{gray!25} & & & \cellcolor{gray!25} & \cellcolor{gray!25} & \cellcolor{gray!25} & \cellcolor{gray!25} & \cellcolor{gray!25} & & & & \cellcolor{gray!25} & & & & \cellcolor{gray!25} & \\
    SHLine \cite{sketchhiddenline2024} & & \cellcolor{gray!25} & & & \cellcolor{gray!25} & & & \cellcolor{gray!25} & & & \cellcolor{gray!25} & \cellcolor{gray!25} & \cellcolor{gray!25} & & & & & & \cellcolor{gray!25} & & & \cellcolor{gray!25} & & \\
    M3DSketch \cite{zang2024magic3dsketchcreatecolorful3d} & & & \cellcolor{gray!25} & & \cellcolor{gray!25} & & \cellcolor{gray!25} & & & & & & & & \cellcolor{gray!25} & \cellcolor{gray!25} & & & \cellcolor{gray!25} & & & & \cellcolor{gray!25} & \\
    Sketch2NeRF \cite{chen2024sketch2nerfmultiviewsketchguidedtextto3d} & & & \cellcolor{gray!25} & & & \cellcolor{gray!25} & & & \cellcolor{gray!25} & & \cellcolor{gray!25} & \cellcolor{gray!25} & & \cellcolor{gray!25} & \cellcolor{gray!25} & \cellcolor{gray!25} & & & & \cellcolor{gray!25} & & & & \cellcolor{gray!25} \\ 
    DualShape \cite{dualshape2024} & & \cellcolor{gray!25} & & \cellcolor{gray!25} & & & \cellcolor{gray!25} & & & & \cellcolor{gray!25} & & & & & & & & \cellcolor{gray!25} & & & \cellcolor{gray!25} & & \\
    Sketch3D \cite{zheng2024sketch3d} & & & \cellcolor{gray!25} & & & \cellcolor{gray!25} & & & \cellcolor{gray!25} & & \cellcolor{gray!25} & & & \cellcolor{gray!25} & \cellcolor{gray!25} & \cellcolor{gray!25} & & & \cellcolor{gray!25} & & & & \cellcolor{gray!25} & \\
    \bottomrule
\end{tabular}%
}
\caption{Unified categorization of $\neuralsketchmethods$ methods (Section~\ref{aimodel}). This table integrates input characteristics (Amount, View, Style), Model architectures, and Output capabilities (Part semantic, Options, Geometry). The \colorbox{green!5}{Input} and \colorbox{red!5}{Output} sections are highlighted to distinguish the processing stages. Filled cells indicate the presence of a feature or architecture component.}
\label{tab:model table_supplementary}
\end{table*}

\subsection{Architectures}
\label{architectures}

\begin{figure*}[ht]
    \centering
    \includegraphics[width=\textwidth]{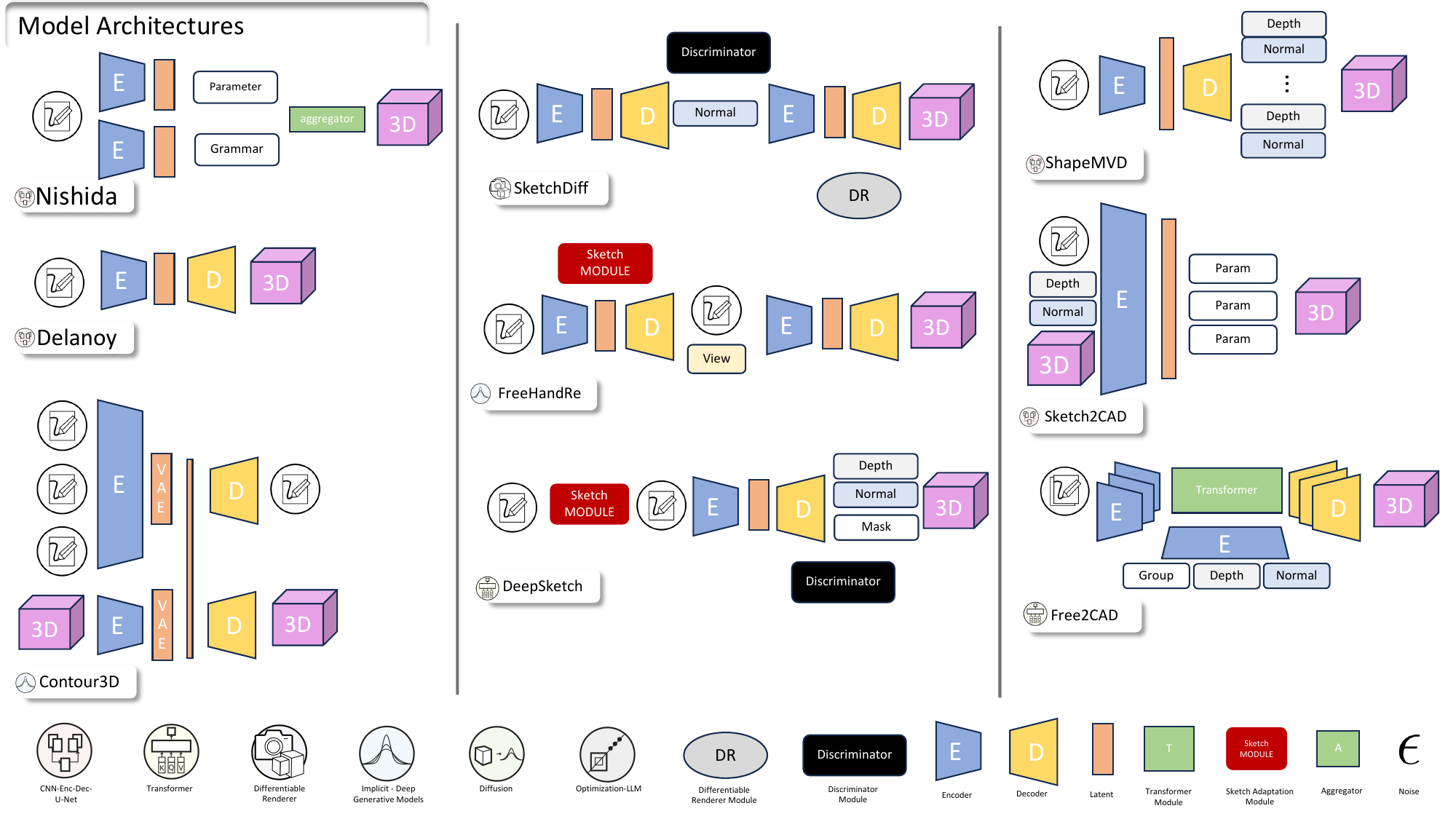}
    \caption{Part 1 illustrates the architectures of AI models for $\neuralsketchmethods$ methods.}
    \label{fig:arch1}
\end{figure*}

\begin{figure*}[ht]
    \centering
    \includegraphics[width=\textwidth]{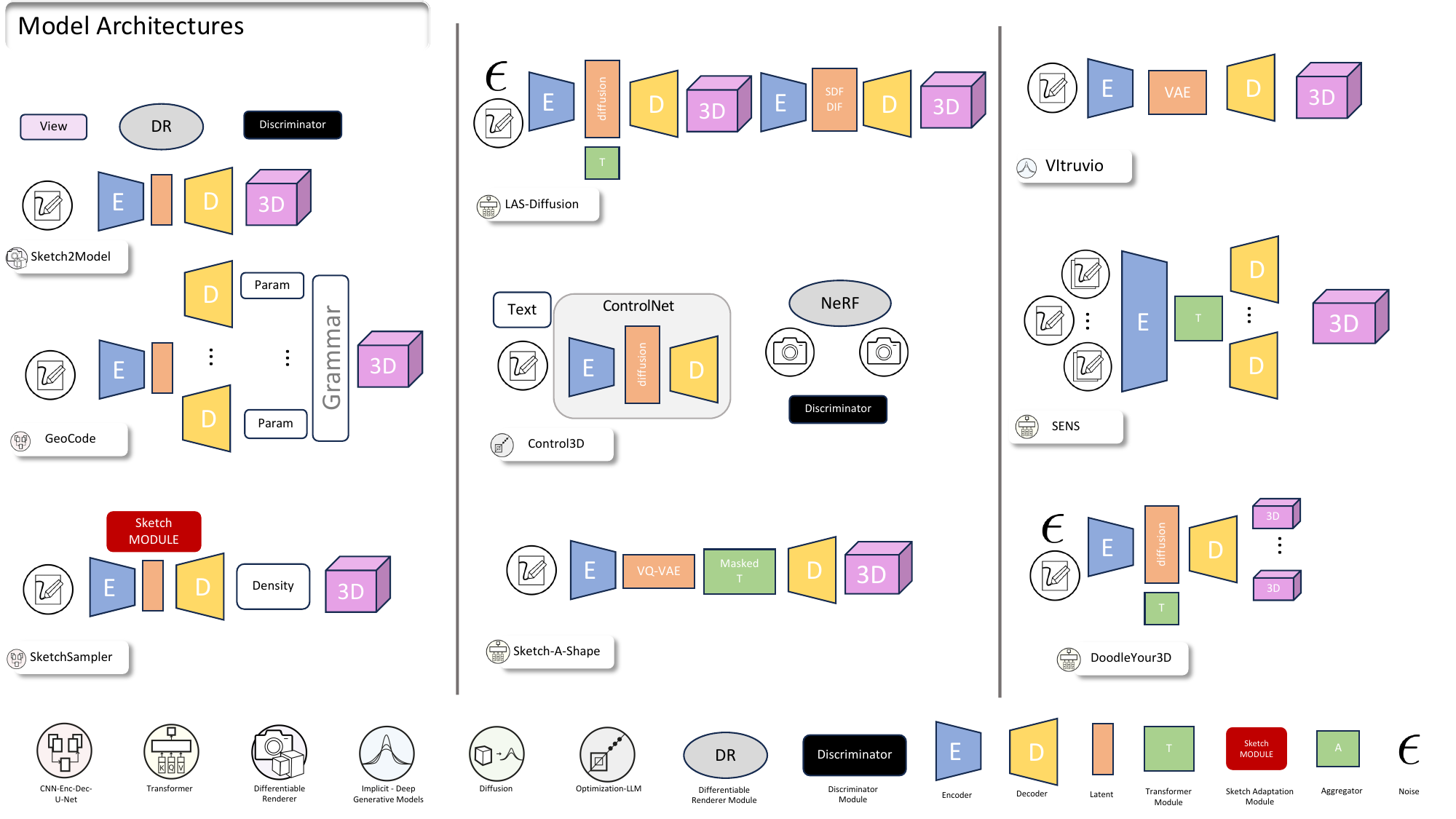}
\caption{Part 2 illustrates the architectures of AI models for $\neuralsketchmethods$ methods.}
\label{fig:arch2}
\end{figure*}

In this section, we display the different model architectures highlighted in Table \ref{tab:model table_supplementary} through different diagrams; see Figures \ref{fig:arch1} and \ref{fig:arch2}

\subsection{Questionnaire}
\label{questionnaire}

Free2CAD evaluated expressiveness and accessibility by asking five novice users to watch a 20-minute tutorial about modeling and then a 15-minute time range to draw from a single isometric view of the object.  

Sketch-a-Shape uses the Amazon SageMaker Ground Truth and crowd workers from the Mechanical Turk workforce. It generates 3D shapes with sketches from different datasets \cite{howhumanssketchobjects2012eitz,sketch2model, wang2019learningrobustglobalrepresentations,quickdraw2017davidha}.

GA-Sketching performed both a usability study and a perceptive study. The usability study was conducted on eight novice 3D modelers aged 18 to 28 years. They watched an 8-minute video and had 15 minutes to familiarize themselves with the system. They had to create two models (a chair and an airplane). After the session, they had to fill in a System Usability Scale (SUS) questionnaire (5-point scale) and a NASA Task Load Index (NASA-TLX) questionnaire to evaluate the mental
demand, physical demand, temporal demand, performance,
effort, frustration.

The perceptive study was conducted through an online questionnaire with 40 novices. 



In Table \ref{questionnaire}, we presented evaluation metrics related to the Quality, Similarity, Alignment, and Usability of the 3D generation methods. Some other methods are also focused on the reconstruction quality. Therefore, they evaluate the Fidelity and Accuracy. Reconstruction-based evaluations focus more on retrieval problems; therefore, the setting and goal differ.

\begin{itemize}
    \item \textbf{Fidelity/Accuracy:}
    \begin{itemize}
        \item How well does the output 3D model match the input sketch? \cite{zang2024magic3dsketchcreatecolorful3d}
        \item Does the generated 3D model represent the sketch? (Y/N) \cite{bhardwaj2022singlesketch2meshgenerating3d}
        \item Are there any features of the sketch missing in the generated 3D model? (Y/N) \cite{bhardwaj2022singlesketch2meshgenerating3d}
        \item Which of the 3D models on the right-hand side best matches the sketch on the left-hand side?
    \end{itemize}
    
    \item \textbf{Quality:}
    \begin{itemize}
        \item How do you think of the quality of the output 3D model? \cite{zang2024magic3dsketchcreatecolorful3d}
        \item What is the quality of the generated 3D model? (on a scale of 1-5) (5-best) \cite{bhardwaj2022singlesketch2meshgenerating3d}
        \item How realistic is the shape? \cite{sens2024binningerpartawaresketchimplicit}
    \end{itemize}
    
    \item \textbf{Similarity:}
    \begin{itemize}
        \item How close is the output to the input sketch (Likert)? \cite{bhardwaj2022singlesketch2meshgenerating3d}
        \item How close to the input sketch does the resulting chair look? \cite{sens2024binningerpartawaresketchimplicit}
    \end{itemize}
    
    \item \textbf{Alignment with Text Prompts:}
    \begin{itemize}
        \item The alignment to the text prompt \cite{Control3D2023}
        \item Text faithfulness \cite{liu2024sketchdreamsketchbasedtextto3dgeneration}
    \end{itemize}
    
    \item \textbf{User Experience and Usability:}
    \begin{itemize}
        \item On a Likert scale from 1 (strongly disagree) to 5 (strongly agree), do you agree that our system allowed them to achieve the buildings they wanted? \cite{Nishida2016}
        \item Do you agree that the system interpreted well the shapes they drew? \cite{Nishida2016}
        \item Choose the better one by jointly considering the following three aspects: (1) the alignment to the text prompt, (2) the fidelity of the visual appearance, and (3) the accuracy of the geometry. \cite{Control3D2023}
        \item Rate the generated shapes based on how they match their expectation using Likert. 
        \item Rate the satisfaction score based on generation speed, shape quality, consistency, and resolution, using Likert.
        \item Rate ''system functional integrity'', ''user interface convenience'', ''generated results satisfaction'', and ''conformity to expectations'' by using Likert .
        \item Answer questions to evaluate the system based on the System Usability Scale (SUS).
        \item Was this approach consistent with your drawing habits? \cite{dualshape2024}
        \item Was it reasonable to use the generation method for car shells?
        \item Did the generated car shell models basically meet your expectations? \cite{dualshape2024}
        \item Are you satisfied with the retrieved models? \cite{dualshape2024}
        \item Answer a System Usability Scale (SUS) questionnaire and a NASA Task Load Index (NASA-TLX) questionnaire to evaluate the usability and workload of our system.
    
    \end{itemize}
\end{itemize}

\begin{table*}[htbp]
\centering
\resizebox{\textwidth}{!}{%
\begin{tabular}{c|ccccccccc}
\textbf{Paper} &
  \textbf{Dataset} &
  \textbf{Category} &
  \textbf{Shapes} &
  \textbf{Sketches} &
  \textbf{Sketch/Shape} &
  \textbf{Viewpoints} &
  \textbf{Usage} &
  \textbf{Format} &
  \textbf{Style} \\ \hline
Nishida 2016 \textit{\cite{Nishida2016}} &
  Parametric Primitives &
  Parametric Primitives &
  procedural &
  grammar &
  - &
  1 &
  TE &
  PNG &
  SSAO \\ \hline
Delanoy 2020 \textit{\cite{Delanoy20203DJohanna}} &
  ShapeCOSEG &
  Chair /vase &
  700 &
  5.600 &
  -- &
  8  &
  TE &
  PNG &
  SC \\
\textit{} &
  Primitives &
  Primitives &
  -- &
  -- &
  -- &
  8 &
  TE &
  PNG &
  SC \\ \hline
Gryaditskaya 2019 \textit{\cite{OpenSketch19}} &
  OpenSketch &
  Product Design &
  12 &
  400 &
  400 / 12 &
  3 &
  -- &
  PNG SVG &
  H 15 PV \\ \hline
Wang 2020 \textit{\cite{freehandreconstruction}} &
  ShapeNet &
  13 categories &
  43.783 &
  390 &
  390 / 130 &
  3 &
  E &
  PNG &
  H 15 wDS \\ \hline
Zhong 2020 \textit{\cite{deepsketchmodeling}} &
  SNCore &
  Chair plane lamp &
  13.141 &
  39.423 &
  3 / 1 &
  48 &
  TE &
  PNG &
  Naive \\
\textit{} &
  SNCore &
  Chair plane lamp &
  13.141 &
  39.423 &
  13 / 1 &
  1 &
  TE &
  SVG &
  Stylize \\
\textit{} &
  SNCore &
  Chair plane lamp &
  13.141 &
  39.423 &
  13 / 1 &
  1 &
  TE &
  PNG  SVG &
  Style-unified \\
\textit{} &
  PS3d &
  Chair plane lamp &
  13.141 &
  PS3d &
  PS3d &
  5 &
  TE &
  PNG  SVG &
  PS3d \\ \hline
Zhang 2021 \textit{\cite{sketch2model}} &
  SN-Sketch &
  13 categories &
  -- &
  1.300 &
  100 / 1 &
  20 &
  E &
  PNG &
  H copy wDS \\
\textit{} &
  SN-Sketch &
  13 categories &
  -- &
  1.300 &
  100 / 1 &
  21 &
  TE &
  PNG &
  Canny \\ \hline
Guillard 2021 \textit{\cite{sketch2mesh}} &
  ShapeNet &
  Car chair &
  6.819 &
  -- &
  -- &
  16 &
  TE &
  PNG &
  Canny \\
\textit{} &
  ShapeNet &
  Car chair &
  6.819 &
  -- &
  -- &
  1 &
  TE &
  PNG &
  SketchFD \\
\textit{} &
  ShapeNet &
  Car chair &
  6.819 &
  -- &
  -- &
  1 &
  E &
  PNG &
  SC \\
\textit{} &
  ShapeNet &
  Car &
  -- &
  113 &
  3 / 1 &
  1 &
  E &
  PNG &
  Students 5 tracing \\
\textit{} &
  PS3d &
  Chair &
  -- &
  177 &
  3 / 1 &
  1 &
  E &
  PNG &
  P wDS \\ \hline
Vitruvio \textit{\cite{TONOVitruvio22}} &
 Manhattan 1k  &
  Building &
  1.000 &
  24k &
  24 / 1 &
  24  &
  TE &
  PNG &
  Canny \\ \hline

  Sketch3D \textit{\cite{zheng2024sketch3d}} &
 ShapeNet-Sketch3D  &
  10 ShapeNet+Text &
  11k &
   220k &
  20 / 1 &
   20 &
   &
   PNG &
   Canny
   \\ \hline

    SketchDream, MVControl  &
    Objaverse &
    3D Obj &
    400k &
    &
    4 / 1&
    30 &
    TE &
    PNG & Canny
    \\ \hline

    Sketch2NeRF \cite{chen2024sketch2nerfmultiviewsketchguidedtextto3d} &
    OmniObject3D-Sketch \cite{wu2023omniobject3d} &
    3D Obj - 20 categories &
    - &
    - &
    24 / 1&
    - &
    TE &
    PNG & 
     HED
   \\ \hline
   
      DoodleYour3D \cite{bandyopadhyay2024doodle} &
    ShapeNet \cite{chang2015shapenet} &
    Chair &
    6,755 &
    81,060 &
    12 / 1 &
    6 &
    TE &
    PNG SVG & 
     CLIPasso + InfDraw \cite{vinker2022clipasso,informativedrawings} 
   \\  \hline

   SENS \cite{sens2024binningerpartawaresketchimplicit} &
    ShapeNet \cite{chang2015shapenet} &
    Chair plane lamp &
    9,363 &
    224,712 &
    24 / 1 &
    6 &
    TE &
    PNG SVG & 
     CLIPasso + others \cite{sens2024binningerpartawaresketchimplicit} 
   \\  \hline

     LAS-Diffusion \cite{zheng2023lasdiffusion} &
    ShapeNet \cite{chang2015shapenet} &
    5 Cat. &
    -&
    - &
    50 / 1 &
    10 &
    TE &
    PNG & 
     Canny
   \\  \hline

\end{tabular}%
}
\caption{Datasets for monocular sketch reconstruction. Few of these works involved professionals ($P$) or students in sketching on different digital surfaces ($wDS$), such as ISKN Slate 2 and iPad Pro. Some of them were also capable of tracking strokes. $H$ indicates that a heterogeneous pool of people have been asked to sketch: students, professionals, and others. $Ps3d$: ProSketch3d \cite{deepsketchmodeling}, $SC$: suggestive contours, $T$: Test , $TE$: Test and Evaluation.}
\label{datasets_table2} 
\end{table*}


\end{document}